# Using Electronegativity and Hardness to Test Density Functional Universality

Klaus A. Moltved and Kasper P. Kepp*

*Technical University of Denmark, DTU Chemistry, Building 206, 2800 Kgs. Lyngby, DK – Denmark.*

* Corresponding author: Phone: +045 45 25 24 09.  E-mail: kpj@kemi.dtu.dk

**Abstract**

Density functional theory (DFT) is used in thousands of papers each year, yet lack of universality reduces DFT's predictive capacity, and functionals may produce energy-density imbalances. The absolute electronegativity ($\chi$) and hardness ($\eta$) directly reflect the energy-density relationship via the chemical potential $\partial E/\partial N$ and we thus hypothesized that they probe universality. We studied $\chi$ and $\eta$ for atoms Z = 1−36 using 50 diverse functionals covering all major classes. Very few functionals describe both $\chi$ and $\eta$ well. $\eta$ benefits from error cancelation whereas $\chi$ is marred by error propagation from IP and EA; thus almost all standard GGA and hybrid functionals display a plateau in the MAE at ~0.2-0.3 eV for $\eta$. In contrast, variable performance for $\chi$ indicates problems in describing the chemical potential by DFT. The accuracy and precision of a functional is far from linearly related, yet for a universal functional we expect linearity. Popular functionals such as B3LYP, PBE, and revPBE, perform poorly for both properties. Density sensitivity calculations indicate large density-derived errors as occupation of degenerate p- and d-orbitals causes "non-universality" and large dependency on exact exchange. Thus, we argue that performance for $\chi$ for the *same* systems is a hallmark of universality by probing $\partial E/\partial N$. With this metric, B98, B97-1, PW6B95D3, APFD are the most "universal" tested functionals. B98 and B97-1 are accurate for very diverse metal-ligand bonds, supporting that a balanced description of $\partial E/\partial N$ and $\partial E^2/\partial N^2$, via $\chi$ and $\eta$, is probably a first simple probe of universality.

**Introduction**

Due to its combined computational speed and general accuracy, Density Functional Theory (DFT) is the main methodology used to study electronic structure of larger molecular systems, which tens of thousands of papers using the methods every year.[1,2] Often it is the only reasonable option due to the scaling of cpu requirements with system size for more accurate methods.[2–4] However, the acronym "DFT" covers an enormous range of functionals of distinct philosophies and designs, with different physical conditions fulfilled, parameterization ranges, and mathematical forms.[2,3,5] This fragmentation prevents comparison between studies and obscures the significance of conclusions based on a few functionals, since tests of sensitivity to method choice are rare despite the necessity of performing them.[4] Separating error contributions beyond the functional itself from e.g. entropy, zero-point energies, relativistic effects, dispersion effects, or basis set deficiencies is a necessary step in this process.[4,6–8]

After error analysis we can search for a "universal functional".[9–11] One can define universality either mathematically as a pure functional that satisfies all possible fundamental restraints, or practically, as a functional that performs accurately across the broadest ranges of properties of molecular systems by mimicking *both* the ground state energy and density of all relevant external potentials. Straying[12] can be defined as imbalances in a thermochemical cycle of trial densities, either along the density or energy path (**Figure 1A**).[13] Some prefer less parameterized functionals that satisfy physical bounds,[5,12,14] whereas others prefer careful and extensive parameterization towards high-quality thermochemical data.[15,16] Regardless of philosophy, the functional form is central, and different "rungs" represent increasing mathematical complexity.[4,17,18] Universality requires the energy and density to be described in a balanced way.[9] One may assume that less empirical functionals are more energy-density balanced (i.e. closer to the central diagonal line of **Figure 1A**) relatively to empirical functionals that have not considered this balance explicitly, as so far the case.

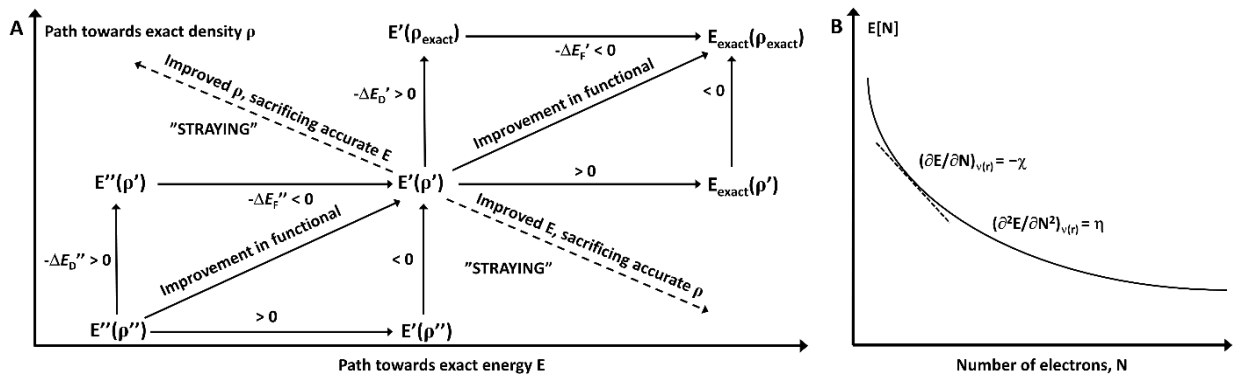

**Figure 1. Energy-density relationships, exactness, and electronegativity.** (**A**) Energy-density relationships via trial functionals towards the exact functional, based on Ref.[13]. The sign of each process follows from the Hohenberg-Kohn variational principle.[9] (**B**) Definitions of electronegativity $\chi$ and hardness $\eta$.[19,20]

Medvedev et al.[12] recently used atomic ions to identify energy-density imbalances. Although they did not actually compare the energies and densities for the same systems,[21] their work and that of Burke[22,23] show that the balance between density-derived ($\Delta E_D{'}$ in **Figure 1A**) and functional-derived errors ($\Delta E_F{'}$) is important (please note that as written in **Figure 1A**, $\Delta E_D{'} < 0$ and $\Delta E_F{'} > 0$). There are very many ways to estimate density errors: At the nucleus, at some radial distance, at some point in space, at the tail, for a sum of points, or the density-weighted root-mean-squared deviation from the "exact" density over a grid; this makes error estimates[12] very conditional.[21]

Instead, Burke suggested to measure density errors via their impact on the energy relative to the exact density, giving a single, unambiguous value for each density, $\Delta E_D{'}$.[22] Unfortunately, despite its theoretical adequacy, this is impractical since exact densities are mostly obscure: Experimental electron densities from diffraction do not have the accuracy required, with "atomic" resolution (~1 Å) being completely useless, although this could change in the future for small systems. Furthermore, any "exact" estimate from quantum chemistry is beyond reach for almost all systems. Even for the closed-shell $1s^2$ and $1s^22s^2$ systems studied by Medvedev et al.,[12] a high-level quantum chemical density may not be "exact", since errors mainly evaluated at the nucleus will be affected by violations of the nuclear cusp condition of all quantum-chemical methods that use Gaussian basis functions.

To solve this issue, we suggested using density sensitivity analysis by replacing the exact density with one or more densities of distinct known functionals.[13] This simple protocol measures the global effect of variations in densities by computing $E´´[ρ´]−E´´[ρ´´]$ where $ρ´´$ is the studied functional's self-consistent converged density, $E´´$ is an energy evaluation with the functional, and $ρ´$ is a converged Hohenberg-Kohn trial density (bottom-left cycle in **Figure 1A**) rather than the elusive $E[ρ´]−E[ρ_{exact}]$ (top-right cycle of **Figure 1A**). If necessary, the average result for a range of diverse "trial" densities can be computed.[13] This protocol, which is easily applied in programs such as Turbomole, is closely related to the Hohenberg-Kohn variational principle, which determines the signs of the processes in the thermochemical cycles in **Figure 1A**.[13] We recommend using two distinct theories to produce the densities such as a hybrid GGA (e.g. PBE0) and a local density approximation (LDA). $ΔE_D´$ is often small ("normal systems") since the functional's treatment of the density commonly matters more than the finer details of the density itself. This is the case both the systems studied by Medvedev et al. once the energy impact of the reported density errors is evaluated, but also for chemically relevant systems.[13,21] Thus, the interesting question is not whether there are density errors but whether they are chemically significant.[13,22–24]

To move forward on how to obtain energy-density balanced DFT, we note that in any chemical process, electrons are moved from one place to another such the chemical potential is equal everywhere in a molecular system at equilibrium.[20,25] The electronegativity χ is defined as the negative of the chemical potential at constant external potential v(r), as seen in **Figure 1B**:[20]

$$χ = −(∂E/∂N)_{v(r)} \sim (IP + EA)/2 \qquad (1)$$

The last expression with the ionization potential (IP) and electron affinity (EA) is the finite-difference approximation for discrete electron numbers, which is identical to the Mulliken electronegativity.[26] The hardness can be defined similarly as:[3,25,27,28]

$$\eta = (\partial E^2/\partial N^2)_{v(r)} \sim (IP - EA) \qquad (2)$$

Since both the uptake and removal of electrons must be well-described at the equilibrium chemical potential, and since these changes relate to the energy-density balance, we hypothesized that DFT may feature imbalanced chemical potentials. In the finite-difference approximation, this amounts to imbalances in $\chi = (IP + EA)/2$ and $\eta = IP - EA$. This translates to the requirement that DFT should work well for the EA and IP for the *same* systems simultaneously, rather than for different systems often tested during DFT development. $\chi$ may then capture the energy-density relationship that has been so debated recently.[12,13,21,24] Since $\chi$ excellently describes many types of reactivity and diverse chemical bond strengths,[29,30] and since the Pauling and Mulliken electronegativities correlate strongly,[31] we propose that performance for $\chi$ is a "descriptor of universality", which we think may be of interest to further development of DFT.

To explore this idea, we computed the hardness and electronegativity for all elements from $Z = 1-36$ for 50 diverse density functionals. Because of their importance, the atomic IPs and EAs have been studied massively with DFT.[8,16,32–39] The absolute $\chi$ and $\eta$ for a range of molecules and four atoms were studied by Dixon and co-workers using several functionals.[35] Hybrid DFT is somewhat less accurate for IPs and EAs than the G2 method, but generally performs well for both on average. There has been confusion regarding DFT's ability to describe anions and thereby EAs.[40,41] With large basis sets including diffuse functions DFT is fully capable to do so in broad benchmarks.[33,41] However, the required balance of IP and EA for the *same* systems is typically not obtained despite average good performance: For example, the IP of oxygen, of major importance to catalysis, displays errors up to 0.4−1 eV,[33] which can only be remedied by special corrections.[16]

The novelty in the present work lies mainly in linking $\chi$ and $\eta$ to the energy-density balance, suggesting their use as probes of universality, and offering a systematic study of 50 functionals, including both modern empirical and non-empirical density functionals, to test these ideas. We particularly emphasize the relationship between the accuracy and precision of DFT, which turns out to vary greatly when describing $\chi$ and $\eta$.

**Methods**

**Energy computations and functionals studied**

We used the Gaussian 16[42] software for all computations. IPs and EAs were computed for the 36 elements Z = 1−36 (H-Kr). This choice was made because relativistic spin-orbit coupling affects the EA and IP of heavier atoms, and these are hard to evaluate accurately, both for energies[43–45] but in particular their effect on densities. Our data set covers both the s- and p-elements as well as the 3d series, where major deviations from universality are expected to occur.[46,47] Spin states for the atoms and atomic ions were taken from NIST[48] and are summarized in **Table S1.** For a complete discussion of electron affinities, see the review by Schaefer and co-workers.[32] Experimental values for the IPs and EAs were taken from the CRC Handbook of Chemistry and Physics[49] and are summarized in **Table S2.**

Energies were computed using the 50 functionals of **Table 1**, with literature references, type, and amount of HF exchange (for the hybrid functionals) noted. These 50 functionals were chosen to 1) include many popular DFT functionals[50], 2) span many design types, and 3) include both older and newer functionals, as time has been claimed to work against universality because the energy is increasingly over-emphasized.[12] We note that we use basis sets that provide chemical accuracy for CCSD(T).[13] The electron affinities depend on loosely bound anion states which may be difficult to describe,[38,51,52] probably because of basis set limitations rather than failure of DFT itself.[40,41] We used aug-cc-pV5Z[53] except for K and Ca, which used def2-QZVPPD.[54] All electronic energies are summarized in **Tables S3-S11.** The IPs were calculated as:

$$IP = E(X^+) - E(X) \qquad (3)$$

$E(X)$ and $E(X^+)$ are the single-point energies of neutral X and its mono-cation $X^+$. Correspondingly, the EAs were calculated as:

$$EA = -\bigl(E(X^-) - E(X)\bigr) \qquad (4)$$

Here, $E(X^-)$ is the single-point energy of the monoanion of element X.

**Table 1. Overview of the 50 exchange-correlation functionals studied in this work.**

| Functional | Type | % HF exchange | References |
|---|---|---|---|
| APFD | Hybrid GGA | 23 | 55 |
| B1B95 | Hybrid GGA | 28 | 56 |
| B2PLYP | Double Hybrid | 53 | 57 |
| B3LYP | Hybrid GGA | 20 | 58–60 |
| B3P86 | Hybrid GGA | 20 | 61,62 |
| B97-1 | Hybrid GGA | 21 | 63 |
| B97-2 | Hybrid GGA | 21 | 64 |
| B97-D | GGA | 0 | 65 |
| B98 | Hybrid GGA | 22 | 66 |
| BHandH | Hybrid GGA | 50 | 67 |
| BHandHLYP | Hybrid GGA | 50 | 67 |
| BLYP | GGA | 0 | 61,62 |
| BMK | Hybrid meta GGA | 42 | 68 |
| BP86 | GGA | 0 | 61,62 |
| CAM-B3LYP | Range-separated | 19-65 | 69 |
| G96PBE | GGA | 0 | 14,70,71 |
| HCTH407 | GGA | 0 | 63,72,73 |
| HSE06 | Range-separated | 0-25 | 74–80 |
| LC-wHPBE | Range-separated | 0-100 | 76,81–83 |
| M06 | Hybrid meta GGA | 27 | 84 |
| M06-L | Meta GGA | 0 | 85 |
| M11 | Range-separated | 43-100 | 86 |
| M11-L | Meta GGA | 0 | 86 |
| MN15 | Hybrid meta NGA | 44 | 87 |
| MN15-L | Meta NGA | 0 | 88 |
| mPW1PW91 | Hybrid GGA | 25 | 89–91 |
| mPW3PBE | Hybrid GGA | 25 | 14,89 |
| N12-SX | Screened exchange |  | 92 |
| O3LYP | Hybrid GGA | 12 | 93 |
| OLYP | GGA | 0 | 59,94 |
| OP86 | GGA | 0 | 62,94 |
| OPBE | GGA | 0 | 14,94 |
| OVWN | GGA | 0 | 94,95 |
| PBE | GGA | 0 | 14 |
| PBE0 | Hybrid GGA | 25 | 14,96 |
| PW6B95 | Hybrid meta GGA | 28 | 97 |
| PW6B95D3 | Hybrid meta GGA | 28 | 97 |
| RevPBE0 | Hybrid GGA | 25 | 96,98 |
| RevTPSS | Meta GGA | 0 | 99 |
| RPBE | GGA | 0 | 100 |
| SLYP | GGA | 0 | 61,101 |
| SVWN | LSDA | 0 | 95,101 |
| SVWN5 | LSDA | 0 | 95,101 |
| tHCTH | meta GGA | 0 | 102 |
| tHCTHhyb | Hybrid meta GGA | 15 | 102 |
| TPSS | Meta GGA | 0 | 103 |
| TPSSh | Hybrid meta GGA | 10 | 103 |
| VSXC | Meta GGA | 0 | 98 |
| wB97XD | Range-separated | 22-100 | 104 |
| wPBEhPBE | GGA | 0 | 14,75,76,105 |
| X3LYP | Hybrid GGA | 22 | 92 |

**Computing density-derived errors**

Burke's group has defined systems with large density-derived errors $\Delta E_D'$ as "abnormal".[22–24] A practical threshold for abnormality was suggested to be chemical accuracy, i.e. 4 kJ/mol.[13] Burke has advocated[23,38] using a DFT non-consistent single-point energy calculation on the self-consistent HF density (HF-DFT) for abnormal electronic systems, and found it to work well for EAs.[38] Since pathological HF densities are common to systems with static correlation, as is in particular the case for the 3d-series, using HF densities may overestimate DFT abnormality.[46] Instead we follow the protocol of using density sensitivity analysis by testing how reasonable variations in density affect the total energy.[13] As reasonable variations, we only consider DFT-derived densities, spanning from LDA to hybrid DFT, and favor non-empirical functionals.

We followed the protocol described previously.[13] Specifically, we used PBE as a widely used non-empirical functional, and evaluated the non-consistent PBE energy on the converged HF densities (referred to as HF-PBE), SVWN densities (SVWN-PBE), and PBE0 densities (PBE0-PBE) by setting the number of iterations for the SCF procedure to 1 and changing the density convergence threshold to $10^7$ au in Turbomole. The obtained energies can be seen in **Table S12**. Sim et al.[24] suggested a simple specific metric, $S(E_{xc})$, for quantifying abnormality that involves only the non-consistent energies computed by PBE on the densities from HF and SVWN[95,101]:

$$S(E_{xc}) = E_{PBE(SVWN)} - E_{PBE(HF)} \qquad (5)$$

The first acronym in subscript represents the energy calculations, and acronyms in parenthesis represent the methods used to compute the self-consistent densities. As example, for EAs,

$$S^{EA}_{PBE(SVWN,HF)} = E_{PBE(SVWN),X^-} - E_{PBE(HF),X^-} - E_{PBE(SVWN),X} + E_{PBE(HF),X} \qquad (6)$$

which can also be written as:

$$S^{EA}_{PBE(SVWN,HF)} = EA_{PBE(SVWN)} - EA_{PBE(HF)} \qquad (7)$$

Since HF densities may overestimate density sensitivities, we use our preferred metric:

$$S^{EA}_{PBE(SVWN,PBE0)} = EA_{PBE(SVWN)} - EA_{PBE(PBE0)} \qquad (8)$$

These values applied to both IPs and EAs are summarized in **Tables S13-S14**.

**Results and Discussion**

**Hypothesis: Using χ = -∂E/∂N to probe universality**

The main hypothesis of the present work is that the chemical potential, $\partial E/\partial N$ may not be balanced in most density functionals due to an over-emphasis on energies of neutral and cationic systems (enthalpies of formation and ionization energies), rather than densities and electron-rich systems. If so, the gain of electrons will be less well described than the loss of electrons. The general performance for anions has been claimed to be challenged,[40,51] and diffuse densities of anions may in some cases be abnormal (i.e. contribute large density-derived errors to the result).[23,38,51] However, DFT performance for anions in broader benchmarks is generally quite good, as measured by EAs.[32,33,86] Cancellation of errors in the density and the functional (i.e. $\Delta E_D'$ and $\Delta E_F'$ of **Figure 1A**) caused by parameterization could possibly obscure the problem. Since we cannot generally measure the errors in the densities, using density sensitivity analysis is a possible alternative,[13,24] as applied below, revealing large density errors in the p- and d-blocks.

However, even with cancellation of errors in energy and density, systems may experience energy-density imbalance in terms of adding or removing electrons near the chemical potential $\partial E/\partial N$ (**Figure 1B**) and its variation with N, $\partial E^2/\partial N^2$, simply because the X, $X^+$, and $X^-$ systems were not considered together when developing and parameterizing DFT. Since the finite difference approximations to $\partial E/\partial N$ and $\partial E^2/\partial N^2$ are $-\chi$ and $\eta$,[20,106] these properties may probe DFT universality. To test this, we studied the atoms of a significant part of the periodic table (Z = 1−36) to avoid artefacts of relativistic energy but still probe the s-, p-, and d-block, and assess errors in χ and η directly against high-quality experimental data. An inspiration for our hypothesis is that DFT errors tend to scale monotonically with effective nuclear charge, i.e. functionals are usually more accurate for either the left or right side of both the p- and d-blocks.[47,107] Considering that χ relates to effective nuclear charge, this implies an imbalance in the chemical potential, $\partial E/\partial N$. Even beyond this linear error effect, different functionals are accurate for different types of bonds and electronic configurations,[17,46,47,108] and errors from d-orbital occupation and spin scatter massively.[109–114] These fundamental issues prevent DFT from reaching universality for the periodic table and even for different electronic processes involving only a few types of atoms.[114]

**Distinct performance of 50 functionals for electronegativity and hardness**

**Figure 2** summarizes the mean absolute errors (MAE) and mean signed errors (MSE) in the computed χ (**Figure 2A**) and η (**Figure 2B**) of the 36 atoms, for the 50 studied density functionals, ranked according to MAE. We note that all values are without division of IP + EA by 2, to compare χ and η more fairly. LDA performs much worse for χ than for η (MAEs of 1.41 and 1.52 eV for SVWN and OVWN not shown due to scale in **Figure 2A**), which we interpret as a major cancellation of errors in η, as EA is subtracted from the IP. Although uninteresting in terms of performance, the LDA functionals illustrate this most clearly. The corresponding performance for the IPs and EAs is summarized in **Figure S1**. The MSEs of both IPs and EAs are approximately normal-distributed around zero, although slightly skewed towards too large EAs (**Figure S2**), which indicates that we have sampled the DFT world well for the problem at hand.

Similarly, even functionals that perform decently for both IP and EA for the same systems may show larger errors in χ due to error propagation, since the IP and EA are added when calculating χ. For example, BMK is the best functional for η but average for χ, because the decent MAEs for IPs and EAs (0.13 eV) carry systematic errors that add in χ. Specifically, since η = $E(X^+) + E(X^-) - 2\ E(X)$ and χ = $E(X^+) - E(X^-)$, η reflects a disproportionation reaction, i.e. two single-electron transfers, whereas χ reflects two-electron transfer (the sum of the first and second ionization energy of the anion state). The error cancellation is therefore expected to be larger in η. Indeed, we see that many standard functionals plateau at a MAE of ~0.2-0.3 eV (**Figure 2B**). In contrast, DFT performance for χ varies considerably: Typical GGAs show errors from 0.25 to 0.50 eV for χ, whereas some hybrids and meta functionals display lower MAEs all the way down to

~0.1 eV. Similarly, the standard deviation of the errors (gray thin bars in **Figure 2**) oscillates wildly for χ (**Figure 2A**) but is approximately symmetric around zero for η (**Figure 2B**).

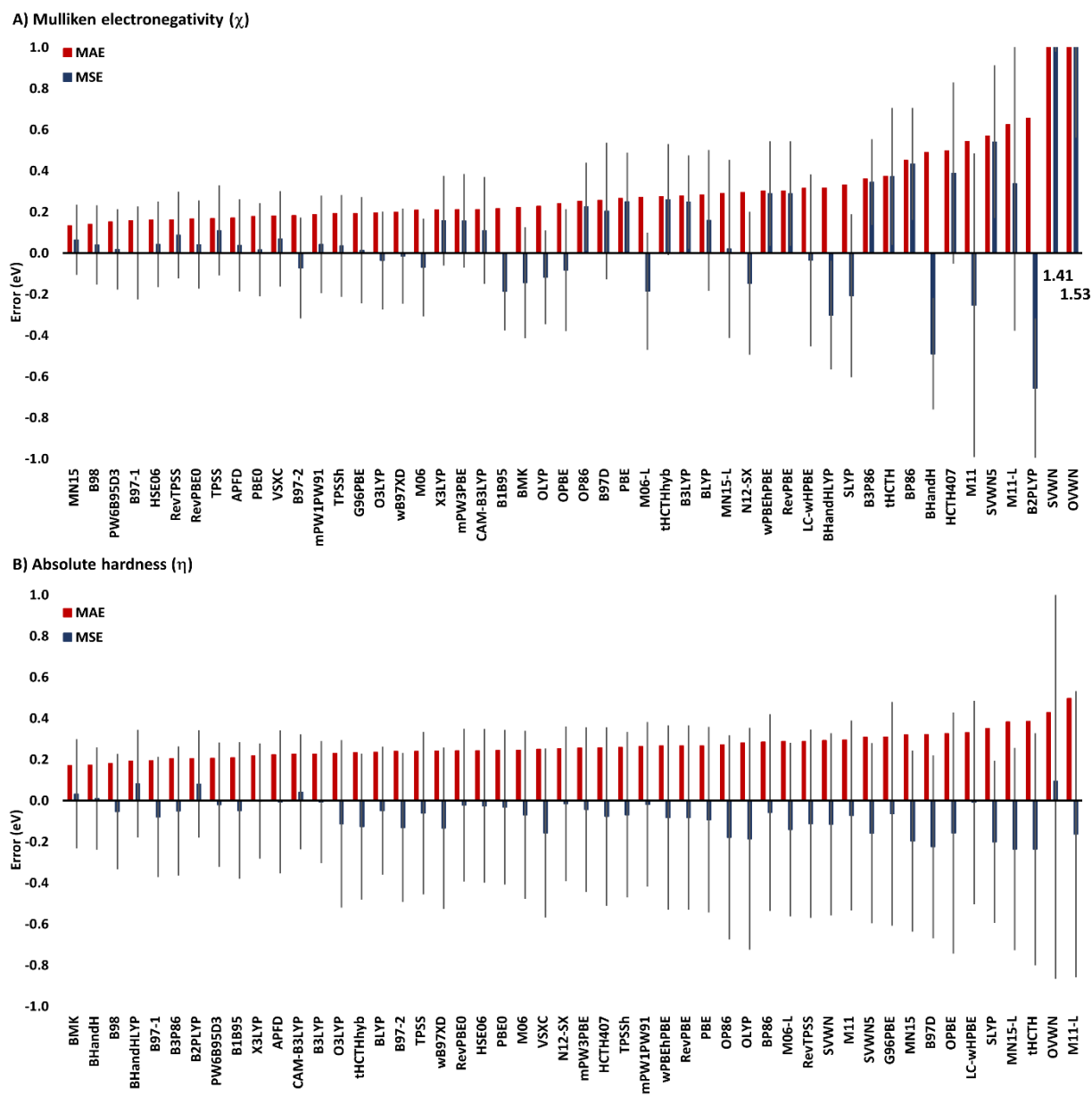

**Figure 2. Errors vs. experiment (in eV) of 50 density functionals applied to atoms Z = 1−36.** The histograms show the MAEs (red bars), the MSEs (blue bars), and the standard deviations of the errors (thin gray bars). (**A**) Mulliken electronegativity (χ). (**B**) absolute hardness (η). In (**A**), the maxima of SVWN and OVWN (MAEs = 1.41 and 1.52 eV) have been left out for better viewing and for putting the errors in χ and η on the same scale.

Thus, apparent good performance for IPs and EAs, as commonly tested, hides major error cancellations from treating most of the electronic structure in the same way, yet $\chi$ reveals these cancellations. Since $\chi \approx -\partial E/\partial N$ we relate this to a commonly poor description of the chemical potential, which again implies a weakness in the incremental energy-density relationship, i.e. change in energy with small changes in electron count at fixed external potential, or similarly, the electron density within a confined part of the system assuming a fixed volume. As the derivative of the E/N relationship in **Figure 1B**, it implies that DFT is unbalanced in terms of adding or subtracting electrons from the same systems. A simple way to solve this problem is to ensure that the systematic errors (not MAEs) are close to zero for both IPs and EAs of the *same* systems (not different systems, as commonly applied in benchmark and parameterization data sets), by minimizing errors in $\chi$ which is much more sensitive than EA and IP separately. Still, this only ensures that functional-derived errors are minimized, i.e. that $\Delta E_F´ \rightarrow 0$; it does not ensure that the density-derived errors $\Delta E_D´$ are not propagated at the same time.

We note from **Figure 2** that a handful of functional that perform particularly well for both $\chi$ and $\eta$. Functionals that feature in top-10 for both properties are B98, B97-1, and PW6B95D3. Since the latter functional is corrected by empirical dispersion, it may be a matter of taste if it should be listed. In the top-15 of both, we only find one more functional, APFD. In the top-20 of both, we find O3LYP, TPSS, B97-2, wB97XD, X3LYP, revPBE0, with HSE06 (#5 for $\chi$ for #21 for $\eta$) and PBE0 (#10 for $\chi$ for #22 for $\eta$) not far behind.

**The chemical potential is poorly described in the d-block**

The analysis above considered the total performance across the s-, p-, and d-elements of the first four periods of the periodic table. In order to know which part of the periodic table produces most of the errors in $\partial E/\partial N$ and $\partial E^2/\partial N^2$ as probed by $\chi$ and $\eta$, the performance vs. experiment was separated into elements for all the studied atoms (**Supplementary excel data sheet**). The main errors for IPs and EAs are summarized in the Supporting information pdf file, **Table S15** and **Table S16**, respectively. Generally, trend prediction is extremely good for IPs for all functionals ($R^2 > 0.99$), but this is largely due to the spread of energies across 25 eV, which obscures chemically significant errors. For EAs, which are numerically much smaller, trend prediction falls to typically $R^2 \sim 0.95$ and down to 0.7 for some LDA methods. Importantly, the full MAEs for IPs and EAs are largely comparable (**Table S15/S16**). Thus, DFT does not have a *general* problem with EAs over IPs when averaged over all 36 atoms, but this tendency hides certain very pathological cases.

To identify the main pathological systems, we use B98, one of the very best performing functionals, as example in **Figure 3**. Other well-performing functionals such as B97-1 gave similar results (**Figure S3**). Despite the excellent trend predictions, errors in the computed hardness for some 3d metals (**Figure 3A**) arise from corresponding errors in the EAs (**Figure 3B**). The pathology of the 3d series is also seen from the comparison of computed and experimental $\chi$ (**Figure 3C**), in particular when zooming in on a smaller energy range (**Figure 3D**). From **Table S15**, where the maximum errors are listed for method, we see that the IPs of oxygen and boron are particularly pathological, confirming the old benchmark study by Pople and co-workers.[33] The error for boron even for B98 is also evident from **Figure 3D**. For the typically very similarly performing GGA functionals BP86 and PBE and their derived methods, Cr is interestingly particularly pathological.

For the EAs, titanium is a major cause of error, followed by some other 3d-metals (**Table S16**). Since DFT across the board overestimates the EA of Ti, and predicts it to trend with the EAs of Sc and V (**Figure 3B**) as we might expect from the continued occupation of similar d-orbitals, one could question the experimental value[32] in this single case. However, considering the error's magnitude (~0.5 eV error for the best functionals), the possible uncertainty in this experimental value out of 36 does not change the performance reported above. Even excluding Ti, the maximal EA errors of other functionals were exclusively seen for 3d metals, except one case, MN11, which had largest errors for Br.

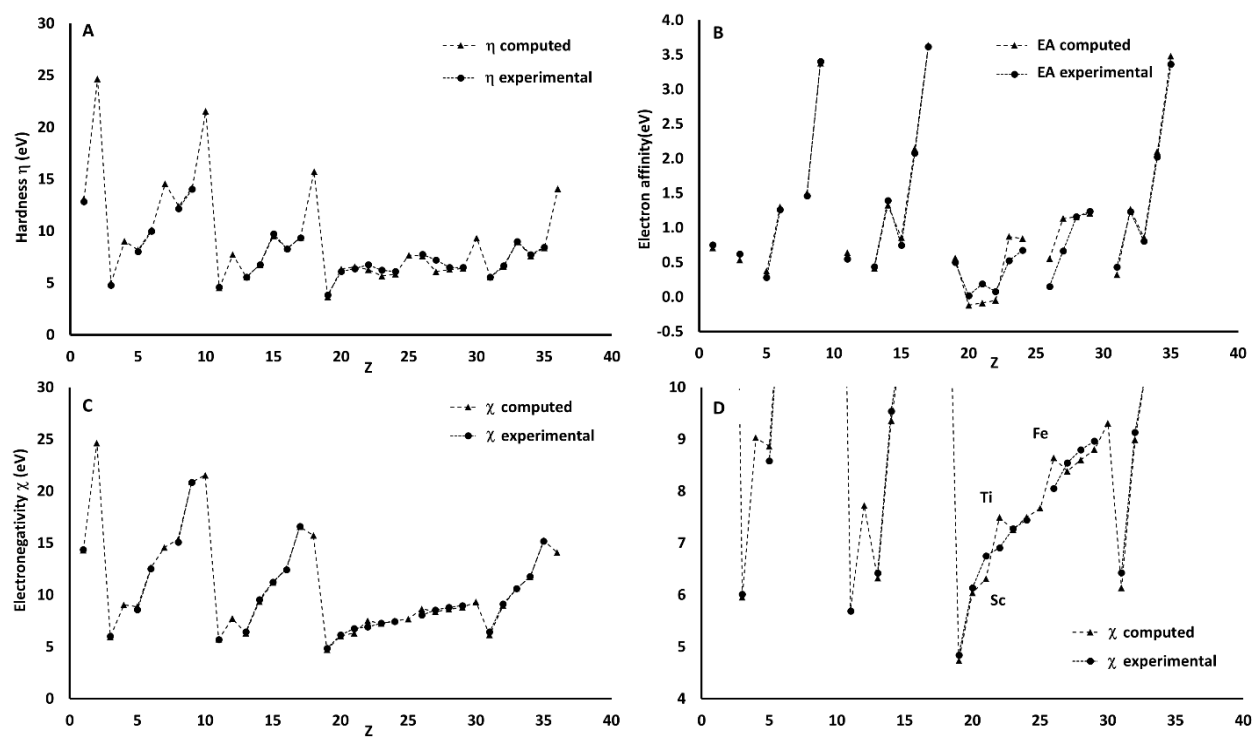

**Figure 3. Example of performance for one of the 50 studied functionals, B98.** **(A)** Computed vs. experimental η. **(B)** Computed vs. experimental EAs. **(C)** Computed vs. experimental χ. **(D)** Computed vs. experimental χ as in (C) but zoomed on values < 10 eV, with notable outliers marked. All values are in eV.

**Accuracy and precision of the chemical potential probed by χ**

We have seen large differences in the description of η and χ by DFT as evaluated by 50 distinct exchange-correlation functionals, and found that functionals that are good for one property are rarely good for the other, because η cancels errors much better in the electronic structures of X, $X^+$, and $X^-$. Most importantly, χ, due to error propagation from the two-electron transfer process $E^+ - E^-$, when applied to the *same* system X, is a sensitive probe of the description of the chemical potential both with respect to loss and gain of electrons. Even the best performing functional have problems with the gradual occupation of degenerate p- and d-orbitals. We also noted that the standard deviations of the errors are much more random for χ than for η, which could also imply that the precision differs from the accuracy of DFT applied to these simple processes.

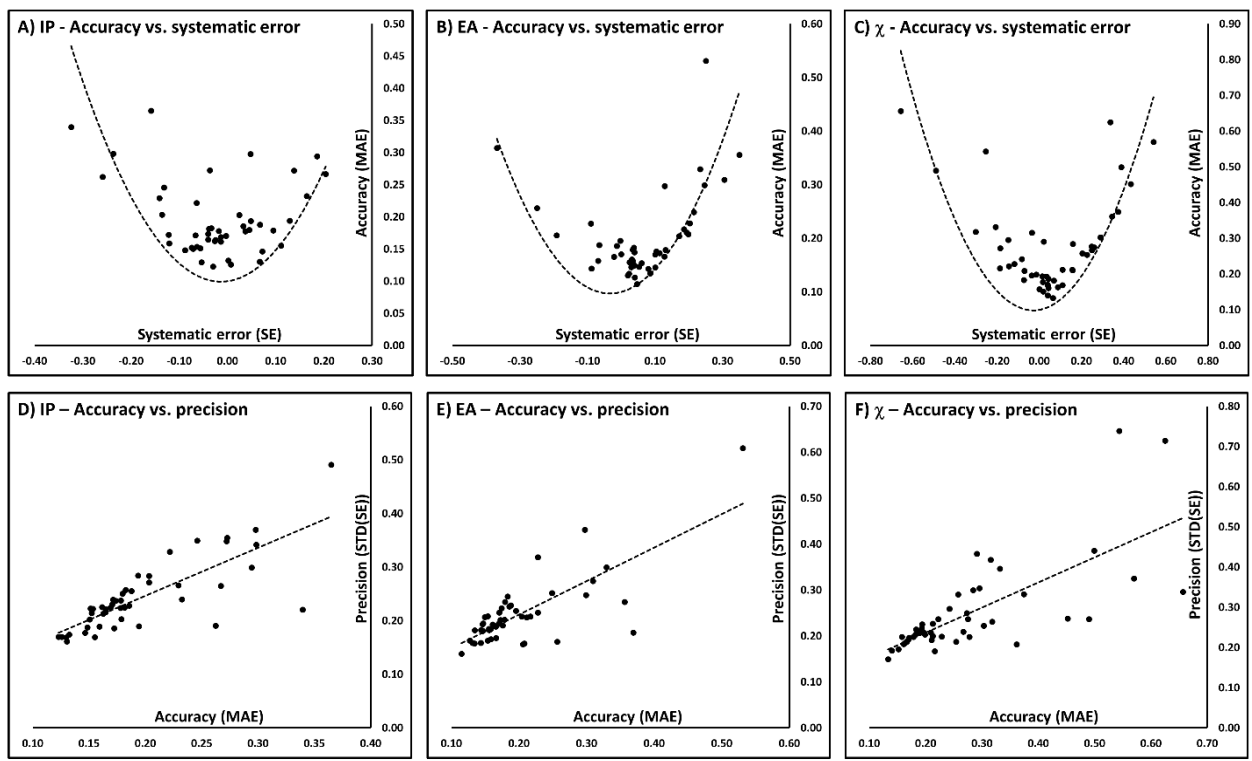

**Figure 4. Accuracy, precision, and systematic errors of density functionals (in eV).** The LDA functionals OVWN and SVWN have been removed for better viewing. (**A**) Accuracy vs. systematic errors (computed – experiment) for IPs. (**B**) Accuracy vs. systematic errors for EAs. (**C**) Accuracy vs. systematic errors for χ. (**D**) Accuracy vs. precision (standard deviations of the errors) for IPs. (**E**) Accuracy vs. precision for EAs. (**F**) Accuracy vs. precision for χ.

To understand whether this is the case, **Figure 4** compares the accuracy, as measured by the MAE, to the systematic errors, measured by the MSE. **Figure 4A** show the relationship for IPs, whereas **Figure 4B** shows it for EAs, and **Figure 4C** shows it for χ = IP + EA. The MSE reflects the tendency to over-stabilize one of the states, either X, X$^+$, or X$^−$ (typically one of the two latter). We have suggested the use of these curves previously as a tool to pinpoint error relationships and room for improvement. The well-shapes or "inverse volcanos" represent the expectation that one would have if the MSEs were monotonously related to the MAEs. The reason for using curves rather than lines is that we expect a "hard" limit of accuracy to smooth out the bottom of the well, simply due to statistical noise. As seen from **Figure 4A-4C**, the hard limit of accuracy of 0.1 eV is indeed seen in the bottom of the wells.

The IPs (**Figure 4A**) show the largest deviation from the expected well shape, because systematic errors are on average small and favor both X and $X^+$, and thus their relationship with overall accuracy (MAE) is weaker, producing more scatter in **Figure 4A**. This tendency can also be seen from direct comparison of the methods in the histograms for IPs and EAs in **Figure S1**. In contrast, the well shape is substantially more pronounced for the EAs (**Figure 4B,** please note the slight change of scale). A majority of the 50 studied functionals have positive MSEs, i.e. they overestimate EA by favoring $X^-$ relative to X, which is particularly true for the local functionals (not shown; 0.63 and 0.76 eV MSE for SVWN and OVWN, **Figure S1**) and GGA functionals such as PBE and BP86 (~0.17 and 0.25 eV MSE). This is also seen in the skewed distribution of errors (**Figure S2**). Thus, anion states are in fact too stabilized when using sufficiently large basis sets with GGA DFT. For hybrid functionals, the balance is better, as reflected in the many functionals with small MSEs near the center of **Figure 4B**. **Figure 4C** shows the corresponding error plot for $\chi$; it is symmetric and follows the well shape, with the best functionals from the ranking in **Figure 2A** also tending to have small systematic errors. For $\eta$, we found no well-shape but only a weak linear relationship (**Figure S4**), again illustrating that systematic errors largely cancel for $\eta$.

**Figure 4D-4F** show the relationship between the accuracy, measured by the MAE, and the precision, measured by the standard deviation of the errors (gray bars in **Figure 2**). The precision and accuracy of density functionals are not generally as strongly related as one could expect, as shown previously for spin-crossover systems.[115] For the more generic IPs and EAs (**Figure 4D** and **4E**) and also for the computed $\chi$, we confirm this finding. The plots show a triangular shape with the highest accuracy (lowest MAE) being consistently associated also with higher precision. Functionals that stray from this relationship can be said to exhibit less predictable behavior, as the expected error becomes more uncertain. Thus, there are functionals that are quite accurate but not very precise (on the lower side of the line in **Figures 4D-4F**), while there are also functionals that are not so precise but of decent accuracy. We think these types of curves can be important as they may indicate potential overfitting of functionals in energy space, and for balanced nearly universal functionals, we expect a linear relationship to be fulfilled. The individual cases can be deduced from **Tables S15-S16**, but for example, BHandH is more precise (0.19 eV) than accurate (0.26 eV) for EAs, whereas MN15-L is more imprecise (0.43 eV) than it is accurate (0.30 eV).

$\chi$ is the primary interest of this work, both via its excellent predictive power in broad chemistry,[29] its close relationship to the theory via the chemical potential,[20] and as a sensitive probe of imbalances in the energy-density relationship. Accordingly, the accuracy-precision relationship for this property (**Figure 4F**) was clarified in more detail, with notable outliers shown in **Figure 5A**, and a zoom-in on the best-performing functionals in **Figure 5B**. Even for these, the triangular shape is maintained. We note that MN15[87] performs excellently for $\chi$. This functional is the most broadly accurate of the Minnesota class and was parameterized to a very diverse range of data that included both IPs and EAs, explaining its success here. It demonstrates how far one can go with careful parameterization, yet all other functionals in the lower green quadrant of **Figure 5B** represent nearly "non-empirical" functionals with very few parameters. We also note that MN15 performs markedly worse for $\eta$ by error propagation, i.e. it has a negative MSE for IPs (**Table S15**) but a positive MSE for EA (**Table S16**), and these systematic errors add in $\eta$ to make $\partial E^2/\partial N^2$ poorly described, if this property is probed by $\eta$ as proposed by Parr and Pearson[25] (**Figure 2B**).

Thus, we emphasize that "universality" requires *both* $\chi$ and $\eta$ to be well-described for the *same* systems broadly. If doing so, as not done yet, since the chemical potential describes changing electron densities universally as reflected in the Pauling electronegativity, we hope that DFT can become more universal. In support of this assumption, we observe that the best performing functionals across both properties, B98, PW6B95D3, and B97-1, have been recently shown to be particularly accurate in detailed benchmarks of very diverse metal-ligand bonds.[46,47,116] Accordingly, a balanced description of $\partial E/\partial N$ and $\partial E^2/\partial N^2$, via $\chi$ and $\eta$, is an important test descriptor of DFT, and probably a first simple probe of DFT universality.

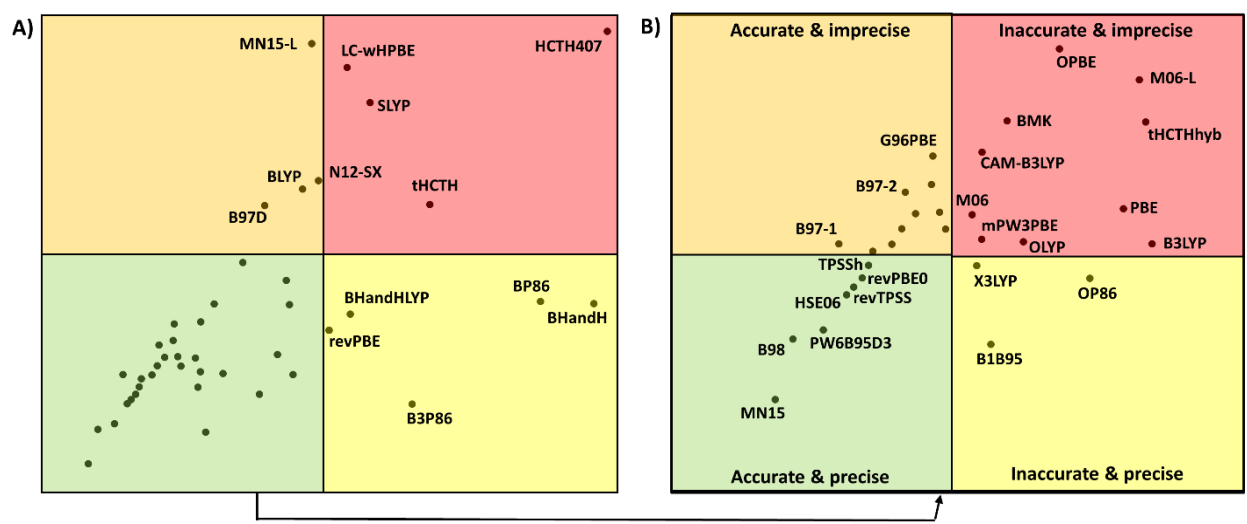

**Figure 5. Clustering of density functionals in accuracy-precision space.** The plot is quantified by MAE (horizontal axis) and standard deviation of errors (vertical axis): A) Cutoff: 0.5 eV (i.e. the two local DFT methods SVWN and OVWN were excluded). B) Cutoff: 0.3 eV.

**Contribution of densities to errors in the chemical potential probed by $\chi$**

As the last question of interest, we wanted to understand whether the errors observed are largely reflected by the energy-functional space, or whether the densities contribute to some of the pathology, as claimed in some cases for e.g. anion states, which can be described better in some cases by HF densities.[23,51] Since the exact densities corresponding to the experimental IPs and EAs are unknown, and since quantum chemistry cannot produce them accurately enough (except perhaps by full-CI using Slater orbitals to mimic the nuclear cusp), any comparison of densities is conditional, and we therefore proposed using density sensitivity analysis,[13] as explained in the introduction. Here, we consider the two simplest metrics of abnormality (systems with large $\Delta E_D´$ in **Figure 1A**) that we consider most useful, the specific version suggested by Burke and co-workers[24] that compares PBE on SVWN and HF densities, and our version that compares PBE on SVWN and PBE0 densities. The advantage of these methods is their simplicity; in comparison to more elaborate metrics that average over more rungs of the DFT ladder,[13] they tend to maximize estimates of abnormality since the applied theories to compute the densities are quite diverse.

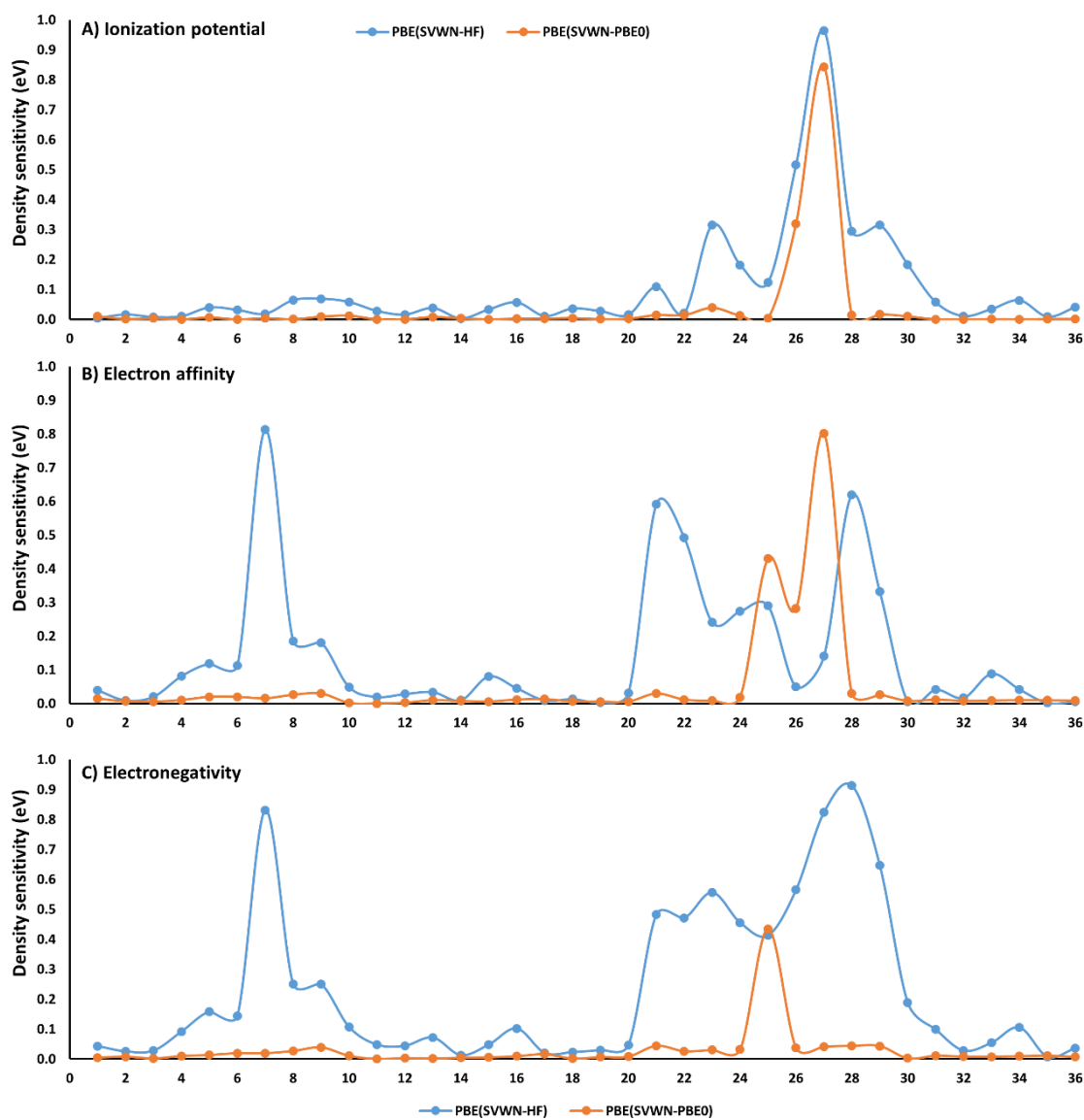

**Figure 6. Density sensitivities calculated using PBE on SVWN, HF, and PBE0 densities.** (**A**) Ionization potentials for all atoms Z = 1-36. (**B**) Electron affinities for Z = 1-36. (**C**) Electronegativities for Z = 1-36. All values are in eV and describe how much a change in density from HF to SVWN or from PBE0 to SVWN affect the PBE-computed property.

**Figure 6** compares the values of these two metrics for the IPs (**Figure 6A**), EAs (**Figure 6B**), and χ (**Figure 6C**). For the IPs, we see that the HF-based metric $S^{IP}_{PBE(SVWN,HF)}$ suggested by Burke,[24] Equation (7), is always larger than our PBE0-based metric,[13] equation (8), as expected because it represents a large difference in applied theory, with HF heavily favoring loose densities and weaker binding, whereas LDA strongly overbinds.

For the IPs (**Figure 6A**) the two metrics are in qualitative agreement, and both confirm that the 3d-series is extremely density-sensitive, with effects as large as 1 eV (100 kJ/mol) most notably for Co and Fe, the first clear data showing this for the 3d series. The p-block is also sensitive but only at a level that maximally contributes 0.1 eV (10 kJ/mol) to total errors, and when using our DFT-based metric, these effects are negligible. The d-block harbors large density sensitivity partly due to distinct electronic configurations of the HF and DFT treatment, with the 4s-3d energetics being a main difficulty for these systems. We tried to optimize PBE and HF in alternative configurations to test the electronic configuration dependence but failed in both cases, indicating that the "excited" configurations are too unstable in both theories. For the DFT-based metric, the sensitivity at the early d-block mostly disappears, because the theories converge to the same configurations. Since DFT is a ground state theory, we must accept the change in lowest energy configuration, which arises from the near-degeneracy and static correlation, which makes DFT and HF very different in the 3d series.

For the EAs (**Figure 6B**), which may be more abnormal (i.e. displaying larger density-derived errors) and sometimes improved by using HF densities,[38,51] we indeed see more density sensitivity. Surprisingly, the HF metric predicts extreme density sensitivity for the 2p-series (up to 0.8 eV), probably due to the static correlation effects of nearly degenerate p-orbitals. Confirming this interpretation, s-block elements, where degenerate occupation issues are absent, are unproblematic in all cases for both metrics. The effect only emerges in the anion state, not the neutral and cation state, thus confirming earlier work.[38,51] Importantly, although both metrics indicate that the d-block produces large density-derived errors, they disagree on whether the same is the case for the p-block and the affected 3d metals differ, i.e. the metrics give different results, even qualitatively, for EAs. As discussed previously,[13,46] we do not believe that HF theory estimates density errors well due to its pathologies relative to DFT, as largely confirmed by **Figure 6**. Thus, we still favor the metric of Equation (8),[46] shown in orange color, for estimating density sensitivity of DFT. For the same reason, HF-DFT (where HF densities are used with DFT energies) should be treated with care and may work mainly in cases of error cancellation. The PBE0 and SVWN theories are almost maximally different within DFT and favor loose and tight binding, respectively, yet do not suffer the pathology that causes large effects in HF densities.

Finally, the density sensitivity for the most important property studied in this work, the finite-difference approximation to the chemical potential $-\partial E/\partial N$ (i.e. $\chi$) is shown in **Figure 6C**. Remarkably, while the HF metric suggests that density plays a major role in the errors across the p- and d-blocks (blue), with errors again reaching 1 eV. However, correlated DFT experiences much more efficient cancellation of errors of having mostly similar orbitals occupied in the $X^+$ and $X^-$ states, and thus the density sensitivity is much smaller for $\partial E/\partial N$ for the DFT-based than for the HF-based metric, consistent with the expectation that errors in the energy-density relationship cancel in the finite difference $E(X^+) - E(X^-)$ when correlation effects of the "double electron transfer" are well accounted for. The only exception is seen for Mn in the middle of the d-block, which has a heptet $4s^13d^5$ cation state, a neutral $4s^23d^5$ state, and is the only state not showing having error cancellation in the DFT metric. In summary, although the configuration dependence complicates the picture. Figure 6 clearly shows the density sensitivity of the 2p series and 3d series being outstanding in relation to other parts of the periodic table, and these parts coincide with the largest errors in the computed energies, as discussed above.

**Conclusions**

The quest towards universal functionals dates back to the original formulations by Hohenberg and Kohn.[9] Transferability across the periodic table is lacking, as is commonly a balanced description of the electronic energy and density.[12,21,23]. There are two main approaches to this challenge: the use of extensively parameterized functionals[15,16] and the use of fundamental bounds, which may aid universal behavior with few parameters.[5,14] There are no simple metrics of universality to aid this quest, yet we expect that such metrics could help rationalizing and improving DFT results.

We hypothesized that the chemical potential, $\partial E/\partial N$, is not balanced in current DFT. $\partial E/\partial N$ is of major importance to chemical reactions: In any chemical process, electrons are moved from one place to another such that $\partial E/\partial N$ is equal in all parts of the molecular system at equilibrium. We hypothesized that DFT is unbalanced because both the uptake and removal of electrons need to be probed well around the chemical potential of the same systems, which has typically not been a requirement in functional development. In the finite difference approximation, this amounts to performing well both for absolute electronegativities $\chi$ = IP + EA and hardness $\eta$ = IP – EA for the same systems specifically, rather than for IPs and EAs of different systems more broadly.

We have studied how 50 different density functionals describe $\chi$ and $\eta$. We find that: 1) Pathological cases are not due to anions *per se*, but to adding electrons to degenerate p- and d-orbitals. 2) Popular functionals such as B3LYP, PBE, and revPBE, perform poorly for both properties. 3) Functionals that are good for $\chi$ are commonly not good for $\eta$ and *vice versa*. For example, B3LYP, MN15, and MN15-L are quite good for IPs, but not for EAs, and therefore, produce poor $\chi$ and $\eta$. 4) The accuracy and precision of a functional is not generally linearly related, yet for a universal functional we expect linearity, and the best-performing functionals support this notion. Deviations from the accuracy-precision relationship are often seen for highly parameterized functionals. 5) The pathology of $\partial E/\partial N$ in the d-block as probed by $\chi$ is accompanied by large density-derived errors, as revealed by density sensitivity analysis.[13]

Based on these results, we argue that balanced good performance for $\chi$ is a hallmark of universality by probing both sides of $\partial E/\partial N$. With this metric, B98, B97-1, PW6B95D3, APFD are the most "universal" among the tested functionals. In previous work, B98 and B97-1, and to some extent PW6B95D3, excellently described very diverse metal-ligand bonds across 3d-, 4d-, and 5d-metals and ligands such as oxide, halides, and hydride.[46,47,116] This seems to support that a balanced description of $\partial E/\partial N$ and $\partial E^2/\partial N^2$, via $\chi$ and $\eta$, is important to chemistry broadly and thus be a very simple probe of universality, as it requires only the calculation of IP+EA for the same system, compared to experiment.

**Acknowledgments**

We kindly acknowledge the use of the High-Performance Computing Cluster at DTU for the computations carried out in this work.

**Supporting Information available**

The Supporting information file contains all the computed electronic energies and details of the analysis, including the density sensitivity metrics.

**Data availability statement**

The data required to reproduce the findings of this study are available from the corresponding author upon reasonable request.

**Funding**

There is no funding to be acknowledged in this study.

**SUPPORTING INFORMATION**

**Using Electronegativity and Hardness to Test Density Functional Universality**

Klaus A. Moltved and Kasper P. Kepp*

*Technical University of Denmark, DTU Chemistry, Building 206, 2800 Kgs. Lyngby, DK – Denmark*

*Phone: +045 45 25 24 09. E-mail: kpj@kemi.dtu.dk

**Table 1.** Experimental spin multiplicity and electron configuration from NIST[4] used for atoms and atomic ions in all computations.

| Atom | Spin multiplicity of neutral atom | Spin multiplicity of monocation | Spin multiplicity of monoanion |
|---|---|---|---|
| **H**  | 2 | 1 | 1 |
| **He** | 1 | 2 | 2 |
| **Li** | 2 | 1 | 1 |
| **Be** | 1 | 2 | 2 |
| **B**  | 2 | 1 | 3 |
| **C**  | 3 | 2 | 4 |
| **N**  | 4 | 3 | 3 |
| **O**  | 3 | 4 | 2 |
| **F**  | 2 | 3 | 1 |
| **Ne** | 1 | 2 | 2 |
| **Na** | 2 | 1 | 1 |
| **Mg** | 1 | 2 | 2 |
| **Al** | 2 | 1 | 3 |
| **Si** | 3 | 2 | 4 |
| **P**  | 4 | 3 | 3 |
| **S**  | 3 | 4 | 2 |
| **Cl** | 2 | 3 | 1 |
| **Ar** | 1 | 2 | 2 |
| **K**  | 2 | 1 | 1 |
| **Ca** | 1 | 2 | 2 |
| **Sc** | 2 | 3 | 3 |
| **Ti** | 3 | 4 | 4 |
| **V**  | 4 | 5 | 5 |
| **Cr** | 7 | 6 | 6 |
| Mn | 6 | 7 | 5 |
| Fe | 5 | 6 | 4 |
| Co | 4 | 3 | 3 |
| Ni | 3 | 2 | 2 |
| Cu | 2 | 1 | 1 |
| Zn | 1 | 2 | 2 |
| Ga | 2 | 1 | 3 |
| Ge | 3 | 2 | 4 |
| As | 4 | 3 | 3 |
| Se | 3 | 4 | 2 |
| Br | 2 | 3 | 1 |
| Kr | 1 | 2 | 2 |

**Table 2.** Experimental IPs and EAs (eV) used in this work. The data are from the CRC Handbook of Chemistry and Physics[58].

| Atom | IP (eV) | EA (eV) |
|------|---------|---------|
| H  | 13.60 | 0.75 |
| He | 24.59 | NA   |
| Li | 5.39  | 0.62 |
| Be | 9.32  | NA   |
| B  | 8.30  | 0.28 |
| C  | 11.26 | 1.26 |
| N  | 14.53 | NA   |
| O  | 13.62 | 1.46 |
| F  | 17.42 | 3.40 |
| Ne | 21.56 | NA   |
| Na | 5.14  | 0.55 |
| Mg | 7.65  | NA   |
| Al | 5.99  | 0.43 |
| Si | 8.15  | 1.39 |
| P  | 10.49 | 0.75 |
| S  | 10.36 | 2.08 |
| Cl | 12.97 | 3.61 |
| Ar | 15.76 | NA   |
| K  | 4.34  | 0.50 |
| Ca | 6.11  | 0.02 |
| Sc | 6.56  | 0.19 |
| Ti | 6.83  | 0.08 |
| V  | 6.75  | 0.53 |
| Cr | 6.77  | 0.68 |
| Mn | 7.43  | NA   |
| Fe | 7.90  | 0.15 |
| Co | 7.88  | 0.66 |
| Ni | 7.64  | 1.16 |
| Cu | 7.73  | 1.24 |
| Zn | 9.39  | NA   |
| Ga | 6.00  | 0.43 |
| Ge | 7.90  | 1.23 |
| As | 9.79  | 0.81 |
| Se | 9.75  | 2.02 |
| Br | 11.81 | 3.36 |
| Kr | 14.00 | NA   |

**Table 3.     Computed electronic energies for the 36 atoms, 36 cations and 36 anions, in a.u.**

| Atom | APFD | B1B95 | B2PLYP | B3LYP | B3P86 | B97D |
|---|---|---|---|---|---|---|
| H | -0.502516 | -0.498348 | -0.498841 | -0.502438 | -0.500920 | -0.502991 |
| He | -2.900899 | -2.904196 | -2.893843 | -2.915165 | -2.908129 | -2.916553 |
| Li | -7.474503 | -7.483686 | -7.469522 | -7.492892 | -7.482413 | -7.487120 |
| Be | -14.645507 | -14.660557 | -14.638507 | -14.673238 | -14.658685 | -14.664869 |
| B | -24.630649 | -24.647677 | -24.621079 | -24.665450 | -24.648680 | -24.644708 |
| C | -37.820570 | -37.841923 | -37.806439 | -37.861540 | -37.843927 | -37.830159 |
| N | -54.562293 | -54.590902 | -54.541178 | -54.606772 | -54.591245 | -54.574757 |
| O | -75.041928 | -75.077690 | -75.010780 | -75.100613 | -75.077845 | -75.057383 |
| F | -99.704758 | -99.752530 | -99.661874 | -99.775967 | -99.747943 | -99.724925 |
| Ne | -128.898756 | -128.963183 | -128.843251 | -128.980438 | -128.949573 | -128.934220 |
| Na | -162.212068 | -162.289777 | -162.164543 | -162.299615 | -162.266588 | -162.243365 |
| Mg | -200.000079 | -200.091453 | -199.954491 | -200.099915 | -200.063330 | -200.056402 |
| Al | -242.289820 | -242.391427 | -242.241912 | -242.394516 | -242.359642 | -242.355241 |
| Si | -289.295108 | -289.408108 | -289.242469 | -289.401919 | -289.371024 | -289.372181 |
| P | -341.184731 | -341.310925 | -341.125622 | -341.291825 | -341.266935 | -341.278634 |
| S | -398.028706 | -398.170254 | -397.963226 | -398.145348 | -398.119339 | -398.142731 |
| Cl | -460.057684 | -460.217261 | -459.983545 | -460.181635 | -460.156155 | -460.192985 |
| Ar | -527.437329 | -527.617113 | -527.353330 | -527.566335 | -527.543355 | -527.598011 |
| K | -599.804366 | -600.001432 | -599.721476 | -599.938679 | -599.917166 | -599.981145 |
| Ca | -677.444860 | -677.659527 | -677.357150 | -677.590164 | -677.568333 | -677.670422 |
| Sc | -760.495665 | -760.724609 | -760.391095 | -760.648870 | -760.630582 | -760.767247 |
| Ti | -849.220619 | -849.465151 | -849.101115 | -849.380524 | -849.365913 | -849.552185 |
| V | -943.769588 | -944.030656 | -943.636356 | -943.935159 | -943.927383 | -944.166319 |
| Cr | -1044.326774 | -1044.598834 | -1044.147078 | -1044.477094 | -1044.489275 | -1044.756027 |
| Mn | -1150.865963 | -1151.164437 | -1150.697954 | -1151.034586 | -1151.041831 | -1151.405032 |
| Fe | -1263.550593 | -1263.868025 | -1263.357020 | -1263.728624 | -1263.738473 | -1264.169767 |
| Co | -1382.617535 | -1382.951467 | -1382.371674 | -1382.802245 | -1382.798958 | -1383.313699 |
| Ni | -1508.176140 | -1508.535176 | -1507.903520 | -1508.367685 | -1508.386704 | -1508.982430 |
| **Cu** | -1640.394541 | -1640.778009 | -1640.097887 | -1640.590917 | -1640.616719 | -1641.312080 |
| **Zn** | -1779.297534 | -1779.707166 | -1779.006160 | -1779.500607 | -1779.529511 | -1780.313179 |
| **Ga** | -1924.706843 | -1925.136974 | -1924.425044 | -1924.906050 | -1924.945161 | -1925.801131 |
| **Ge** | -2076.815286 | -2077.265470 | -2076.534851 | -2077.011923 | -2077.061576 | -2078.002375 |
| **As** | -2235.705409 | -2236.176012 | -2235.423231 | -2235.897988 | -2235.959773 | -2236.991902 |
| **Se** | -2401.378328 | -2401.874170 | -2401.093737 | -2401.577009 | -2401.642885 | -2402.769759 |
| **Br** | -2573.992047 | -2574.513778 | -2573.702122 | -2574.194482 | -2574.266099 | -2575.488800 |
| **Kr** | -2753.639410 | -2754.187722 | -2753.342713 | -2753.843436 | -2753.922555 | -2755.244210 |
| **H$^+$** | 0.000000 | 0.000000 | 0.000000 | 0.000000 | 0.000000 | 0.000000 |
| **He$^+$** | -1.998348 | -1.996312 | -1.997500 | -1.998599 | -1.992829 | -2.000301 |
| **Li$^+$** | -7.269864 | -7.282658 | -7.268548 | -7.286098 | -7.277398 | -7.290735 |
| **Be$^+$** | -14.315397 | -14.333775 | -14.318234 | -14.338445 | -14.323360 | -14.326593 |
| **B$^+$** | -24.311826 | -24.337490 | -24.311110 | -24.344181 | -24.325957 | -24.330907 |
| **C$^+$** | -37.396414 | -37.425112 | -37.394602 | -37.437359 | -37.416099 | -37.409466 |
| **N$^+$** | -54.020953 | -54.054960 | -54.015630 | -54.067997 | -54.046498 | -54.032354 |
| **O$^+$** | -74.531352 | -74.573675 | -74.520097 | -74.580940 | -74.562426 | -74.551781 |
| **F$^+$** | -99.061706 | -99.112217 | -99.039665 | -99.124541 | -99.099917 | -99.082636 |
| **Ne$^+$** | -128.108415 | -128.172521 | -128.074952 | -128.182714 | -128.154190 | -128.135696 |
| **Na$^+$** | -162.017711 | -162.100077 | -161.973458 | -162.100111 | -162.070885 | -162.065067 |
| **Mg$^+$** | -199.722943 | -199.818060 | -199.685941 | -199.816053 | -199.780681 | -199.761921 |
| **Al$^+$** | -242.065240 | -242.175309 | -242.028432 | -242.173087 | -242.132006 | -242.135163 |
| **Si$^{iv}$** | -288.992641 | -289.112752 | -288.952979 | -289.103710 | -289.065799 | -289.074037 |
| **P$^+$** | -340.797461 | -340.929577 | -340.752641 | -340.910179 | -340.877144 | -340.894596 |
| **S$^+$** | -397.645112 | -397.791059 | -397.593662 | -397.757479 | -397.731068 | -397.763705 |
| **Cl$^+$** | -459.580500 | -459.742488 | -459.521330 | -459.701479 | -459.674893 | -459.719778 |
| **Ar$^+$** | -526.858881 | -527.039744 | -526.790103 | -526.986158 | -526.961253 | -527.021698 |
| **K$^+$** | -599.643721 | -599.845152 | -599.564904 | -599.773297 | -599.753307 | -599.835225 |
| **Ca$^+$** | -677.224612 | -677.443007 | -677.145986 | -677.364503 | -677.342274 | -677.430207 |
| **Sc$^+$** | -760.262218 | -760.494262 | -760.167196 | -760.408056 | -760.395049 | -760.529173 |
| **Ti$^+$** | -848.980854 | -849.226906 | -848.868043 | -849.130577 | -849.118407 | -849.312553 |
| **V$^+$** | -943.546777 | -943.807862 | -943.404854 | -943.695327 | -943.693943 | -943.921378 |
| **Cr$^+$** | -1044.070414 | -1044.348377 | -1043.910254 | -1044.219294 | -1044.226762 | -1044.513240 |
| **Mn$^+$** | -1150.607241 | -1150.903396 | -1150.442764 | -1150.759087 | -1150.772614 | -1151.099786 |
| **Fe$^+$** | -1263.272510 | -1263.589289 | -1263.087133 | -1263.437141 | -1263.450280 | -1263.847425 |
| **Co⁺** | -1382.340963 | -1382.682033 | -1382.113646 | -1382.519388 | -1382.533122 | -1383.043787 |
| **Ni⁺** | -1507.893330 | -1508.259308 | -1507.639398 | -1508.077590 | -1508.096377 | -1508.701042 |
| **Cu⁺** | -1640.107881 | -1640.497702 | -1639.829188 | -1640.295823 | -1640.321867 | -1641.026871 |
| **Zn⁺** | -1778.962686 | -1779.374232 | -1778.686758 | -1779.154823 | -1779.186501 | -1779.960567 |
| **Ga⁺** | -1924.484817 | -1924.922297 | -1924.211816 | -1924.684410 | -1924.718562 | -1925.583617 |
| **Ge⁺** | -2076.522471 | -2076.980125 | -2076.252988 | -2076.721619 | -2076.764681 | -2077.713819 |
| **As⁺** | -2235.338577 | -2235.816454 | -2235.068736 | -2235.535763 | -2235.589373 | -2236.628775 |
| **Se⁺** | -2401.021871 | -2401.520434 | -2400.749344 | -2401.215139 | -2401.281023 | -2402.416925 |
| **Br⁺** | -2573.556314 | -2574.080671 | -2573.279584 | -2573.755517 | -2573.825839 | -2575.056834 |
| **Kr⁺** | -2753.120646 | -2753.671449 | -2752.837310 | -2753.323232 | -2753.399850 | -2754.728864 |
| **H⁻** | -0.527343 | -0.522051 | -0.514695 | -0.536177 | -0.537841 | -0.538007 |
| **He⁻** | -2.821859 | -2.822716 | -2.807051 | -2.835023 | -2.193806 | -2.833042 |
| **Li⁻** | -7.492894 | -7.498022 | -7.479981 | -7.513388 | -7.506167 | -7.512995 |
| **Be⁻** | -14.644094 | -14.653515 | -14.628036 | -14.671783 | -14.665537 | -14.662370 |
| **B⁻** | -24.649003 | -24.659381 | -24.624368 | -24.682933 | -24.672198 | -24.659011 |
| **C⁻** | -37.873470 | -37.888805 | -37.840452 | -37.912249 | -37.901564 | -37.880291 |
| **N⁻** | -54.562081 | -54.584404 | -54.520216 | -54.616033 | -54.599910 | -54.578957 |
| **O⁻** | -75.094840 | -75.126599 | -75.039530 | -75.162646 | -75.137517 | -75.115719 |
| **F⁻** | -99.825806 | -99.872021 | -99.756433 | -99.905600 | -99.875688 | -99.856938 |
| **Ne⁻** | -128.741433 | -128.802186 | -128.677826 | -128.826971 | -128.802943 | -128.799405 |
| **Na⁻** | -162.230807 | -162.304352 | -162.175777 | -162.321140 | -162.289960 | -162.277800 |
| **Mg⁻** | -199.987126 | -200.070175 | -199.931176 | -200.082832 | -200.055424 | -200.038471 |
| **Al⁻** | -242.309732 | -242.405250 | -242.249918 | -242.411652 | -242.384471 | -242.369220 |
| **Si⁻** | -289.348908 | -289.455892 | -289.282154 | -289.451340 | -289.428374 | -289.419816 |
| **P⁻** | -341.214656 | -341.336369 | -341.142518 | -341.327247 | -341.303590 | -341.308573 |
| **S⁻** | -398.105548 | -398.243996 | -398.025301 | -398.226301 | -398.201432 | -398.218969 |
| **Cl⁻** | -460.189643 | -460.347244 | -460.099908 | -460.316565 | -460.292827 | -460.324649 |
| **Ar⁻** | -527.356859 | -527.533277 | -527.265363 | -527.488551 | -527.475683 | -527.504928 |
| **K⁻** | -599.820182 | -600.012290 | -599.728503 | -599.955564 | -599.937008 | -600.012253 |
| **Ca⁻** | -677.445585 | -677.652506 | -677.347933 | -677.586115 | -677.572476 | -677.665728 |
| **Sc⁻** | -760.494185 | -760.716291 | -760.364524 | -760.648311 | -760.635068 | -760.781984 |
| **Ti⁻** | -849.222408 | -849.485451 | -849.104011 | -849.407051 | -849.399952 | -849.586046 |
| **V⁻** | -943.807265 | -944.062166 | -943.642849 | -943.968785 | -943.969680 | -944.208874 |
| **Cr⁻** | -1044.338260 | -1044.610694 | -1044.155743 | -1044.500932 | -1044.511775 | -1044.812483 |
| Mn- | -1150.854781 | -1151.150873 | -1150.677661 | -1151.026688 | -1151.043055 | -1151.396001 |
| Fe- | -1263.570104 | -1263.881245 | -1263.339527 | -1263.754908 | -1263.756012 | -1264.195804 |
| Co- | -1382.648162 | -1382.980369 | -1382.391359 | -1382.841052 | -1382.855496 | -1383.368889 |
| Ni- | -1508.210382 | -1508.567580 | -1507.926028 | -1508.409928 | -1508.429409 | -1509.033725 |
| Cu- | -1640.432563 | -1640.813566 | -1640.123026 | -1640.636613 | -1640.662005 | -1641.365985 |
| Zn- | -1779.286814 | -1779.691334 | -1778.988390 | -1779.489673 | -1779.528276 | -1780.299768 |
| Ga- | -1924.723724 | -1925.147848 | -1924.431540 | -1924.921954 | -1924.967074 | -1925.812040 |
| Ge- | -2076.867462 | -2077.310690 | -2076.574568 | -2077.060852 | -2077.117894 | -2078.048259 |
| As- | -2235.735789 | -2236.203417 | -2235.442251 | -2235.934938 | -2235.996889 | -2237.024056 |
| Se- | -2401.454796 | -2401.947438 | -2401.157150 | -2401.657834 | -2401.724728 | -2402.846101 |
| Br- | -2574.119553 | -2574.638094 | -2573.815996 | -2574.324650 | -2574.398369 | -2575.615214 |
| Kr- | -2753.581052 | -2754.126341 | -2753.276725 | -2753.787064 | -2753.875710 | -2755.174369 |

**Table 4.** Computed electronic energies for the 36 atoms, 36 cations and 36 anions, in a.u.

| Atom | B98 | B97-1 | B97-2 | BHandH | BHandHLYP | BLYP |
|---|---|---|---|---|---|---|
| H | -0.503001 | -0.502920 | -0.504396 | -0.478142 | -0.498778 | -0.497908 |
| He | -2.909939 | -2.906999 | -2.910036 | -2.835288 | -2.905703 | -2.907013 |
| Li | -7.486741 | -7.485690 | -7.487855 | -7.365632 | -7.483946 | -7.482557 |
| Be | -14.665124 | -14.665153 | -14.665936 | -14.491707 | -14.663996 | -14.661364 |
| B | -24.651614 | -24.653347 | -24.650725 | -24.423693 | -24.654865 | -24.653463 |
| C | -37.843043 | -37.846140 | -37.841129 | -37.559681 | -37.849271 | -37.849139 |
| N | -54.586565 | -54.589865 | -54.586674 | -54.246079 | -54.592997 | -54.592810 |
| O | -75.071333 | -75.076134 | -75.071828 | -74.662450 | -75.080664 | -75.090285 |
| F | -99.740714 | -99.746224 | -99.744022 | -99.263359 | -99.749507 | -99.767678 |
| Ne | -128.942714 | -128.947178 | -128.953479 | -128.396910 | -128.948012 | -128.972342 |
| Na | -162.255486 | -162.259880 | -162.272241 | -161.653799 | -162.275611 | -162.287406 |
| Mg | -200.053519 | -200.057460 | -200.077379 | -199.385944 | -200.079174 | -200.087066 |
| Al | -242.343121 | -242.346226 | -242.374105 | -241.608183 | -242.375716 | -242.382117 |
| Si | -289.348065 | -289.350084 | -289.387883 | -288.544978 | -289.385739 | -289.388396 |
| P | -341.237238 | -341.237788 | -341.288794 | -340.364965 | -341.278479 | -341.276356 |
| S | -398.086071 | -398.084668 | -398.149499 | -397.141104 | -398.132818 | -398.130273 |
| Cl | -460.119898 | -460.116372 | -460.196942 | -459.101370 | -460.169672 | -460.166032 |
| Ar | -527.503918 | -527.497814 | -527.597715 | -526.411764 | -527.555579 | -527.548839 |
| K | -599.868292 | -599.860149 | -599.976774 | -598.710529 | -599.931600 | -599.921791 |
| Ca | -677.518371 | -677.507589 | -677.643020 | -676.282315 | -677.581344 | -677.577541 |
| Sc | -760.571297 | -760.555910 | -760.713505 | -759.248793 | -760.629959 | -760.646226 |
| Ti | -849.300082 | -849.279217 | -849.463538 | -847.891681 | -849.354641 | -849.386940 |
| V | -943.853795 | -943.826256 | -944.041918 | -942.360549 | -943.904314 | -943.947113 |
| Cr | -1044.388998 | -1044.350598 | -1044.609986 | -1042.811307 | -1044.423773 | -1044.493803 |
| Mn | -1150.959738 | -1150.914855 | -1151.214116 | -1149.298524 | -1150.993851 | -1151.046358 |
| Fe | -1263.653044 | -1263.600526 | -1263.935725 | -1261.891178 | -1263.673694 | -1263.761521 |
| Co | -1382.720179 | -1382.657090 | -1383.035171 | -1380.843178 | -1382.711925 | -1382.845526 |
| Ni | -1508.293514 | -1508.220120 | -1508.649449 | -1506.317763 | -1508.263806 | -1508.417594 |
| Cu | -1640.525638 | -1640.440234 | -1640.926339 | -1638.453039 | -1640.477162 | -1640.643836 |
| Zn | -1779.444995 | -1779.347979 | -1779.886685 | -1777.291933 | -1779.400406 | -1779.544363 |
| Ga | -1924.857644 | -1924.748465 | -1925.341947 | -1922.633085 | -1924.828631 | -1924.934135 |
| Ge | -2076.972624 | -2076.849788 | -2077.498027 | -2074.668343 | -2076.948511 | -2077.030632 |
| As | -2235.870418 | -2235.733436 | -2236.437995 | -2233.482802 | -2235.846892 | -2235.908007 |
| Se | -2401.556379 | -2401.405905 | -2402.169020 | -2399.082940 | -2401.534428 | -2401.582069 |
| Br | -2574.182331 | -2574.017861 | -2574.840690 | -2571.622100 | -2574.159371 | -2574.194408 |
| Kr | -2753.841771 | -2753.662648 | -2754.546967 | -2751.193855 | -2753.815958 | -2753.837388 |
| $H^+$ | 0.000000 | 0.000000 | 0.000000 | 0.000000 | 0.000000 | 0.000000 |
| $He^+$ | -2.000568 | -2.001923 | -2.004284 | -1.955859 | -1.997353 | -1.995080 |
| $Li^+$ | -7.285230 | -7.284582 | -7.288704 | -7.169251 | -7.281274 | -7.279352 |
| $Be^+$ | -14.331548 | -14.333487 | -14.334538 | -14.170360 | -14.334646 | -14.331455 |
| $B^+$ | -24.336686 | -24.340353 | -24.339421 | -24.117260 | -24.339742 | -24.336371 |
| $C^+$ | -37.422451 | -37.427814 | -37.423521 | -37.146470 | -37.432140 | -37.430073 |
| $N^+$ | -54.048662 | -54.054842 | -54.050214 | -53.716032 | -54.062335 | -54.059919 |
| $O^+$ | -74.561413 | -74.566841 | -74.567750 | -74.170713 | -74.575597 | -74.570179 |
| $F^+$ | -99.095876 | -99.102812 | -99.102855 | -98.633010 | -99.113553 | -99.117033 |
| $Ne^+$ | -128.149046 | -128.156067 | -128.160558 | -127.615803 | -128.166501 | -128.176074 |
| $Na^+$ | -162.066823 | -162.071868 | -162.089287 | -161.464087 | -162.081712 | -162.090380 |
| $Mg^+$ | -199.768799 | -199.773708 | -199.794378 | -199.113466 | -199.801723 | -199.806714 |
| $Al^+$ | -242.123292 | -242.128028 | -242.156067 | -241.396586 | -242.158753 | -242.166259 |
| $Si^v$ | -289.050233 | -289.053844 | -289.091769 | -288.254950 | -289.092435 | -289.096478 |
| $P^+$ | -340.854762 | -340.856912 | -340.906957 | -339.989917 | -340.901592 | -340.902094 |
| $S^+$ | -397.701801 | -397.702094 | -397.768195 | -396.766931 | -397.752100 | -397.747421 |
| $Cl^+$ | -459.641539 | -459.639561 | -459.720642 | -458.631888 | -459.696571 | -459.691819 |
| $Ar^+$ | -526.924554 | -526.920047 | -527.018753 | -525.839593 | -526.981780 | -526.975766 |
| $K^+$ | -599.713279 | -599.705959 | -599.827354 | -598.553801 | -599.772281 | -599.758850 |
| $Ca^+$ | -677.290118 | -677.280124 | -677.415615 | -676.066491 | -677.362443 | -677.355082 |
| $Sc^+$ | -760.334702 | -760.319512 | -760.481818 | -759.019374 | -760.397325 | -760.412972 |
| $Ti^+$ | -849.046915 | -849.026243 | -849.213366 | -847.652246 | -849.111660 | -849.141614 |
| $V^+$ | -943.617425 | -943.588896 | -943.814667 | -942.131172 | -943.666344 | -943.707151 |
| $Cr^+$ | -1044.142648 | -1044.105732 | -1044.370921 | -1042.574043 | -1044.185127 | -1044.232004 |
| $Mn^+$ | -1150.677949 | -1150.633090 | -1150.931479 | -1149.035301 | -1150.726291 | -1150.772063 |
| $Fe^+$ | -1263.355533 | -1263.302472 | -1263.637893 | -1261.613209 | -1263.392335 | -1263.458959 |
| $Co^+$ | -1382.452737 | -1382.391268 | -1382.775340 | -1380.581650 | -1382.448516 | -1382.556121 |
| $Ni^+$ | -1508.018557 | -1507.946579 | -1508.381602 | -1506.049996 | -1507.993720 | -1508.120942 |
| Cu⁺ | -1640.245317 | -1640.161109 | -1640.653101 | -1638.180084 | -1640.201953 | -1640.342257 |
| Zn⁺ | -1779.102676 | -1779.005905 | -1779.547208 | -1776.965263 | -1779.070975 | -1779.195489 |
| Ga⁺ | -1924.640431 | -1924.533457 | -1925.127292 | -1922.419445 | -1924.611292 | -1924.718008 |
| Ge⁺ | -2076.684687 | -2076.564274 | -2077.212840 | -2074.384494 | -2076.662591 | -2076.746839 |
| As⁺ | -2235.508062 | -2235.373657 | -2236.078299 | -2233.125444 | -2235.488505 | -2235.553643 |
| Se⁺ | -2401.199710 | -2401.050812 | -2401.813403 | -2398.731751 | -2401.179052 | -2401.225464 |
| Br⁺ | -2573.746536 | -2573.583881 | -2574.405894 | -2571.191274 | -2573.726371 | -2573.761854 |
| Kr⁺ | -2753.322951 | -2753.145931 | -2754.029003 | -2750.679339 | -2753.300563 | -2753.325020 |
| H⁻ | -0.528941 | -0.525203 | -0.527137 | -0.494149 | -0.523317 | -0.529518 |
| He⁻ | -2.826260 | -2.823138 | -2.826643 | -2.743755 | -2.820000 | -2.823804 |
| Li⁻ | -7.506345 | -7.503351 | -7.506315 | -7.378234 | -7.499957 | -7.499399 |
| Be⁻ | -14.660794 | -14.660110 | -14.660273 | -14.479841 | -14.655136 | -14.658639 |
| B⁻ | -24.665336 | -24.666005 | -24.661159 | -24.429091 | -24.661408 | -24.670653 |
| C⁻ | -37.890862 | -37.892443 | -37.885637 | -37.598265 | -37.887123 | -37.899372 |
| N⁻ | -54.586813 | -54.589740 | -54.582267 | -54.228711 | -54.583560 | -54.607414 |
| O⁻ | -75.126353 | -75.130683 | -75.122610 | -74.700136 | -75.121480 | -75.157867 |
| F⁻ | -99.864821 | -99.869155 | -99.865496 | -99.370794 | -99.856080 | -99.902963 |
| Ne⁻ | -128.777925 | -128.781182 | -128.782492 | -128.231235 | -128.786258 | -128.839662 |
| Na⁻ | -162.279070 | -162.282198 | -162.295772 | -161.667365 | -162.292302 | -162.305875 |
| Mg⁻ | -200.035448 | -200.038041 | -200.057458 | -199.360903 | -200.057434 | -200.065296 |
| Al⁻ | -242.358287 | -242.360352 | -242.386479 | -241.616763 | -242.385732 | -242.396439 |
| Si⁻ | -289.396786 | -289.397657 | -289.433776 | -288.586614 | -289.427972 | -289.433635 |
| P⁻ | -341.268686 | -341.267846 | -341.317052 | -340.386587 | -341.304793 | -341.309831 |
| S⁻ | -398.164773 | -398.162210 | -398.225244 | -397.210308 | -398.204341 | -398.208499 |
| Cl⁻ | -460.253366 | -460.248679 | -460.328337 | -459.226500 | -460.295417 | -460.297248 |
| Ar⁻ | -527.416845 | -527.409501 | -527.504974 | -526.324420 | -527.470450 | -527.468704 |
| K⁻ | -599.888977 | -599.879817 | -599.997514 | -598.720224 | -599.943741 | -599.934918 |
| Ca⁻ | -677.513928 | -677.501991 | -677.636868 | -676.271384 | -677.573195 | -677.568768 |
| Sc⁻ | -760.568115 | -760.551370 | -760.710859 | -759.224117 | -760.608090 | -760.655081 |
| Ti⁻ | -849.298227 | -849.301524 | -849.489029 | -847.897363 | -849.361721 | -849.415618 |
| V⁻ | -943.885843 | -943.855604 | -944.077236 | -942.374614 | -943.914696 | -943.982268 |
| Cr⁻ | -1044.419929 | -1044.381229 | -1044.642533 | -1042.825025 | -1044.441186 | -1044.517104 |
| Mn⁻ | -1150.947528 | -1150.902697 | -1151.201445 | -1149.276080 | -1150.974064 | -1151.055371 |
| Fe⁻ | -1263.673263 | -1263.619955 | -1263.953456 | -1261.878462 | -1263.667041 | -1263.784009 |
| **Co⁻** | -1382.761853 | -1382.699259 | -1383.076402 | -1380.867877 | -1382.738944 | -1382.885806 |
| **Ni⁻** | -1508.336009 | -1508.262752 | -1508.690488 | -1506.345179 | -1508.293465 | -1508.461646 |
| **Cu⁻** | -1640.570013 | -1640.484476 | -1640.968888 | -1638.482629 | -1640.509272 | -1640.692672 |
| **Zn⁻** | -1779.431239 | -1779.333560 | -1779.871745 | -1777.272517 | -1779.383852 | -1779.530904 |
| **Ga⁻** | -1924.869351 | -1924.758844 | -1925.350611 | -1922.640323 | -1924.837202 | -1924.947740 |
| **Ge⁻** | -2077.019186 | -2076.894532 | -2077.541089 | -2074.710003 | -2076.990714 | -2077.075347 |
| **As⁻** | -2235.902021 | -2235.763960 | -2236.468483 | -2233.507492 | -2235.875235 | -2235.942929 |
| **Se⁻** | -2401.633545 | -2401.481837 | -2402.244764 | -2399.153714 | -2401.606954 | -2401.659635 |
| **Br⁻** | -2574.310141 | -2574.144144 | -2574.966898 | -2571.744335 | -2574.282121 | -2574.319913 |
| **Kr⁻** | -2753.777299 | -2753.597110 | -2754.477415 | -2751.128401 | -2753.752426 | -2753.778566 |

**Table 5.** Computed electronic energies for the 36 atoms, 36 cations and 36 anions, in a.u.

| Atom | BMK | BP86 | CAM-B3LYP | G96PBE | HCTH407 | HSE06 |
|---|---|---|---|---|---|---|
| H | -0.498914 | -0.500322 | -0.499089 | -0.504886 | -0.510414 | -0.501496 |
| He | -2.907327 | -2.906302 | -2.901381 | -2.907361 | -2.922149 | -2.896417 |
| Li | -7.481907 | -7.480881 | -7.471422 | -7.482681 | -7.494309 | -7.468829 |
| Be | -14.655032 | -14.659363 | -14.650305 | -14.651354 | -14.673896 | -14.638822 |
| B | -24.641514 | -24.652041 | -24.642249 | -24.638870 | -24.655932 | -24.622976 |
| C | -37.834171 | -37.848867 | -37.839095 | -37.832581 | -37.846519 | -37.811733 |
| N | -54.577165 | -54.596047 | -54.585437 | -54.579803 | -54.598787 | -54.552016 |
| O | -75.064612 | -75.090901 | -75.077629 | -75.065504 | -75.075359 | -75.030336 |
| F | -99.736501 | -99.767830 | -99.752741 | -99.736341 | -99.742730 | -99.691264 |
| Ne | -128.937055 | -128.974433 | -128.957668 | -128.940413 | -128.956214 | -128.882590 |
| Na | -162.242193 | -162.291128 | -162.275301 | -162.253543 | -162.271578 | -162.195288 |
| Mg | -200.028835 | -200.090946 | -200.076879 | -200.040405 | -200.085150 | -199.982391 |
| Al | -242.310444 | -242.391504 | -242.374979 | -242.337722 | -242.384957 | -242.270337 |
| Si | -289.304005 | -289.406346 | -289.386899 | -289.350965 | -289.405018 | -289.273767 |
| P | -341.180730 | -341.304527 | -341.281850 | -341.249276 | -341.316934 | -341.161332 |
| S | -398.016519 | -398.163423 | -398.139902 | -398.103074 | -398.178057 | -398.002751 |
| Cl | -460.037077 | -460.205892 | -460.181019 | -460.142770 | -460.228568 | -460.028909 |
| Ar | -527.407248 | -527.597288 | -527.570543 | -527.533631 | -527.637456 | -527.405428 |
| K | -599.748257 | -599.975687 | -599.944319 | -599.908302 | -600.026966 | -599.770908 |
| Ca | -677.370583 | -677.635581 | -677.598346 | -677.556717 | -677.715165 | -677.409483 |
| Sc | -760.383287 | -760.712654 | -760.659018 | -760.634845 | -760.818492 | -760.456466 |
| Ti | -849.062184 | -849.458162 | -849.394559 | -849.385612 | -849.604426 | -849.176552 |
| V | -943.559472 | -944.031637 | -943.953946 | -943.957989 | -944.222928 | -943.719799 |
| Cr | -1044.020975 | -1044.603006 | -1044.498296 | -1044.551642 | -1044.836147 | -1044.268429 |
| Mn | -1150.528660 | -1151.153069 | -1151.064901 | -1151.088554 | -1151.474691 | -1150.802325 |
| Fe | -1263.157882 | -1263.881597 | -1263.763388 | -1263.823381 | -1264.239414 | -1263.479543 |
| Co | -1382.135451 | -1382.975958 | -1382.839602 | -1382.921724 | -1383.375504 | -1382.536013 |
| Ni | -1507.624654 | -1508.559422 | -1508.411699 | -1508.509232 | -1509.062064 | -1508.085230 |
| Cu | -1639.762667 | -1640.799631 | -1640.641094 | -1640.755947 | -1641.397017 | -1640.293544 |
| Zn | -1778.608774 | -1779.709929 | -1779.556116 | -1779.667211 | -1780.392392 | -1779.189579 |
| **Ga** | -1923.931588 | -1925.116496 | -1924.970245 | -1925.080405 | -1925.884786 | -1924.592493 |
| **Ge** | -2075.952125 | -2077.229796 | -2077.084983 | -2077.202354 | -2078.086654 | -2076.692601 |
| **As** | -2234.750409 | -2236.125502 | -2235.979659 | -2236.108364 | -2237.076720 | -2235.573653 |
| **Se** | -2400.338325 | -2401.811250 | -2401.666929 | -2401.798640 | -2402.849422 | -2401.238888 |
| **Br** | -2572.863484 | -2574.437106 | -2574.292610 | -2574.431503 | -2575.565557 | -2573.844094 |
| **Kr** | -2752.418980 | -2754.095353 | -2753.949538 | -2754.099047 | -2755.320134 | -2753.481979 |
| **H$^+$** | 0.000000 | 0.000000 | 0.000000 | 0.000000 | 0.000000 | 0.000000 |
| **He$^+$** | -1.997591 | -1.992701 | -1.989875 | -2.003872 | -2.006336 | -1.997301 |
| **Li$^+$** | -7.287445 | -7.276378 | -7.265570 | -7.281742 | -7.289428 | -7.264484 |
| **Be$^+$** | -14.330984 | -14.324507 | -14.315546 | -14.328185 | -14.328549 | -14.309321 |
| **B$^+$** | -24.330003 | -24.330471 | -24.321591 | -24.321417 | -24.333517 | -24.304919 |
| **C$^+$** | -37.417563 | -37.423082 | -37.414097 | -37.409286 | -37.413903 | -37.388647 |
| **N$^+$** | -54.042949 | -54.053895 | -54.045344 | -54.038510 | -54.041896 | -54.011943 |
| **O$^+$** | -74.550548 | -74.568601 | -74.559412 | -74.554922 | -74.569023 | -74.520585 |
| **F$^+$** | -99.090401 | -99.114800 | -99.101101 | -99.091862 | -99.093355 | -99.049777 |
| **Ne$^+$** | -128.145177 | -128.175484 | -128.159173 | -128.147085 | -128.146826 | -128.094416 |
| **Na$^+$** | -162.056455 | -162.094922 | -162.077108 | -162.065008 | -162.081413 | -162.000758 |
| **Mg$^+$** | -199.753875 | -199.807698 | -199.793813 | -199.771949 | -199.786010 | -199.705327 |
| **Al$^+$** | -242.096353 | -242.165223 | -242.154094 | -242.113855 | -242.157443 | -242.046589 |
| **Siv** | -289.010432 | -289.102967 | -289.088891 | -289.049758 | -289.097419 | -288.972144 |
| **P$^+$** | -340.800237 | -340.917606 | -340.900165 | -340.863883 | -340.920984 | -340.774898 |
| **S$^+$** | -397.631052 | -397.773600 | -397.752514 | -397.721054 | -397.795777 | -397.620308 |
| **Cl$^+$** | -459.557142 | -459.724298 | -459.700771 | -459.667561 | -459.749256 | -459.552983 |
| **Ar$^+$** | -526.825073 | -527.016422 | -526.990066 | -526.957723 | -527.052083 | -526.828300 |
| **K$^+$** | -599.594331 | -599.811366 | -599.781448 | -599.752986 | -599.870706 | -599.609878 |
| **Ca$^+$** | -677.150341 | -677.408404 | -677.374934 | -677.343785 | -677.472865 | -677.189145 |
| **Sc$^+$** | -760.148002 | -760.476835 | -760.420766 | -760.417387 | -760.572534 | -760.222979 |
| **Ti$^+$** | -848.816296 | -849.227620 | -849.147037 | -849.169936 | -849.361526 | -848.936397 |
| **V$^+$** | -943.312836 | -943.793441 | -943.718729 | -943.742718 | -943.978688 | -943.494530 |
| **Cr$^+$** | -1043.773176 | -1044.330973 | -1044.247888 | -1044.286274 | -1044.580341 | -1044.011547 |
| **Mn$^+$** | -1150.258606 | -1150.881865 | -1150.790588 | -1150.840261 | -1151.176184 | -1150.543931 |
| **Fe$^+$** | -1262.868236 | -1263.575738 | -1263.473321 | -1263.532082 | -1263.922008 | -1263.201737 |
| **Co$^+$** | -1381.863631 | -1382.681564 | -1382.562136 | -1382.637214 | -1383.109861 | -1382.258707 |
| **Ni$^+$** | -1507.345196 | -1508.259041 | -1508.126699 | -1508.218699 | -1508.771998 | -1507.801737 |
| Cu⁺ | -1639.477700 | -1640.494541 | -1640.350617 | -1640.461659 | -1641.103259 | -1640.006110 |
| Zn⁺ | -1778.262397 | -1779.359151 | -1779.213472 | -1779.331348 | -1780.043681 | -1778.854262 |
| Ga⁺ | -1923.720203 | -1924.891308 | -1924.748080 | -1924.860151 | -1925.661905 | -1924.370884 |
| Ge⁺ | -2075.670802 | -2076.935025 | -2076.794267 | -2076.911337 | -2077.791744 | -2076.400595 |
| As⁺ | -2234.394965 | -2235.758393 | -2235.617079 | -2235.743637 | -2236.705898 | -2235.208068 |
| Se⁺ | -2399.982859 | -2401.448486 | -2401.304807 | -2401.444463 | -2402.493865 | -2400.882018 |
| Br⁺ | -2572.429775 | -2573.997138 | -2573.853300 | -2573.998231 | -2575.129250 | -2573.408757 |
| Kr⁺ | -2751.902835 | -2753.574554 | -2753.429047 | -2753.583242 | -2754.798723 | -2752.964339 |
| H⁻ | -0.522134 | -0.537908 | -0.530895 | -0.530054 | -0.547969 | -0.524296 |
| He⁻ | -2.811305 | -2.829368 | -2.817566 | -2.829294 | -2.845792 | -2.816768 |
| Li⁻ | -7.494723 | -7.504186 | -7.490058 | -7.496535 | -7.525892 | -7.486842 |
| Be⁻ | -14.646304 | -14.663503 | -14.643491 | -14.651735 | -14.676287 | -14.637561 |
| B⁻ | -24.651231 | -24.677820 | -24.655909 | -24.659054 | -24.679876 | -24.640785 |
| C⁻ | -37.880123 | -37.909373 | -37.888104 | -37.888084 | -37.909271 | -37.863747 |
| N⁻ | -54.574644 | -54.613615 | -54.591088 | -54.583698 | -54.604436 | -54.550641 |
| O⁻ | -75.116901 | -75.161370 | -75.138480 | -75.123513 | -75.140227 | -75.081692 |
| F⁻ | -99.857225 | -99.905969 | -99.882561 | -99.864398 | -99.884651 | -99.810345 |
| Ne⁻ | -128.774318 | -128.846960 | -128.801473 | -128.783774 | -128.828875 | -128.724569 |
| Na⁻ | -162.253820 | -162.314824 | -162.294429 | -162.266327 | -162.311741 | -162.214243 |
| Mg⁻ | -200.005754 | -200.080480 | -200.057672 | -200.027546 | -200.073113 | -199.969001 |
| Al⁻ | -242.321625 | -242.415663 | -242.388128 | -242.357248 | -242.406812 | -242.290147 |
| Si⁻ | -289.351524 | -289.463396 | -289.433755 | -289.403874 | -289.462713 | -289.327175 |
| P⁻ | -341.210294 | -341.344068 | -341.314558 | -341.278214 | -341.348994 | -341.190452 |
| S⁻ | -398.094596 | -398.248321 | -398.219164 | -398.178518 | -398.259022 | -398.078711 |
| Cl⁻ | -460.172073 | -460.344260 | -460.314972 | -460.273141 | -460.367517 | -460.159870 |
| Ar⁻ | -527.321024 | -527.523213 | -527.488946 | -527.452227 | -527.556086 | -527.324784 |
| K⁻ | -599.758893 | -599.995811 | -599.958907 | -599.919319 | -600.061664 | -599.786909 |
| Ca⁻ | -677.361855 | -677.638053 | -677.592647 | -677.557222 | -677.717074 | -677.409809 |
| Sc⁻ | -760.358284 | -760.727399 | -760.647322 | -760.644292 | -760.816836 | -760.454264 |
| Ti⁻ | -849.065173 | -849.499193 | -849.413309 | -849.416600 | -849.586264 | -849.202783 |
| V⁻ | -943.570424 | -944.078218 | -943.980943 | -944.000283 | -944.224136 | -943.755804 |
| Cr⁻ | -1044.038733 | -1044.626067 | -1044.519716 | -1044.554641 | -1044.891371 | -1044.279902 |
| Mn⁻ | -1150.490185 | -1151.162684 | -1151.046304 | -1151.099127 | -1151.483591 | -1150.796679 |
| Fe⁻ | -1263.163935 | -1263.862814 | -1263.781421 | -1263.835582 | -1264.269845 | -1263.497418 |
| **Co⁻** | -1382.175032 | -1383.019876 | -1382.873228 | -1382.950212 | -1383.441927 | -1382.566942 |
| **Ni⁻** | -1507.669320 | -1508.607766 | -1508.448593 | -1508.543297 | -1509.110866 | -1508.119707 |
| **Cu⁻** | -1639.809988 | -1640.852565 | -1640.681389 | -1640.795264 | -1641.448575 | -1640.331738 |
| **Zn⁻** | -1778.592128 | -1779.703576 | -1779.541004 | -1779.657134 | -1780.384041 | -1779.179256 |
| **Ga⁻** | -1923.938819 | -1925.138864 | -1924.981822 | -1925.096805 | -1925.902377 | -1924.609715 |
| **Ge⁻** | -2075.993499 | -2077.285829 | -2077.131332 | -2077.253462 | -2078.139738 | -2076.744424 |
| **As⁻** | -2234.779666 | -2236.165447 | -2236.014173 | -2236.136928 | -2237.110480 | -2235.604823 |
| **Se⁻** | -2400.412877 | -2401.895224 | -2401.746079 | -2401.872973 | -2402.929194 | -2401.315540 |
| **Br⁻** | -2572.988393 | -2574.570108 | -2574.421605 | -2574.556668 | -2575.696810 | -2573.971073 |
| **Kr⁻** | -2752.355250 | -2754.043105 | -2753.889228 | -2754.039663 | -2755.261087 | -2753.423558 |

**Table 6.** Computed electronic energies for the 36 atoms, 36 cations and 36 anions, in a.u.

| Atom | LC-wHPBE | M06 | M06-L | M11 | M11-L | MN15 |
|---|---|---|---|---|---|---|
| H | -0.506261 | -0.500184 | -0.503763 | -0.498385 | -0.506452 | -0.499490 |
| He | -2.904765 | -2.910218 | -2.914187 | -2.905546 | -2.919199 | -2.921282 |
| Li | -7.471607 | -7.486143 | -7.488542 | -7.480624 | -7.523391 | -7.491648 |
| Be | -14.640721 | -14.661362 | -14.663293 | -14.661904 | -14.706731 | -14.664292 |
| B | -24.629214 | -24.643127 | -24.654206 | -24.650806 | -24.684022 | -24.633855 |
| C | -37.824382 | -37.830603 | -37.850272 | -37.841023 | -37.861623 | -37.816966 |
| N | -54.570044 | -54.579600 | -54.603503 | -54.586238 | -54.594679 | -54.551629 |
| O | -75.053785 | -75.065317 | -75.090058 | -75.081058 | -75.079143 | -75.031881 |
| F | -99.720657 | -99.739746 | -99.759266 | -99.758729 | -99.746875 | -99.700620 |
| Ne | -128.916447 | -128.951392 | -128.959019 | -128.972927 | -128.944287 | -128.904931 |
| Na | -162.215082 | -162.263040 | -162.261506 | -162.276553 | -162.274744 | -162.202374 |
| Mg | -199.989695 | -200.062610 | -200.060860 | -200.069281 | -200.093877 | -200.008186 |
| Al | -242.275062 | -242.356792 | -242.355347 | -242.363047 | -242.378660 | -242.273173 |
| Si | -289.278458 | -289.363669 | -289.366327 | -289.372563 | -289.380478 | -289.276681 |
| P | -341.167228 | -341.257027 | -341.268704 | -341.266484 | -341.272571 | -341.172980 |
| S | -398.010664 | -398.108089 | -398.117120 | -398.117136 | -398.112298 | -398.030710 |
| Cl | -460.039201 | -460.143335 | -460.152986 | -460.153834 | -460.152697 | -460.085878 |
| Ar | -527.417298 | -527.532127 | -527.544442 | -527.540556 | -527.551600 | -527.496988 |
| K | -599.773177 | -599.893654 | -599.897356 | -599.895578 | -599.927743 | -599.871027 |
| Ca | -677.400485 | -677.549554 | -677.544986 | -677.534704 | -677.600162 | -677.556250 |
| Sc | -760.444611 | -760.592667 | -760.586305 | -760.579306 | -760.664894 | -760.620340 |
| Ti | -849.165039 | -849.309740 | -849.317241 | -849.286661 | -849.404390 | -849.369822 |
| V | -943.708393 | -943.853611 | -943.866194 | -943.814763 | -943.973105 | -943.940074 |
| Cr | -1044.263659 | -1044.396059 | -1044.396031 | -1044.362999 | -1044.488790 | -1044.525025 |
| Mn | -1150.791633 | -1150.965408 | -1150.959437 | -1150.878450 | -1151.032515 | -1151.109115 |
| Fe | -1263.472397 | -1263.642227 | -1263.640335 | -1263.556534 | -1263.779956 | -1263.839421 |
| Co | -1382.550405 | -1382.687103 | -1382.696188 | -1382.604552 | -1382.891868 | -1382.948522 |
| Ni | -1508.106266 | -1508.260804 | -1508.258127 | -1508.154101 | -1508.431598 | -1508.576847 |
| Cu | -1640.317729 | -1640.490248 | -1640.468870 | -1640.356494 | -1640.629842 | -1640.850361 |
| Zn | -1779.193920 | -1779.407931 | -1779.382057 | -1779.222228 | -1779.521420 | -1779.793538 |
| Ga | -1924.585579 | -1924.812563 | -1924.792942 | -1924.616553 | -1924.887696 | -1925.160133 |
| Ge | -2076.678710 | -2076.911775 | -2076.898233 | -2076.712619 | -2076.977162 | -2077.286529 |
| As | -2235.552998 | -2235.791553 | -2235.784171 | -2235.590475 | -2235.850723 | -2236.195100 |
| Se | -2401.210904 | -2401.462333 | -2401.456891 | -2401.263554 | -2401.528911 | -2401.882318 |
| Br | -2573.809489 | -2574.070902 | -2574.067138 | -2573.873689 | -2574.145952 | -2574.513245 |
| Kr | -2753.440528 | -2753.710531 | -2753.707926 | -2753.513355 | -2753.800032 | -2754.179875 |
| $H^+$ | 0.000000 | 0.000000 | 0.000000 | 0.000000 | 0.000000 | 0.000000 |
| $He^+$ | -1.999333 | -1.999577 | -2.006974 | -1.989646 | -2.013089 | -2.006017 |
| $Li^+$ | -7.266594 | -7.290816 | -7.300422 | -7.276768 | -7.311364 | -7.299736 |
| $Be^+$ | -14.309120 | -14.332752 | -14.335458 | -14.327061 | -14.381427 | -14.335268 |
| $B^+$ | -24.306999 | -24.333044 | -24.337456 | -24.336439 | -24.374899 | -24.330561 |
| $C^+$ | -37.395269 | -37.413945 | -37.431051 | -37.425782 | -37.449929 | -37.400798 |
| $N^+$ | -54.023373 | -54.037313 | -54.062805 | -54.047868 | -54.065066 | -54.016894 |
| $O^+$ | -74.534638 | -74.559124 | -74.586637 | -74.557872 | -74.568265 | -74.520359 |
| $F^+$ | -99.068035 | -99.095067 | -99.124568 | -99.109493 | -99.112085 | -99.054156 |
| $Ne^+$ | -128.116672 | -128.154828 | -128.177789 | -128.171774 | -128.171388 | -128.109118 |
| $Na^+$ | -162.023644 | -162.086288 | -162.096282 | -162.093474 | -162.079085 | -162.017260 |
| $Mg^+$ | -199.715254 | -199.784889 | -199.784059 | -199.791765 | -199.802444 | -199.728814 |
| $Al^+$ | -242.047352 | -242.142523 | -242.138894 | -242.140719 | -242.155995 | -242.060860 |
| Siv | -288.973214 | -289.071496 | -289.070044 | -289.071906 | -289.077447 | -288.978635 |
| $P^+$ | -340.777511 | -340.876808 | -340.880304 | -340.881276 | -340.884841 | -340.788996 |
| $S^+$ | -397.624226 | -397.728347 | -397.741067 | -397.732137 | -397.740369 | -397.649892 |
| $Cl^+$ | -459.559374 | -459.668681 | -459.679134 | -459.674709 | -459.674384 | -459.606779 |
| $Ar^+$ | -526.836876 | -526.952262 | -526.963690 | -526.960456 | -526.966763 | -526.917195 |
| $K^+$ | -599.618114 | -599.746829 | -599.762478 | -599.746258 | -599.758853 | -599.717096 |
| $Ca^+$ | -677.186054 | -677.321268 | -677.320636 | -677.317254 | -677.354615 | -677.326368 |
| $Sc^+$ | -760.218030 | -760.348025 | -760.363687 | -760.353215 | -760.423417 | -760.373681 |
| $Ti^+$ | -848.936060 | -849.062093 | -849.093484 | -849.060799 | -849.183914 | -849.121483 |
| $V^+$ | -943.502240 | -943.625293 | -943.633902 | -943.618814 | -943.729638 | -943.724933 |
| $Cr^+$ | -1044.019939 | -1044.148887 | -1044.148255 | -1044.129388 | -1044.247867 | -1044.283518 |
| $Mn^+$ | -1150.537643 | -1150.680812 | -1150.682156 | -1150.640916 | -1150.774172 | -1150.840194 |
| $Fe^+$ | -1263.201579 | -1263.332231 | -1263.339038 | -1263.298440 | -1263.467266 | -1263.549946 |
| $Co^+$ | -1382.283092 | -1382.413900 | -1382.427980 | -1382.356585 | -1382.626732 | -1382.679258 |
| $Ni^+$ | -1507.831624 | -1507.977126 | -1507.973206 | -1507.901219 | -1508.158509 | -1508.301173 |
| **Cu⁺** | -1640.038330 | -1640.200237 | -1640.187217 | -1640.099912 | -1640.352805 | -1640.571352 |
| **Zn⁺** | -1778.865927 | -1779.053772 | -1779.040775 | -1778.915338 | -1779.187432 | -1779.453148 |
| **Ga⁺** | -1924.360285 | -1924.599943 | -1924.580272 | -1924.391466 | -1924.655709 | -1924.936990 |
| **Ge⁺** | -2076.383590 | -2076.629483 | -2076.616103 | -2076.418969 | -2076.678433 | -2076.992502 |
| **As⁺** | -2235.184675 | -2235.437236 | -2235.429443 | -2235.222237 | -2235.477586 | -2235.828134 |
| **Se⁺** | -2400.851735 | -2401.111071 | -2401.109145 | -2400.892738 | -2401.147659 | -2401.531405 |
| **Br⁺** | -2573.372134 | -2573.642204 | -2573.641007 | -2573.427713 | -2573.688694 | -2574.081858 |
| **Kr⁺** | -2752.921514 | -2753.200879 | -2753.200745 | -2752.987693 | -2753.260137 | -2753.666249 |
| **H⁻** | -0.529817 | -0.524789 | -0.525831 | -0.531769 | -0.559413 | -0.531113 |
| **He⁻** | -2.822275 | -2.824557 | -2.823808 | -2.825738 | -2.660325 | -2.820977 |
| **Li⁻** | -7.489248 | -7.504753 | -7.509403 | -7.502283 | -7.554797 | -7.519090 |
| **Be⁻** | -14.632914 | -14.654923 | -14.657473 | -14.651227 | -14.636112 | -14.657372 |
| **B⁻** | -24.644037 | -24.654695 | -24.664855 | -24.659357 | -24.713008 | -24.650785 |
| **C⁻** | -37.877111 | -37.879106 | -37.898518 | -37.891310 | -37.919192 | -37.865899 |
| **N⁻** | -54.570592 | -54.574599 | -54.596391 | -54.587886 | -54.602432 | -54.550904 |
| **O⁻** | -75.111963 | -75.117475 | -75.136356 | -75.135477 | -75.126510 | -75.085736 |
| **F⁻** | -99.849965 | -99.861297 | -99.871886 | -99.888690 | -99.851094 | -99.822993 |
| **Ne⁻** | -128.754578 | -128.783673 | -128.784430 | -128.803085 | -128.829143 | -128.738310 |
| **Na⁻** | -162.231912 | -162.280678 | -162.278838 | -162.297189 | -162.321427 | -162.232022 |
| **Mg⁻** | -199.977623 | -200.040219 | -200.041155 | -200.055777 | -200.084636 | -199.983695 |
| **Al⁻** | -242.292191 | -242.369927 | -242.367279 | -242.377860 | -242.409155 | -242.293374 |
| **Si⁻** | -289.330843 | -289.409480 | -289.417228 | -289.423524 | -289.442995 | -289.329453 |
| **P⁻** | -341.194965 | -341.285015 | -341.291518 | -341.295412 | -341.304218 | -341.200915 |
| **S⁻** | -398.086223 | -398.183391 | -398.190474 | -398.194542 | -398.197766 | -398.108997 |
| **Cl⁻** | -460.170477 | -460.275022 | -460.284649 | -460.288127 | -460.294358 | -460.218318 |
| **Ar⁻** | -527.330006 | -527.442819 | -527.445788 | -527.449511 | -527.477802 | -527.411232 |
| **K⁻** | -599.787139 | -599.909151 | -599.913857 | -599.913915 | -599.896086 | -599.895659 |
| **Ca⁻** | -677.403817 | -677.540198 | -677.537856 | -677.535490 | -677.603220 | -677.548227 |
| **Sc⁻** | -760.433803 | -760.590839 | -760.595481 | -760.558623 | -760.674542 | -760.616224 |
| **Ti⁻** | -849.187322 | -849.330489 | -849.341865 | -849.295076 | -849.459231 | -849.392748 |
| **V⁻** | -943.744631 | -943.892043 | -943.895776 | -943.841222 | -943.953091 | -943.988260 |
| **Cr⁻** | -1044.271170 | -1044.425030 | -1044.425497 | -1044.355919 | -1044.490876 | -1044.551501 |
| **Mn⁻** | -1150.774341 | -1150.955729 | -1150.952688 | -1150.849153 | -1151.015421 | -1151.098391 |
| **Fe⁻** | -1263.495224 | -1263.631286 | -1263.655381 | -1263.561660 | -1263.828192 | -1263.862770 |
| **Co⁻** | -1382.570996 | -1382.739676 | -1382.739744 | -1382.611383 | -1382.940925 | -1382.994159 |
| **Ni⁻** | -1508.130252 | -1508.309294 | -1508.290710 | -1508.161915 | -1508.481339 | -1508.624138 |
| **Cu⁻** | -1640.345898 | -1640.540853 | -1640.515580 | -1640.365731 | -1640.681094 | -1640.901846 |
| **Zn⁻** | -1779.178372 | -1779.393371 | -1779.366210 | -1779.204274 | -1779.521768 | -1779.779683 |
| **Ga⁻** | -1924.598067 | -1924.822759 | -1924.800677 | -1924.626965 | -1924.918784 | -1925.178671 |
| **Ge⁻** | -2076.728699 | -2076.951577 | -2076.937321 | -2076.761577 | -2077.040900 | -2077.338878 |
| **As⁻** | -2235.581028 | -2235.819154 | -2235.809618 | -2235.630490 | -2235.910584 | -2236.223855 |
| **Se⁻** | -2401.285489 | -2401.533344 | -2401.526226 | -2401.348338 | -2401.627522 | -2401.956915 |
| **Br⁻** | -2573.935079 | -2574.189083 | -2574.183627 | -2574.008139 | -2574.292568 | -2574.637452 |
| **Kr⁻** | -2753.375724 | -2753.644309 | -2753.632397 | -2753.446047 | -2753.750426 | -2754.118820 |

**Table 7.** Computed electronic energies for the 36 atoms, 36 cations and 36 anions, in a.u.

| Atom | MN15-L | mPW1PW91 | mPW3PBE | N12SX | O3LYP | OLYP |
|---|---|---|---|---|---|---|
| H | -0.496357 | -0.503934 | -0.503003 | -0.500676 | -0.500019 | -0.498652 |
| He | -2.914701 | -2.905466 | -2.903856 | -2.906737 | -2.910428 | -2.907560 |
| Li | -7.494750 | -7.483605 | -7.477367 | -7.497884 | -7.488180 | -7.486211 |
| Be | -14.678310 | -14.659158 | -14.648944 | -14.677155 | -14.668012 | -14.667334 |
| B | -24.655719 | -24.649902 | -24.634431 | -24.656148 | -24.654925 | -24.655331 |
| C | -37.833361 | -37.846141 | -37.825166 | -37.841222 | -37.847129 | -37.848459 |
| N | -54.562984 | -54.594886 | -54.568069 | -54.583004 | -54.593421 | -54.595754 |
| O | -75.035462 | -75.081825 | -75.049663 | -75.063195 | -75.075981 | -75.080300 |
| F | -99.692902 | -99.752571 | -99.715216 | -99.744545 | -99.743808 | -99.749608 |
| Ne | -128.893458 | -128.955082 | -128.912470 | -128.965258 | -128.948513 | -128.956099 |
| Na | -162.215240 | -162.277575 | -162.225920 | -162.275699 | -162.269032 | -162.277437 |
| Mg | -200.030409 | -200.074168 | -200.015161 | -200.072082 | -200.066973 | -200.076973 |
| Al | -242.316120 | -242.373892 | -242.306025 | -242.362657 | -242.360517 | -242.372623 |
| Si | -289.326863 | -289.389447 | -289.312457 | -289.362548 | -289.370558 | -289.384845 |
| P | -341.225564 | -341.289616 | -341.203292 | -341.245293 | -341.267081 | -341.283812 |
| S | -398.076962 | -398.144252 | -398.049624 | -398.078977 | -398.121610 | -398.141308 |
| Cl | -460.118713 | -460.184150 | -460.081122 | -460.098333 | -460.162026 | -460.184703 |
| Ar | -527.520445 | -527.574978 | -527.463260 | -527.465824 | -527.555935 | -527.581821 |
| K | -599.899258 | -599.952668 | -599.831454 | -599.785628 | -599.937769 | -599.968770 |
| Ca | -677.559977 | -677.603455 | -677.474298 | -677.402295 | -677.594417 | -677.631168 |
| Sc | -760.591032 | -760.665899 | -760.529326 | -760.423436 | -760.667366 | -760.715817 |
| Ti | -849.307069 | -849.402808 | -849.258039 | -849.111561 | -849.416025 | -849.473558 |
| V | -943.848430 | -943.964127 | -943.810805 | -943.612385 | -943.991995 | -944.071775 |
| Cr | -1044.399953 | -1044.533878 | -1044.373507 | -1044.057974 | -1044.581185 | -1044.660345 |
| Mn | -1150.970785 | -1151.085039 | -1150.915100 | -1150.542417 | -1151.161916 | -1151.246121 |
| Fe | -1263.638114 | -1263.782167 | -1263.607153 | -1263.167944 | -1263.883666 | -1263.988011 |
| Co | -1382.687750 | -1382.862226 | -1382.681187 | -1382.166224 | -1382.991375 | -1383.084279 |
| Ni | -1508.246309 | -1508.433151 | -1508.245754 | -1507.620561 | -1508.598741 | -1508.728731 |
| Cu | -1640.496427 | -1640.664420 | -1640.469814 | -1639.704881 | -1640.870201 | -1641.014179 |
| Zn | -1779.434125 | -1779.581625 | -1779.375596 | -1778.463199 | -1779.812333 | -1779.960592 |
| Ga | -1924.862033 | -1925.006513 | -1924.785641 | -1923.685877 | -1925.257931 | -1925.409319 |
| Ge | -2076.994191 | -2077.130383 | -2076.896477 | -2075.611101 | -2077.405901 | -2077.563570 |
| As | -2235.913320 | -2236.035946 | -2235.789202 | -2234.311505 | -2236.337989 | -2236.503018 |
| Se | -2401.604884 | -2401.723891 | -2401.466048 | -2399.785815 | -2402.055881 | -2402.229051 |
| Br | -2574.237682 | -2574.352790 | -2574.083870 | -2572.194299 | -2574.715523 | -2574.897314 |
| Kr | -2753.912092 | -2754.015491 | -2753.735296 | -2751.629843 | -2754.411589 | -2754.602538 |
| H$^+$ | 0.000000 | 0.000000 | 0.000000 | 0.000000 | 0.000000 | 0.000000 |
| He$^+$ | -1.993560 | -2.001854 | -1.998308 | -2.004122 | -1.996765 | -1.996650 |
| Li$^+$ | -7.300016 | -7.278827 | -7.272032 | -7.295968 | -7.282744 | -7.281514 |
| Be$^+$ | -14.347176 | -14.329226 | -14.317205 | -14.354837 | -14.334898 | -14.336235 |
| B$^+$ | -24.351428 | -24.330506 | -24.314168 | -24.345694 | -24.340072 | -24.343111 |
| C$^+$ | -37.421962 | -37.421524 | -37.399300 | -37.419600 | -37.425874 | -37.429907 |
| N$^+$ | -54.032812 | -54.053175 | -54.024801 | -54.041889 | -54.053607 | -54.058565 |
| O$^+$ | -74.538022 | -74.571312 | -74.536443 | -74.552961 | -74.570799 | -74.576822 |
| F$^+$ | -99.060245 | -99.109784 | -99.068817 | -99.092382 | -99.100931 | -99.108197 |
| Ne$^+$ | -128.107194 | -128.165161 | -128.118385 | -128.166792 | -128.152426 | -128.160662 |
| Na$^+$ | -162.033007 | -162.083751 | -162.030733 | -162.090755 | -162.074752 | -162.084315 |
| Mg$^+$ | -199.750281 | -199.797836 | -199.736673 | -199.796662 | -199.788100 | -199.799967 |
| Al$^+$ | -242.093784 | -242.148779 | -242.080662 | -242.143432 | -242.140567 | -242.154969 |
| Si$^{iv}$ | -289.027956 | -289.086516 | -289.009091 | -289.063418 | -289.073026 | -289.089971 |
| P$^+$ | -340.840344 | -340.901911 | -340.815140 | -340.860088 | -340.884406 | -340.904172 |
| S$^+$ | -397.701145 | -397.760368 | -397.664008 | -397.698348 | -397.741074 | -397.763866 |
| Cl$^+$ | -459.643075 | -459.706749 | -459.601885 | -459.620784 | -459.686797 | -459.712636 |
| Ar$^+$ | -526.935952 | -526.996333 | -526.882877 | -526.886611 | -526.977474 | -527.006537 |
| K$^+$ | -599.747855 | -599.792607 | -599.669856 | -599.633948 | -599.778534 | -599.810925 |
| Ca$^+$ | -677.329246 | -677.384020 | -677.252794 | -677.179572 | -677.374726 | -677.413377 |
| Sc$^+$ | -760.350917 | -760.433412 | -760.299943 | -760.195502 | -760.440814 | -760.491438 |
| Ti$^+$ | -849.071635 | -849.163973 | -849.017137 | -848.862277 | -849.176180 | -849.258547 |
| V$^+$ | -943.621401 | -943.742330 | -943.587153 | -943.377469 | -943.775776 | -943.845081 |
| Cr$^+$ | -1044.153239 | -1044.278161 | -1044.114429 | -1043.822846 | -1044.332819 | -1044.411298 |
| Mn$^+$ | -1150.701058 | -1150.827697 | -1150.654217 | -1150.263238 | -1150.894024 | -1150.979250 |
| Fe$^+$ | -1263.348773 | -1263.505791 | -1263.325190 | -1262.873519 | -1263.597320 | -1263.694662 |
| Co$^+$ | -1382.406402 | -1382.586572 | -1382.401748 | -1381.908383 | -1382.719374 | -1382.834626 |
| Ni$^+$ | -1507.967938 | -1508.151331 | -1508.151331 | -1507.352932 | -1508.320191 | -1508.449540 |
| Cu$^+$ | -1640.216079 | -1640.378754 | -1640.180163 | -1639.429827 | -1640.587050 | -1640.729412 |
| Zn$^+$ | -1779.096927 | -1779.248148 | -1779.037983 | -1778.120011 | -1779.480482 | -1779.629346 |
| Ga$^+$ | -1924.642153 | -1924.783951 | -1924.562541 | -1923.472381 | -1925.041677 | -1925.196159 |
| Ge$^+$ | -2076.703995 | -2076.836978 | -2076.602542 | -2075.325061 | -2077.119492 | -2077.280594 |
| As$^+$ | -2235.546307 | -2235.668484 | -2235.421341 | -2233.948881 | -2235.977424 | -2236.146230 |
| Se$^+$ | -2401.262360 | -2401.367229 | -2401.107377 | -2399.433353 | -2401.705057 | -2401.882296 |
| Br$^+$ | -2573.812676 | -2573.916784 | -2573.646023 | -2571.761036 | -2574.284511 | -2574.470332 |
| Kr$^+$ | -2753.396236 | -2753.496374 | -2753.214637 | -2751.112578 | -2753.895813 | -2754.090804 |
| H$^-$ | -0.537412 | -0.527907 | -0.530116 | -0.515579 | -0.531362 | -0.528879 |
| He$^-$ | -2.837690 | -2.826491 | -2.825266 | -2.824763 | -2.829958 | -2.826611 |
| Li$^-$ | -7.528206 | -7.501888 | -7.497046 | -7.506383 | -7.507849 | -7.504604 |
| Be$^-$ | -14.678416 | -14.657456 | -14.648397 | -14.669836 | -14.664344 | -14.663578 |
| B$^-$ | -24.674913 | -24.667705 | -24.654182 | -24.664610 | -24.670062 | -24.670380 |
| C$^-$ | -37.881062 | -37.898424 | -37.879843 | -37.886381 | -37.896813 | -37.897869 |
| N$^-$ | -54.566160 | -54.593337 | -54.570470 | -54.571803 | -54.593620 | -54.597938 |
| O$^-$ | -75.086580 | -75.133210 | -75.105915 | -75.113811 | -75.131279 | -75.137898 |
| F$^-$ | -99.811036 | -99.871961 | -99.840106 | -99.867245 | -99.871096 | -99.879919 |
| Ne$^-$ | -128.759389 | -128.797241 | -128.756088 | -128.793783 | -128.789431 | -128.824710 |
| Na$^-$ | -162.251492 | -162.295893 | -162.245708 | -162.290081 | -162.289897 | -162.297429 |
| Mg$^-$ | -200.014773 | -200.061392 | -200.002626 | -200.049543 | -200.048011 | -200.055948 |
| Al$^-$ | -242.337840 | -242.393657 | -242.326764 | -242.374649 | -242.375434 | -242.386205 |
| Si$^-$ | -289.379946 | -289.443149 | -289.367141 | -289.410285 | -289.418591 | -289.430880 |
| P$^-$ | -341.256614 | -341.319139 | -341.235131 | -341.268674 | -341.295614 | -341.310947 |
| S$^-$ | -398.156635 | -398.220677 | -398.128477 | -398.152772 | -398.196918 | -398.214998 |
| Cl$^-$ | -460.256808 | -460.315705 | -460.215068 | -460.229186 | -460.293543 | -460.314478 |
| Ar$^-$ | -527.439737 | -527.493999 | -527.383678 | -527.373357 | -527.472077 | -527.496561 |
| K$^-$ | -599.928720 | -599.967992 | -599.847985 | -599.797862 | -599.952528 | -599.982028 |
| Ca$^-$ | -677.559870 | -677.604384 | -677.475298 | -677.392972 | -677.588929 | -677.623665 |
| Sc$^-$ | -760.537408 | -760.663455 | -760.530560 | -760.414103 | -760.669967 | -760.722508 |
| Ti$^-$ | -849.297261 | -849.429787 | -849.288397 | -849.124933 | -849.447870 | -849.507490 |
| V$^-$ | -943.899316 | -944.000875 | -943.851774 | -943.626788 | -944.036599 | -944.104999 |
| Cr$^-$ | -1044.438111 | -1044.544196 | -1044.386811 | -1044.081895 | -1044.602050 | -1044.681072 |
| Mn$^-$ | -1150.968284 | -1151.072733 | -1150.906808 | -1150.535563 | -1151.154755 | -1151.239778 |
| Fe$^-$ | -1263.657238 | -1263.800482 | -1263.629561 | -1263.208012 | -1263.903668 | -1264.004217 |
| **Co⁻** | -1382.735173 | -1382.891225 | -1382.713568 | -1382.206261 | -1383.022093 | -1383.139037 |
| **Ni⁻** | -1508.300858 | -1508.465859 | -1508.281932 | -1507.661219 | -1508.631915 | -1508.763459 |
| **Cu⁻** | -1640.554919 | -1640.700969 | -1640.509931 | -1639.746798 | -1640.906756 | -1641.051062 |
| **Zn⁻** | -1779.426409 | -1779.570716 | -1779.365475 | -1778.447221 | -1779.798528 | -1779.946012 |
| **Ga⁻** | -1924.885713 | -1925.023235 | -1924.803520 | -1923.690555 | -1925.269283 | -1925.419164 |
| **Ge⁻** | -2077.049919 | -2077.182625 | -2076.949677 | -2075.653148 | -2077.451190 | -2077.606396 |
| **As⁻** | -2235.943400 | -2236.065912 | -2235.821637 | -2234.334097 | -2236.364818 | -2236.527664 |
| **Se⁻** | -2401.678702 | -2401.800095 | -2401.544567 | -2399.856628 | -2402.128195 | -2402.298843 |
| **Br⁻** | -2574.365393 | -2574.480160 | -2574.213307 | -2572.318022 | -2574.839676 | -2575.018704 |
| **Kr⁻** | -2753.857609 | -2753.956628 | -2753.677762 | -2751.561464 | -2754.349519 | -2754.539071 |

**Table 8.** Computed electronic energies for the 36 atoms, 36 cations and 36 anions, in a.u.

| Atom | OP86 | OPBE | OVWN | PBE0 | RevPBE0 | RevPBE |
|---|---|---|---|---|---|---|
| H | -0.501098 | -0.504458 | -0.538551 | -0.501317 | -0.501671 | -0.500469 |
| He | -2.906843 | -2.905010 | -3.013652 | -2.895125 | -2.896601 | -2.894872 |
| Li | -7.484841 | -7.483660 | -7.638595 | -7.467103 | -7.469299 | -7.464921 |
| Be | -14.666011 | -14.658312 | -14.871107 | -14.636519 | -14.639484 | -14.633709 |
| B | -24.654672 | -24.643013 | -24.912278 | -24.619673 | -24.623642 | -24.617301 |
| C | -37.848896 | -37.834025 | -38.159927 | -37.807261 | -37.812448 | -37.805431 |
| N | -54.599669 | -54.582773 | -54.963952 | -54.546362 | -54.552796 | -54.544052 |
| O | -75.081589 | -75.056864 | -75.510712 | -75.022625 | -75.031079 | -75.025962 |
| F | -99.750522 | -99.718752 | -100.245258 | -99.681557 | -99.692021 | -99.689678 |
| Ne | -128.959038 | -128.921017 | -129.518980 | -128.871228 | -128.883386 | -128.882129 |
| Na | -162.281779 | -162.238917 | -162.892867 | -162.182811 | -162.196196 | -162.186843 |
| Mg | -200.081774 | -200.028032 | -200.747821 | -199.968693 | -199.983325 | -199.970911 |
| Al | -242.383437 | -242.323572 | -243.101824 | -242.255674 | -242.271469 | -242.255908 |
| Si | -289.404112 | -289.338625 | -290.174023 | -289.258295 | -289.275095 | -289.255248 |
| P | -341.313156 | -341.242708 | -342.134261 | -341.145219 | -341.162888 | -341.137927 |
| S | -398.175672 | -398.095795 | -399.055202 | -397.985553 | -398.004401 | -397.976488 |
| Cl | -460.225829 | -460.137744 | -461.163874 | -460.010852 | -460.030705 | -459.999411 |
| Ar | -527.631580 | -527.536261 | -528.627750 | -527.386826 | -527.407344 | -527.371343 |
| K | -600.023488 | -599.921361 | -601.068398 | -599.751535 | -599.772641 | -599.732910 |
| Ca | -677.689995 | -677.575292 | -678.786678 | -677.389187 | -677.411071 | -677.372374 |
| Sc | -760.783965 | -760.662160 | -761.938033 | -760.435129 | -760.457910 | -760.427846 |
| Ti | -849.550796 | -849.421536 | -850.760988 | -849.155561 | -849.178033 | -849.150842 |
| V | -944.153881 | -944.017902 | -945.422211 | -943.699985 | -943.721377 | -943.700420 |
| Cr | -1044.769983 | -1044.636588 | -1046.088430 | -1044.250742 | -1044.269943 | -1044.251993 |
| Mn | -1151.353892 | -1151.204802 | -1152.731845 | -1150.787222 | -1150.804098 | -1150.774083 |
| Fe | -1264.107724 | -1263.950239 | -1265.554455 | -1263.465402 | -1263.481245 | -1263.473451 |
| Co | -1383.240149 | -1383.073582 | -1384.752561 | -1382.524275 | -1382.537811 | -1382.538209 |
| Ni | -1508.872573 | -1508.697474 | -1510.447371 | -1508.077054 | -1508.087113 | -1508.092484 |
| Cu | -1641.170883 | -1640.987578 | -1642.807725 | -1640.289700 | -1640.295496 | -1640.302174 |
| Zn | -1780.126767 | -1779.932379 | -1781.824071 | -1779.188940 | -1779.191443 | -1779.182112 |
| Ga | -1925.592379 | -1925.391152 | -1927.342530 | -1924.595684 | -1924.594650 | -1924.560210 |
| Ge | -2077.763599 | -2077.555572 | -2079.567075 | -2076.700527 | -2076.695015 | -2076.642480 |
| As | -2236.721463 | -2236.507279 | -2238.577600 | -2235.586928 | -2235.576307 | -2235.506611 |
| Se | -2402.459305 | -2402.234803 | -2404.375009 | -2401.255194 | -2401.241691 | -2401.159508 |
| Br | -2575.141149 | -2574.907603 | -2577.115962 | -2573.864215 | -2573.847046 | -2573.752490 |
| Kr | -2754.861679 | -2754.620090 | -2756.894926 | -2753.506903 | -2753.485088 | -2753.377168 |
| H$^+$ | 0.000000 | 0.000000 | 0.000000 | 0.000000 | 0.000000 | 0.000000 |
| He$^+$ | -1.994319 | -2.002786 | -2.045736 | -1.996681 | -1.997368 | -1.994653 |
| Li$^+$ | -7.278575 | -7.278014 | -7.408195 | -7.262448 | -7.264584 | -7.259554 |
| Be$^+$ | -14.329616 | -14.328803 | -14.505752 | -14.306837 | -14.309667 | -14.302922 |
| B$^+$ | -24.337885 | -24.328867 | -24.566096 | -24.301855 | -24.305400 | -24.297925 |
| C$^+$ | -37.423565 | -37.410444 | -37.706389 | -37.384127 | -37.389137 | -37.381042 |
| N$^+$ | -54.053151 | -54.037222 | -54.389884 | -54.006101 | -54.012505 | -54.002660 |
| O$^+$ | -74.575863 | -74.558195 | -74.965435 | -74.513758 | -74.521250 | -74.507857 |
| F$^+$ | -99.106624 | -99.080575 | -99.562675 | -99.040562 | -99.050433 | -99.040081 |
| Ne$^+$ | -128.160814 | -128.127222 | -128.683072 | -128.083164 | -128.095125 | -128.086040 |
| Na$^+$ | -162.089130 | -162.048872 | -162.676131 | -161.988267 | -162.001568 | -161.989707 |
| Mg$^+$ | -199.801471 | -199.755830 | -200.441565 | -199.691548 | -199.706145 | -199.691112 |
| Al$^+$ | -242.155155 | -242.097414 | -242.851927 | -242.031543 | -242.047407 | -242.032588 |
| Si$^{iv}$ | -289.097604 | -289.033998 | -289.845677 | -288.956227 | -288.973184 | -288.954158 |
| P$^+$ | -340.920741 | -340.851799 | -341.719645 | -340.758294 | -340.776176 | -340.752743 |
| S$^+$ | -397.791047 | -397.717522 | -398.640382 | -397.603231 | -397.621788 | -397.592790 |
| Cl$^+$ | -459.746258 | -459.663464 | -460.654460 | -459.534825 | -459.554593 | -459.522650 |
| Ar$^+$ | -527.048397 | -526.957426 | -528.015059 | -526.809419 | -526.830012 | -526.794396 |
| K$^+$ | -599.863948 | -599.765720 | -600.887630 | -599.590810 | -599.611755 | -599.569298 |
| Ca$^+$ | -677.467230 | -677.360703 | -678.541682 | -677.168951 | -677.190830 | -677.149310 |
| Sc$^+$ | -760.556719 | -760.445054 | -761.689519 | -760.202016 | -760.224570 | -760.195900 |
| Ti$^+$ | -849.331985 | -849.214798 | -850.521626 | -848.915743 | -848.938031 | -848.924746 |
| V$^+$ | -943.932346 | -943.809369 | -945.175052 | -943.476846 | -943.496448 | -943.467770 |
| Cr$^+$ | -1044.511176 | -1044.383237 | -1045.808152 | -1043.996196 | -1044.013514 | -1043.982591 |
| Mn$^+$ | -1151.089505 | -1150.954518 | -1152.439502 | -1150.529075 | -1150.545710 | -1150.510458 |
| Fe$^+$ | -1263.811912 | -1263.664925 | -1265.227556 | -1263.188450 | -1263.203580 | -1263.174330 |
| Co$^+$ | -1382.961100 | -1382.802207 | -1384.451678 | -1382.249535 | -1382.260801 | -1382.246441 |
| Ni$^+$ | -1508.588801 | -1508.421238 | -1510.140895 | -1507.796173 | -1507.803891 | -1507.794637 |
| Cu⁺ | -1640.882926 | -1640.707204 | -1642.496375 | -1640.004948 | -1640.008297 | -1639.999421 |
| Zn⁺ | -1779.794012 | -1779.609405 | -1781.463585 | -1778.855607 | -1778.856354 | -1778.834804 |
| Ga⁺ | -1925.369823 | -1925.172777 | -1927.098827 | -1924.374309 | -1924.372850 | -1924.338979 |
| Ge⁺ | -2077.469412 | -2077.264842 | -2079.251971 | -2076.408281 | -2076.402831 | -2076.351503 |
| As⁺ | -2236.351753 | -2236.140224 | -2238.186950 | -2235.220554 | -2235.210529 | -2235.143260 |
| Se⁺ | -2402.106178 | -2401.888486 | -2403.993314 | -2400.900026 | -2400.884687 | -2400.801369 |
| Br⁺ | -2574.706671 | -2574.478721 | -2576.653637 | -2573.429594 | -2573.411576 | -2573.317005 |
| Kr⁺ | -2754.341464 | -2754.104502 | -2756.347457 | -2752.989083 | -2752.967300 | -2752.861048 |
| H⁻ | -0.537296 | -0.529499 | -0.603781 | -0.524103 | -0.524835 | -0.527258 |
| He⁻ | -2.832309 | -2.829575 | -2.959291 | -2.815703 | -2.817188 | -2.816682 |
| Li⁻ | -7.509983 | -7.502729 | -7.681842 | -7.485276 | -7.487587 | -7.483954 |
| Be⁻ | -14.669019 | -14.658544 | -14.890539 | -14.634579 | -14.637662 | -14.635099 |
| B⁻ | -24.678494 | -24.663078 | -24.954887 | -24.637448 | -24.641422 | -24.640703 |
| C⁻ | -37.908773 | -37.890573 | -38.240211 | -37.859392 | -37.864584 | -37.864173 |
| N⁻ | -54.604995 | -54.579429 | -54.998289 | -54.544659 | -54.551468 | -54.555925 |
| O⁻ | -75.142246 | -75.109599 | -75.601720 | -75.073761 | -75.082554 | -75.092218 |
| F⁻ | -99.883793 | -99.845362 | -100.409326 | -99.800508 | -99.811241 | -99.824613 |
| Ne⁻ | -128.832984 | -128.793903 | -129.415483 | -128.713188 | -128.725487 | -128.752701 |
| Na⁻ | -162.307594 | -162.258861 | -162.936244 | -162.201688 | -162.215255 | -162.207385 |
| Mg⁻ | -200.072416 | -200.016401 | -200.754834 | -199.955462 | -199.970086 | -199.957414 |
| Al⁻ | -242.407425 | -242.344553 | -243.142523 | -242.275356 | -242.291104 | -242.277573 |
| Si⁻ | -289.462316 | -289.394100 | -290.250407 | -289.311863 | -289.328546 | -289.310115 |
| P⁻ | -341.346587 | -341.268393 | -342.192432 | -341.174123 | -341.192038 | -341.170960 |
| S⁻ | -398.256174 | -398.169571 | -399.160889 | -398.061459 | -398.080450 | -398.056350 |
| Cl⁻ | -460.362839 | -460.268989 | -461.326402 | -460.141910 | -460.161792 | -460.133624 |
| Ar⁻ | -527.552507 | -527.452323 | -528.567509 | -527.305862 | -527.326596 | -527.293660 |
| K⁻ | -600.044553 | -599.937712 | -601.106322 | -599.767485 | -599.788680 | -599.749983 |
| Ca⁻ | -677.694372 | -677.578154 | -678.806438 | -677.389787 | -677.411601 | -677.372363 |
| Sc⁻ | -760.797814 | -760.671286 | -761.972594 | -760.431982 | -760.454926 | -760.439764 |
| Ti⁻ | -849.592473 | -849.458560 | -850.822777 | -849.181340 | -849.203963 | -849.188409 |
| V⁻ | -944.203450 | -944.063280 | -945.488244 | -943.735642 | -943.756697 | -943.744293 |
| Cr⁻ | -1044.791490 | -1044.645645 | -1046.131186 | -1044.261977 | -1044.281424 | -1044.268133 |
| Mn⁻ | -1151.355225 | -1151.210587 | -1152.756174 | -1150.774437 | -1150.791636 | -1150.786348 |
| Fe⁻ | -1264.126471 | -1263.963263 | -1265.441828 | -1263.482439 | -1263.498817 | -1263.495807 |
| **Co⁻** | -1383.273770 | -1383.098424 | -1384.807686 | -1382.554439 | -1382.568401 | -1382.578609 |
| **Ni⁻** | -1508.910577 | -1508.726573 | -1510.506869 | -1508.110549 | -1508.121263 | -1508.137905 |
| **Cu⁻** | -1641.211573 | -1641.019309 | -1642.870181 | -1640.326782 | -1640.333381 | -1640.351176 |
| **Zn⁻** | -1780.119224 | -1779.921881 | -1781.832041 | -1779.177799 | -1779.180495 | -1779.172768 |
| **Ga⁻** | -1925.611429 | -1925.405945 | -1927.378166 | -1924.612185 | -1924.611465 | -1924.579712 |
| **Ge⁻** | -2077.818178 | -2077.606555 | -2079.639444 | -2076.752345 | -2076.746745 | -2076.695421 |
| **As⁻** | -2236.751359 | -2236.529061 | -2238.631669 | -2235.616265 | -2235.607403 | -2235.541883 |
| **Se⁻** | -2402.535608 | -2402.304244 | -2404.475543 | -2401.330751 | -2401.318337 | -2401.239376 |
| **Br⁻** | -2575.270095 | -2575.030697 | -2577.269121 | -2573.990893 | -2573.974070 | -2573.881223 |
| **Kr⁻** | -2754.805024 | -2754.558983 | -2756.856102 | -2753.448128 | -2753.426498 | -2753.321454 |

**Table 9.** Computed electronic energies for the 36 atoms, 36 cations and 36 anions, in a.u.

| Atom | PBE | PW6B95D3 | SLYP | RevTPSS | SVWN | SVWN5 |
|---|---|---|---|---|---|---|
| H | -0.499984 | -0.501601 | -0.457073 | -0.500145 | -0.496403 | -0.478665 |
| He | -2.892880 | -2.917379 | -2.767007 | -2.911984 | -2.872119 | -2.834786 |
| Li | -7.461983 | -7.506601 | -7.246970 | -7.489879 | -7.398311 | -7.343890 |
| Be | -14.629749 | -14.697397 | -14.318002 | -14.672240 | -14.520408 | -14.447132 |
| B | -24.611979 | -24.699102 | -24.193069 | -24.667194 | -24.448672 | -24.355919 |
| C | -37.798444 | -37.908820 | -37.272118 | -37.857044 | -37.582320 | -37.470128 |
| N | -54.535387 | -54.673591 | -53.900892 | -54.597473 | -54.267879 | -54.136530 |
| O | -75.014557 | -75.181085 | -74.256429 | -75.082545 | -74.685555 | -74.530976 |
| F | -99.675555 | -99.876559 | -98.798157 | -99.737668 | -99.292517 | -99.114432 |
| Ne | -128.865734 | -129.107449 | -127.872681 | -128.923550 | -128.434222 | -128.232889 |
| Na | -162.168931 | -162.453929 | -161.045132 | -162.224789 | -161.659484 | -161.440070 |
| Mg | -199.951313 | -200.279032 | -198.703274 | -200.005471 | -199.372729 | -199.133794 |
| Al | -242.234683 | -242.600501 | -240.849312 | -242.287634 | -241.576806 | -241.318326 |
| Si | -289.232673 | -289.638714 | -287.709406 | -289.282536 | -288.496820 | -288.218426 |
| P | -341.114208 | -341.563103 | -339.451596 | -341.160714 | -340.300356 | -340.002045 |
| S | -397.951180 | -398.446043 | -396.149826 | -397.996542 | -397.061812 | -396.741159 |
| Cl | -459.972760 | -460.516631 | -458.032702 | -460.012294 | -459.009904 | -458.666532 |
| Ar | -527.343820 | -527.939794 | -525.264205 | -527.377005 | -526.308103 | -525.942023 |
| K | -599.704680 | -600.347633 | -597.482722 | -599.725869 | -598.580835 | -598.196645 |
| Ca | -677.343100 | -678.032331 | -674.982639 | -677.351336 | -676.136748 | -675.733080 |
| Sc | -760.397187 | -761.123385 | -757.885691 | -760.391179 | -759.105299 | -758.680523 |
| Ti | -849.120572 | -849.889978 | -846.467625 | -849.097367 | -847.749978 | -847.303934 |
| V | -943.672332 | -944.481525 | -940.868323 | -943.629410 | -942.219650 | -941.752671 |
| Cr | -1044.226232 | -1045.069430 | -1041.272782 | -1044.157841 | -1042.699197 | -1042.210818 |
| Mn | -1150.751374 | -1151.665018 | -1147.659799 | -1150.666416 | -1149.143840 | -1148.636401 |
| Fe | -1263.451789 | -1264.397650 | -1260.198736 | -1263.345865 | -1261.762358 | -1261.228731 |
| Co | -1382.520448 | -1383.508674 | -1379.112617 | -1382.385552 | -1380.754980 | -1380.197057 |
| Ni | -1508.078932 | -1509.119577 | -1504.529970 | -1507.912053 | -1506.245777 | -1505.663277 |
| Cu | -1640.294282 | -1641.389343 | -1636.600268 | -1640.098266 | -1638.391557 | -1637.786203 |
| Zn | -1779.178354 | -1780.347526 | -1775.332632 | -1778.958146 | -1777.194206 | -1776.566984 |
| Ga | -1924.560993 | -1925.804308 | -1920.546156 | -1924.325184 | -1922.477304 | -1921.829815 |
| Ge | -2076.649367 | -2077.960183 | -2072.473638 | -2076.395080 | -2074.474948 | -2073.806922 |
| As | -2235.520453 | -2236.897991 | -2231.183002 | -2235.248086 | -2233.255375 | -2232.566839 |
| Se | -2401.177222 | -2402.625026 | -2396.682760 | -2400.886249 | -2398.826340 | -2398.115318 |
| Br | -2573.775188 | -2575.293309 | -2569.123713 | -2573.464613 | -2571.339892 | -2570.606066 |
| Kr | -2753.406176 | -2754.995561 | -2748.596772 | -2753.076255 | -2750.886594 | -2750.130000 |
| $H^+$ | 0.000000 | 0.000000 | 0.000000 | 0.000000 | 0.000000 | 0.000000 |
| $He^+$ | -1.993724 | -2.002723 | -1.912537 | -2.000092 | -1.961317 | -1.941651 |
| $Li^+$ | -7.256691 | -7.302422 | -7.056183 | -7.288639 | -7.182153 | -7.142743 |
| $Be^+$ | -14.299138 | -14.365153 | -14.003942 | -14.338267 | -14.172748 | -14.115412 |
| $B^+$ | -24.293195 | -24.384631 | -23.892648 | -24.345327 | -24.114467 | -24.038122 |
| $C^+$ | -37.374323 | -37.488330 | -36.860673 | -37.439950 | -37.136096 | -37.040412 |
| $N^+$ | -53.994064 | -54.134480 | -53.369461 | -54.061251 | -53.699749 | -53.584963 |
| $O^+$ | -74.497812 | -74.669495 | -73.762324 | -74.567526 | -74.149914 | -74.016349 |
| $F^+$ | -99.026812 | -99.230205 | -98.158516 | -99.102740 | -98.611884 | -98.454506 |
| $Ne^+$ | -128.069960 | -128.311935 | -127.077452 | -128.137806 | -127.598740 | -127.417590 |
| $Na^+$ | -161.971925 | -162.260026 | -160.856319 | -162.035822 | -161.447155 | -161.242655 |
| $Mg^+$ | -199.671558 | -199.999543 | -198.432754 | -199.727572 | -199.073212 | -198.850127 |
| $Al^+$ | -242.011261 | -242.382029 | -240.643484 | -242.060490 | -241.339257 | -241.096485 |
| Siv | -288.931374 | -289.340938 | -287.423804 | -288.981882 | -288.178118 | -287.915807 |
| $P^+$ | -340.728742 | -341.179484 | -339.081281 | -340.775990 | -339.895300 | -339.613328 |
| $S^+$ | -397.567916 | -398.062673 | -395.779294 | -397.611456 | -396.654397 | -396.352819 |
| $Cl^+$ | -459.496141 | -460.038027 | -457.565483 | -459.538054 | -458.505618 | -458.181382 |
| $Ar^+$ | -526.766790 | -527.359045 | -524.694720 | -526.802124 | -525.701480 | -525.354422 |
| $K^+$ | -599.541312 | -600.187462 | -597.324843 | -599.570876 | -598.400090 | -598.030311 |
| $Ca^+$ | -677.120058 | -677.810501 | -674.766234 | -677.130929 | -675.893213 | -675.504900 |
| $Sc^+$ | -760.166279 | -760.887305 | -757.657205 | -760.167592 | -758.852306 | -758.442030 |
| $Ti^+$ | -848.896306 | -849.645273 | -846.238363 | -848.876744 | -847.499704 | -847.068647 |
| $V^+$ | -943.441331 | -944.250692 | -940.640592 | -943.402896 | -941.967653 | -941.516034 |
| $Cr^+$ | -1043.959374 | -1044.815936 | -1041.013651 | -1043.898112 | -1042.408497 | -1041.936578 |
| $Mn^+$ | -1150.488182 | -1151.396729 | -1147.394061 | -1150.407575 | -1148.852721 | -1148.361217 |
| $Fe^+$ | -1263.154066 | -1264.111977 | -1259.905297 | -1263.054486 | -1261.435771 | -1260.921278 |
| $Co^+$ | -1382.231404 | -1383.235027 | -1378.827604 | -1382.104750 | -1380.442309 | -1379.899489 |
| $Ni^+$ | -1507.784355 | -1508.839403 | -1504.237942 | -1507.625203 | -1505.926849 | -1505.360557 |
| **Cu⁺** | -1639.994952 | -1641.104386 | -1636.303036 | -1639.807261 | -1638.067467 | -1637.477972 |
| **Zn⁺** | -1778.833718 | -1780.008736 | -1774.988991 | -1778.621636 | -1776.821006 | -1776.210994 |
| **Ga⁺** | -1924.340360 | -1925.586820 | -1920.336985 | -1924.103470 | -1922.237964 | -1921.606320 |
| **Ge⁺** | -2076.358306 | -2077.672110 | -2072.193836 | -2076.103402 | -2074.163310 | -2073.511418 |
| **As⁺** | -2235.156289 | -2236.535950 | -2230.830895 | -2234.882768 | -2232.869612 | -2232.197443 |
| **Se⁺** | -2400.821576 | -2402.267264 | -2396.333999 | -2400.529952 | -2398.443010 | -2397.750621 |
| **Br⁺** | -2573.340834 | -2574.856497 | -2568.695340 | -2573.029820 | -2570.876413 | -2570.161301 |
| **Kr⁺** | -2752.889983 | -2754.476005 | -2748.086415 | -2752.558980 | -2750.340747 | -2749.602790 |
| **H⁻** | -0.526306 | -0.528482 | -0.471765 | -0.532401 | -0.544947 | -0.513062 |
| **He⁻** | -2.814682 | -2.837282 | -2.672279 | -2.831001 | -2.803778 | -2.751865 |
| **Li⁻** | -7.480853 | -7.524113 | -7.257159 | -7.513123 | -7.432544 | -7.365585 |
| **Be⁻** | -14.631005 | -14.691723 | -14.308974 | -14.671016 | -14.534732 | -14.448568 |
| **B⁻** | -24.635357 | -24.713122 | -24.208167 | -24.682126 | -24.490784 | -24.383724 |
| **C⁻** | -37.857141 | -37.958179 | -37.323570 | -37.908828 | -37.664226 | -37.536824 |
| **N⁻** | -54.546694 | -54.672504 | -53.899816 | -54.603642 | -54.298723 | -54.149292 |
| **O⁻** | -75.080309 | -75.235112 | -74.317926 | -75.134603 | -74.780221 | -74.607099 |
| **F⁻** | -99.810068 | -100.000440 | -98.934876 | -99.861171 | -99.462761 | -99.266180 |
| **Ne⁻** | -128.736267 | -128.949091 | -127.725083 | -128.763783 | -128.314151 | -128.097657 |
| **Na⁻** | -162.189254 | -162.472395 | -161.057501 | -162.247301 | -161.694872 | -161.462723 |
| **Mg⁻** | -199.937828 | -200.258905 | -198.675276 | -199.993302 | -199.372274 | -199.119246 |
| **Al⁻** | -242.256382 | -242.616125 | -240.860983 | -242.305527 | -241.614429 | -241.342087 |
| **Si⁻** | -289.287666 | -289.688443 | -287.753336 | -289.334426 | -288.570385 | -288.277008 |
| **P⁻** | -341.146860 | -341.592240 | -339.476336 | -341.193154 | -340.355522 | -340.040255 |
| **S⁻** | -398.030807 | -398.523419 | -396.223629 | -398.072573 | -397.167252 | -396.829134 |
| **Cl⁻** | -460.106894 | -460.649922 | -458.162385 | -460.143038 | -459.172038 | -458.810990 |
| **Ar⁻** | -527.265800 | -527.858361 | -525.179797 | -527.293138 | -526.248520 | -525.867740 |
| **K⁻** | -599.721650 | -600.361733 | -597.491042 | -599.745100 | -598.613185 | -598.215809 |
| **Ca⁻** | -677.343162 | -678.026336 | -674.968760 | -677.353519 | -676.148950 | -675.731814 |
| **Sc⁻** | -760.409082 | -761.117054 | -757.891610 | -760.398312 | -759.135277 | -758.697123 |
| **Ti⁻** | -849.158061 | -849.912130 | -846.492192 | -849.132277 | -847.804949 | -847.344866 |
| **V⁻** | -943.715348 | -944.513440 | -940.906370 | -943.667527 | -942.286488 | -941.805583 |
| **Cr⁻** | -1044.241901 | -1045.087094 | -1041.288685 | -1044.173170 | -1042.736743 | -1042.235480 |
| **Mn⁻** | -1150.708892 | -1151.652849 | -1147.651906 | -1150.662641 | -1149.167015 | -1148.643023 |
| **Fe⁻** | -1263.473429 | -1264.415332 | -1260.216138 | -1263.371107 | -1261.807466 | -1261.258230 |
| **Co⁻** | -1382.559434 | -1383.542451 | -1379.149872 | -1382.419284 | -1380.816735 | -1380.243173 |
| **Ni⁻** | -1508.122118 | -1509.156532 | -1504.570807 | -1507.950150 | -1506.312360 | -1505.715042 |
| **Cu⁻** | -1640.341650 | -1641.429326 | -1636.644365 | -1640.140344 | -1638.461789 | -1637.841079 |
| **Zn⁻** | -1779.168762 | -1780.332859 | -1775.313297 | -1778.946125 | -1777.197880 | -1776.557705 |
| **Ga⁻** | -1924.579983 | -1925.817306 | -1920.557264 | -1924.340907 | -1922.513474 | -1921.852181 |
| **Ge⁻** | -2076.702354 | -2078.007643 | -2072.517202 | -2076.445950 | -2074.547492 | -2073.864464 |
| **As⁻** | -2235.553274 | -2236.929216 | -2231.211050 | -2235.279674 | -2233.312538 | -2232.607369 |
| **Se⁻** | -2401.255589 | -2402.702004 | -2396.756963 | -2400.963210 | -2398.931059 | -2398.202867 |
| **Br⁻** | -2573.903449 | -2575.420987 | -2569.247958 | -2573.592041 | -2571.495670 | -2570.744449 |
| **Kr⁻** | -2753.350169 | -2754.936332 | -2748.534244 | -2753.014547 | -2750.848488 | -2750.077558 |

**Table 10.** Computed electronic energies for the 36 atoms, 36 cations and 36 anions, in a.u.

| Atom | tHCTH | tHCTHhyb | TPSSh | TPSS | VSXC | wB97XD |
|---|---|---|---|---|---|---|
| H | -0.509508 | -0.507497 | -0.500179 | -0.500226 | -0.502882 | -0.502930 |
| He | -2.923121 | -2.917151 | -2.909038 | -2.909600 | -2.916911 | -2.909539 |
| Li | -7.495467 | -7.494214 | -7.488056 | -7.488781 | -7.506131 | -7.490814 |
| Be | -14.674335 | -14.671925 | -14.670155 | -14.671329 | -14.692551 | -14.666811 |
| B | -24.654660 | -24.656925 | -24.666772 | -24.669028 | -24.678514 | -24.652302 |
| C | -37.841880 | -37.847562 | -37.863750 | -37.866529 | -37.875550 | -37.844328 |
| N | -54.591924 | -54.593021 | -54.612652 | -54.615669 | -54.631539 | -54.589479 |
| O | -75.072782 | -75.076686 | -75.103565 | -75.109366 | -75.119231 | -75.074829 |
| F | -99.740445 | -99.744944 | -99.771737 | -99.779312 | -99.797629 | -99.747777 |
| Ne | -128.954956 | -128.948477 | -128.971467 | -128.980319 | -129.019437 | -128.954520 |
| Na | -162.266252 | -162.255707 | -162.288444 | -162.295274 | -162.348465 | -162.270536 |
| Mg | -200.082159 | -200.053541 | -200.083107 | -200.089356 | -200.159268 | -200.065983 |
| Al | -242.383408 | -242.342216 | -242.380599 | -242.386826 | -242.466651 | -242.360355 |
| Si | -289.405522 | -289.347983 | -289.393045 | -289.398770 | -289.494894 | -289.370152 |
| P | -341.319636 | -341.240303 | -341.289847 | -341.294794 | -341.410215 | -341.264572 |
| S | -398.188713 | -398.089499 | -398.142035 | -398.147302 | -398.276450 | -398.119908 |
| Cl | -460.245214 | -460.124079 | -460.177483 | -460.182674 | -460.333339 | -460.160419 |
| Ar | -527.659329 | -527.510355 | -527.563146 | -527.567781 | -527.743047 | -527.551477 |
| K | -600.047882 | -599.872008 | -599.932575 | -599.936634 | -600.132630 | -599.919301 |
| Ca | -677.742392 | -677.525919 | -677.577349 | -677.582127 | -677.802418 | -677.570193 |
| Sc | -760.852711 | -760.586009 | -760.637158 | -760.647069 | -760.883223 | -760.624815 |
| Ti | -849.646391 | -849.322680 | -849.366317 | -849.377881 | -849.648459 | -849.354481 |
| V | -944.287694 | -943.885429 | -943.917631 | -943.935798 | -944.237660 | -943.908247 |
| Cr | -1044.894933 | -1044.434075 | -1044.472066 | -1044.488766 | -1044.834715 | -1044.446692 |
| Mn | -1151.563517 | -1151.017320 | -1151.007070 | -1151.019201 | -1151.440610 | -1151.011270 |
| Fe | -1264.346695 | -1263.719777 | -1263.704971 | -1263.729059 | -1264.178112 | -1263.688900 |
| Co | -1383.514458 | -1382.801189 | -1382.772890 | -1382.799227 | -1383.295602 | -1382.776425 |
| Ni | -1509.210610 | -1508.391096 | -1508.324746 | -1508.353043 | -1508.926398 | -1508.348325 |
| Cu | -1641.570298 | -1640.639700 | -1640.536118 | -1640.565218 | -1641.224469 | -1640.577525 |
| Zn | -1780.587329 | -1779.567475 | -1779.425871 | -1779.448633 | -1780.208124 | -1779.496973 |
| Ga | -1926.101767 | -1924.984621 | -1924.824828 | -1924.838667 | -1925.677968 | -1924.908941 |
| Ge | -2078.330356 | -2077.107691 | -2076.924154 | -2076.931804 | -2077.855484 | -2077.019499 |
| As | -2237.348413 | -2236.014901 | -2235.805811 | -2235.807739 | -2236.820677 | -2235.911327 |
| Se | -2403.152802 | -2401.711751 | -2401.470662 | -2401.468340 | -2402.571278 | -2401.599245 |
| Br | -2575.899039 | -2574.347838 | -2574.076421 | -2574.069994 | -2575.265344 | -2574.224165 |
| Kr | -2755.683434 | -2754.016920 | -2753.715568 | -2753.704793 | -2754.994912 | -2753.880363 |
| $H^+$ | 0.000000 | 0.000000 | 0.000000 | 0.000000 | 0.000000 | 0.000000 |
| $He^+$ | -2.007293 | -2.005891 | -2.000123 | -2.000164 | -2.005015 | -2.001239 |
| $Li^+$ | -7.294608 | -7.291803 | -7.286059 | -7.286554 | -7.302442 | -7.293231 |
| $Be^+$ | -14.331131 | -14.336402 | -14.337371 | -14.338294 | -14.356466 | -14.342080 |
| $B^+$ | -24.335899 | -24.340108 | -24.345102 | -24.346708 | -24.365736 | -24.337900 |
| $C^+$ | -37.414867 | -37.424031 | -37.442720 | -37.445527 | -37.450060 | -37.424073 |
| $N^+$ | -54.040552 | -54.050012 | -54.073743 | -54.076662 | -54.084463 | -54.052436 |
| $O^+$ | -74.567538 | -74.565908 | -74.590782 | -74.593309 | -74.617430 | -74.567930 |
| $F^+$ | -99.095890 | -99.098319 | -99.133531 | -99.138770 | -99.154257 | -99.105629 |
| $Ne^+$ | -128.149796 | -128.150137 | -128.184485 | -128.190896 | -128.219081 | -128.164430 |
| $Na^+$ | -162.086419 | -162.069390 | -162.098575 | -162.104933 | -162.163586 | -162.088282 |
| $Mg^+$ | -199.782541 | -199.764298 | -199.806113 | -199.811895 | -199.878944 | -199.791592 |
| $Al^+$ | -242.156552 | -242.119144 | -242.154200 | -242.160375 | -242.243992 | -242.139757 |
| Siv | -289.100292 | -289.046280 | -289.090857 | -289.096844 | -289.190110 | -289.071808 |
| $P^+$ | -340.927200 | -340.852790 | -340.903272 | -340.908785 | -341.020260 | -340.881605 |
| $S^+$ | -397.805383 | -397.704239 | -397.758293 | -397.762934 | -397.897434 | -397.734299 |
| $Cl^+$ | -459.767021 | -459.644163 | -459.701725 | -459.706712 | -459.855096 | -459.681290 |
| $Ar^+$ | -527.076011 | -526.928051 | -526.986288 | -526.991113 | -527.162734 | -526.971645 |
| $K^+$ | -599.900502 | -599.719143 | -599.776802 | -599.780374 | -599.980248 | -599.768065 |
| $Ca^+$ | -677.496963 | -677.291957 | -677.358050 | -677.362339 | -677.577622 | -677.347567 |
| $Sc^+$ | -760.614838 | -760.348699 | -760.413848 | -760.423465 | -760.654733 | -760.387169 |
| $Ti^+$ | -849.414998 | -849.065643 | -849.127245 | -849.141009 | -849.421955 | -849.109183 |
| $V^+$ | -944.041937 | -943.649968 | -943.696375 | -943.709093 | -944.013312 | -943.678686 |
| $Cr^+$ | -1044.655862 | -1044.186214 | -1044.216460 | -1044.228356 | -1044.587660 | -1044.203337 |
| $Mn^+$ | -1151.251447 | -1150.724893 | -1150.749937 | -1150.760845 | -1151.170186 | -1150.740432 |
| $Fe^+$ | -1264.018009 | -1263.412421 | -1263.420915 | -1263.435237 | -1263.882178 | -1263.417401 |
| $Co^+$ | -1383.249735 | -1382.531440 | -1382.497104 | -1382.518209 | -1383.019764 | -1382.510419 |
| $Ni^+$ | -1508.934221 | -1508.112798 | -1508.042900 | -1508.065887 | -1508.641942 | -1508.073732 |
| Cu⁺ | -1641.290131 | -1640.355531 | -1640.250334 | -1640.273972 | -1640.938438 | -1640.297721 |
| Zn⁺ | -1780.239774 | -1779.217271 | -1779.093192 | -1779.112303 | -1779.860040 | -1779.154484 |
| Ga⁺ | -1925.877008 | -1924.763409 | -1924.603647 | -1924.617734 | -1925.464447 | -1924.691595 |
| Ge⁺ | -2078.034436 | -2076.815781 | -2076.632168 | -2076.640241 | -2077.568182 | -2076.733505 |
| As⁺ | -2236.977145 | -2235.648567 | -2235.439823 | -2235.442554 | -2236.457063 | -2235.552225 |
| Se⁺ | -2402.794498 | -2401.350600 | -2401.114982 | -2401.112430 | -2402.219492 | -2401.238070 |
| Br⁺ | -2575.461333 | -2573.908305 | -2573.641217 | -2573.634852 | -2574.831994 | -2573.787347 |
| Kr⁺ | -2755.161064 | -2753.494937 | -2753.197508 | -2753.187296 | -2754.478134 | -2753.362891 |
| H⁻ | -0.544277 | -0.535454 | -0.528771 | -0.529725 | -0.529067 | -0.532894 |
| He⁻ | -2.843807 | -2.836272 | -2.828344 | -2.829431 | -2.845841 | -2.824657 |
| Li⁻ | -7.524755 | -7.515681 | -7.509282 | -7.510181 | -7.531552 | -7.509730 |
| Be⁻ | -14.674747 | -14.669657 | -14.668047 | -14.670531 | -14.690777 | -14.658563 |
| B⁻ | -24.671659 | -24.672588 | -24.682423 | -24.686302 | -24.699175 | -24.665438 |
| C⁻ | -37.896943 | -37.898866 | -37.915094 | -37.919935 | -37.930016 | -37.892718 |
| N⁻ | -54.595308 | -54.595355 | -54.613928 | -54.621833 | -54.629528 | -54.587392 |
| O⁻ | -75.131231 | -75.134400 | -75.154298 | -75.164982 | -75.176132 | -75.128755 |
| F⁻ | -99.875755 | -99.874122 | -99.892127 | -99.905057 | -99.927850 | -99.870560 |
| Ne⁻ | -128.824675 | -128.782418 | -128.810352 | -128.848122 | -128.889128 | -128.786322 |
| Na⁻ | -162.305735 | -162.283432 | -162.309436 | -162.316569 | -162.371046 | -162.289626 |
| Mg⁻ | -200.069009 | -200.037480 | -200.070296 | -200.076677 | -200.143177 | -200.048170 |
| Al⁻ | -242.400954 | -242.359052 | -242.398903 | -242.405706 | -242.488171 | -242.374167 |
| Si⁻ | -289.458250 | -289.399455 | -289.445390 | -289.451460 | -289.548821 | -289.418593 |
| P⁻ | -341.351938 | -341.272689 | -341.319810 | -341.326212 | -341.437388 | -341.296848 |
| S⁻ | -398.267930 | -398.169661 | -398.217869 | -398.224357 | -398.355002 | -398.199421 |
| Cl⁻ | -460.381507 | -460.260114 | -460.308328 | -460.314485 | -460.466669 | -460.294704 |
| Ar⁻ | -527.566666 | -527.422095 | -527.478434 | -527.483825 | -527.670204 | -527.461148 |
| K⁻ | -600.084399 | -599.897368 | -599.950319 | -599.954643 | -600.155380 | -599.936467 |
| Ca⁻ | -677.742487 | -677.523261 | -677.578663 | -677.583486 | -677.801812 | -677.566063 |
| Sc⁻ | -760.872328 | -760.591192 | -760.639626 | -760.655204 | -760.882334 | -760.613778 |
| Ti⁻ | -849.692555 | -849.355576 | -849.397269 | -849.413286 | -849.678365 | -849.373489 |
| V⁻ | -944.333504 | -943.928081 | -943.955396 | -943.973070 | -944.276403 | -943.935633 |
| Cr⁻ | -1044.958415 | -1044.474402 | -1044.484432 | -1044.502538 | -1044.858861 | -1044.469379 |
| Mn⁻ | -1151.571994 | -1151.012520 | -1151.004638 | -1150.975116 | -1151.430137 | -1150.985487 |
| Fe⁻ | -1264.375506 | -1263.749566 | -1263.727289 | -1263.739719 | -1264.168999 | -1263.726878 |
| **Co⁻** | -1383.569004 | -1382.850689 | -1382.802407 | -1382.831735 | -1383.343747 | -1382.816948 |
| **Ni⁻** | -1509.258293 | -1508.440155 | -1508.358072 | -1508.389670 | -1508.976976 | -1508.391552 |
| **Cu⁻** | -1641.619034 | -1640.690018 | -1640.573356 | -1640.605991 | -1641.278966 | -1640.623508 |
| **Zn⁻** | -1780.576947 | -1779.555585 | -1779.413163 | -1779.436505 | -1780.194852 | -1779.480264 |
| **Ga⁻** | -1926.116155 | -1924.998064 | -1924.840040 | -1924.854655 | -1925.691675 | -1924.917988 |
| **Ge⁻** | -2078.381832 | -2077.156597 | -2076.974935 | -2076.982892 | -2077.903314 | -2077.063012 |
| **As⁻** | -2237.383729 | -2236.050810 | -2235.835682 | -2235.838804 | -2236.849180 | -2235.947299 |
| **Se⁻** | -2403.232832 | -2401.792782 | -2401.546950 | -2401.545580 | -2402.647177 | -2401.678229 |
| **Br⁻** | -2576.030361 | -2574.479018 | -2574.203558 | -2574.197645 | -2575.391530 | -2574.351580 |
| **Kr⁻** | -2755.614145 | -2753.951644 | -2753.653259 | -2753.643184 | -2754.935624 | -2753.812840 |

**Table 11.** Computed electronic energies for the 36 atoms, 36 cations and 36 anions, in a.u.

| Atom | wPBEhPBE | X3LYP |
|------|----------|-------|
| H | -0.500469 | -0.499879 |
| He | -2.894872 | -2.908251 |
| Li | -7.464921 | -7.483394 |
| Be | -14.633709 | -14.661311 |
| B | -24.617301 | -24.650623 |
| C | -37.805431 | -37.843813 |
| N | -54.544052 | -54.586071 |
| O | -75.025962 | -75.076541 |
| F | -99.689678 | -99.748420 |
| Ne | -128.882129 | -128.949339 |
| Na | -162.186843 | -162.266639 |
| Mg | -199.970911 | -200.065045 |
| Al | -242.255908 | -242.356901 |
| Si | -289.255248 | -289.361542 |
| P | -341.137927 | -341.248634 |
| S | -397.976488 | -398.099152 |
| Cl | -459.999411 | -460.132374 |
| Ar | -527.371343 | -527.513970 |
| K | -599.732910 | -599.884204 |
| Ca | -677.372374 | -677.533523 |
| Sc | -760.427846 | -760.588571 |
| Ti | -849.150842 | -849.316791 |
| V | -943.700420 | -943.868063 |
| Cr | -1044.251993 | -1044.404145 |
| Mn | -1150.774083 | -1150.960633 |
| Fe | -1263.473451 | -1263.651079 |
| Co | -1382.538209 | -1382.717591 |
| Ni | -1508.092484 | -1508.279006 |
| Cu | -1640.302174 | -1640.498249 |
| Zn | -1779.182112 | -1779.405771 |
| **Ga** | **-1924.560210** | **-1924.809327** |
| **Ge** | **-2076.642480** | **-2076.912739** |
| **As** | **-2235.506611** | **-2235.796180** |
| **Se** | **-2401.159508** | **-2401.472403** |
| **Br** | **-2573.752490** | **-2574.086968** |
| **Kr** | **-2753.377168** | **-2753.732946** |
| **H$^+$** | **0.000000** | **0.000000** |
| **He$^+$** | **-1.994653** | **-1.995343** |
| **Li$^+$** | **-7.259554** | **-7.277675** |
| **Be$^+$** | **-14.302922** | **-14.327812** |
| **B$^+$** | **-24.297925** | **-24.331166** |
| **C$^+$** | **-37.381042** | **-37.421461** |
| **N$^+$** | **-54.002660** | **-54.049242** |
| **O$^+$** | **-74.507857** | **-74.559278** |
| **F$^+$** | **-99.040081** | **-99.099357** |
| **Ne$^+$** | **-128.086040** | **-128.153976** |
| **Na$^+$** | **-161.989707** | **-162.067941** |
| **Mg$^+$** | **-199.691112** | **-199.782050** |
| **Al$^+$** | **-242.032588** | **-242.137215** |
| **Si$^{iv}$** | **-288.954158** | **-289.065070** |
| **P$^+$** | **-340.752743** | **-340.868756** |
| **S$^+$** | **-397.592790** | **-397.713257** |
| **Cl$^+$** | **-459.522650** | **-459.654105** |
| **Ar$^+$** | **-526.794396** | **-526.935616** |
| **K$^+$** | **-599.569298** | **-599.719642** |
| **Ca$^+$** | **-677.149310** | **-677.308762** |
| **Sc$^+$** | **-760.195900** | **-760.348714** |
| **Ti$^+$** | **-848.924746** | **-849.067439** |
| **V$^+$** | **-943.467770** | **-943.628270** |
| **Cr$^+$** | **-1043.982591** | **-1044.148776** |
| **Mn$^+$** | **-1150.510458** | **-1150.685719** |
| **Fe$^+$** | **-1263.174330** | **-1263.359822** |
| **Co$^+$** | **-1382.246441** | **-1382.436845** |
| **Ni$^+$** | **-1507.794637** | **-1507.991074** |
| Cu⁺ | -1639.999421 | -1640.205201 |
| Zn⁺ | -1778.834804 | -1779.061726 |
| Ga⁺ | -1924.338979 | -1924.589254 |
| Ge⁺ | -2076.351503 | -2076.624036 |
| As⁺ | -2235.143260 | -2235.435606 |
| Se⁺ | -2400.801369 | -2401.112315 |
| Br⁺ | -2573.317005 | -2573.649728 |
| Kr⁺ | -2752.861048 | -2753.214417 |
| H⁻ | -0.527258 | -0.531375 |
| He⁻ | -2.816682 | -2.826480 |
| Li⁻ | -7.483954 | -7.502931 |
| Be⁻ | -14.635099 | -14.658347 |
| B⁻ | -24.640703 | -24.666386 |
| C⁻ | -37.864173 | -37.892594 |
| N⁻ | -54.555925 | -54.593095 |
| O⁻ | -75.092218 | -75.136275 |
| F⁻ | -99.824613 | -99.875646 |
| Ne⁻ | -128.752701 | -128.794287 |
| Na⁻ | -162.207385 | -162.287423 |
| Mg⁻ | -199.957414 | -200.046316 |
| Al⁻ | -242.277573 | -242.372534 |
| Si⁻ | -289.310115 | -289.409321 |
| P⁻ | -341.170960 | -341.282337 |
| S⁻ | -398.056350 | -398.178414 |
| Cl⁻ | -460.133624 | -460.265608 |
| Ar⁻ | -527.293660 | -527.434789 |
| K⁻ | -599.749983 | -599.900159 |
| Ca⁻ | -677.372363 | -677.527946 |
| Sc⁻ | -760.439764 | -760.585457 |
| Ti⁻ | -849.188409 | -849.340782 |
| V⁻ | -943.744293 | -943.898893 |
| Cr⁻ | -1044.268133 | -1044.427528 |
| Mn⁻ | -1150.786348 | -1150.949875 |
| Fe⁻ | -1263.495807 | -1263.673291 |
| **Co⁻** | **-1382.578609** | **-1382.755184** |
| **Ni⁻** | **-1508.137905** | **-1508.319838** |
| **Cu⁻** | **-1640.351176** | **-1640.542380** |
| **Zn⁻** | **-1779.172768** | **-1779.393422** |
| **Ga⁻** | **-1924.579712** | **-1924.823791** |
| **Ge⁻** | **-2076.695421** | **-2076.960101** |
| **As⁻** | **-2235.541883** | **-2235.831554** |
| **Se⁻** | **-2401.239376** | **-2401.551668** |
| **Br⁻** | **-2573.881223** | **-2574.215571** |
| **Kr⁻** | **-2753.321454** | **-2753.675209** |

**Table 12.** Computed PBE electronic energies for the 36 atoms, 36 cations and 36 anions, in a.u, using different densities (the second acronym signifies the method used to compute the density).

| Atom | PBE(SVWN) | PBE(HF) | PBE(PBE0) |
|---|---|---|---|
| H | -0.499553 | -0.499426 | -0.499920 |
| He | -2.892167 | -2.891479 | -2.892714 |
| Li | -7.460938 | -7.460233 | -7.461835 |
| Be | -14.628540 | -14.627513 | -14.629577 |
| B | -24.610280 | -24.607962 | -24.611717 |
| C | -37.796474 | -37.793183 | -37.798036 |
| N | -54.533395 | -54.529813 | -54.534910 |
| O | -75.012253 | -75.007205 | -75.013890 |
| F | -99.673232 | -99.666853 | -99.674731 |
| Ne | -128.863508 | -128.856435 | -128.864786 |
| Na | -162.166104 | -162.163366 | -162.167738 |
| Mg | -199.949144 | -199.945963 | -199.950906 |
| Al | -242.233718 | -242.226114 | -242.234183 |
| Si | -289.231263 | -289.223418 | -289.231955 |
| P | -341.112754 | -341.105982 | -341.113553 |
| S | -397.949408 | -397.940908 | -397.950451 |
| Cl | -459.970834 | -459.962249 | -459.971925 |
| Ar | -527.341869 | -527.335100 | -527.343014 |
| K | -599.702138 | -599.698698 | -599.704353 |
| Ca | -677.340387 | -677.336399 | -677.342758 |
| Sc | -760.395061 | -760.381968 | -760.395250 |
| Ti | -849.118460 | -849.102004 | -849.118183 |
| V | -943.668956 | -943.643748 | -943.667843 |
| Cr | -1044.224737 | -1044.203231 | -1044.224316 |
| Mn | -1150.749802 | -1150.728807 | -1150.749741 |
| Fe | -1263.449887 | -1263.410982 | -1263.438227 |
| Co | -1382.518505 | -1382.458306 | -1382.487196 |
| Ni | -1508.076456 | -1508.039824 | -1508.075291 |
| Cu | -1640.292637 | -1640.253854 | -1640.291106 |
| **Zn** | -1779.176631 | -1779.140607 | -1779.175548 |
| **Ga** | -1924.560226 | -1924.537229 | -1924.559280 |
| **Ge** | -2076.648379 | -2076.628598 | -2076.647844 |
| **As** | -2235.519356 | -2235.503080 | -2235.519174 |
| **Se** | -2401.175928 | -2401.159149 | -2401.175999 |
| **Br** | -2573.773821 | -2573.757939 | -2573.774019 |
| **Kr** | -2753.404843 | -2753.391285 | -2753.405133 |
| **H$^+$** | 0.000000 | 0.000000 | 0.000000 |
| **He$^+$** | -1.993182 | -1.993080 | -1.993671 |
| **Li$^+$** | -7.255792 | -7.255360 | -7.256570 |
| **Be$^+$** | -14.297949 | -14.297292 | -14.298998 |
| **B$^+$** | -24.291840 | -24.290977 | -24.293045 |
| **C$^+$** | -37.372510 | -37.370369 | -37.374062 |
| **N$^+$** | -53.992248 | -53.989341 | -53.993909 |
| **O$^+$** | -74.495775 | -74.493078 | -74.497443 |
| **F$^+$** | -99.024483 | -99.020630 | -99.026318 |
| **Ne$^+$** | -128.067600 | -128.062655 | -128.069336 |
| **Na$^+$** | -161.969312 | -161.967594 | -161.970963 |
| **Mg$^+$** | -199.669464 | -199.666860 | -199.671210 |
| **Al$^+$** | -242.010539 | -242.004319 | -242.010710 |
| **Si$^+$** | -288.930184 | -288.922444 | -288.930753 |
| **P$^+$** | -340.727384 | -340.719403 | -340.728163 |
| **S$^+$** | -397.566393 | -397.559971 | -397.567346 |
| **Cl$^+$** | -459.494240 | -459.486055 | -459.495439 |
| **Ar$^+$** | -526.764705 | -526.756617 | -526.766012 |
| **K$^+$** | -599.538859 | -599.536418 | -599.541035 |
| **Ca$^+$** | -677.117484 | -677.114058 | -677.119768 |
| **Sc$^+$** | -760.164725 | -760.147632 | -760.165445 |
| **Ti$^+$** | -848.877624 | -848.860378 | -848.877876 |
| **V$^+$** | -943.438335 | -943.424710 | -943.438647 |
| **Cr$^+$** | -1043.957884 | -1043.943061 | -1043.957931 |
| **Mn$^+$** | -1150.486723 | -1150.470254 | -1150.486801 |
| **Fe$^+$** | -1263.151424 | -1263.131467 | -1263.151516 |
| **Co⁺** | -1382.229847 | -1382.205092 | -1382.229541 |
| **Ni⁺** | -1507.781972 | -1507.756159 | -1507.781331 |
| **Cu⁺** | -1639.993335 | -1639.966129 | -1639.992440 |
| **Zn⁺** | -1778.832096 | -1778.802790 | -1778.831383 |
| **Ga⁺** | -1924.339795 | -1924.318896 | -1924.338831 |
| **Ge⁺** | -2076.357464 | -2076.338078 | -2076.356936 |
| **As⁺** | -2235.155278 | -2235.137746 | -2235.155058 |
| **Se⁺** | -2400.820510 | -2400.806064 | -2400.820564 |
| **Br⁺** | -2573.339566 | -2573.323996 | -2573.339814 |
| **Kr⁺** | -2752.888641 | -2752.873601 | -2752.888986 |
| **H⁻** | -0.527711 | -0.526137 | -0.527537 |
| **He⁻** | -2.852729 | -2.851706 | -2.853498 |
| **Li⁻** | -7.480306 | -7.478857 | -7.481008 |
| **Be⁻** | -14.634630 | -14.630615 | -14.636049 |
| **B⁻** | -24.635494 | -24.628833 | -24.636204 |
| **C⁻** | -37.855908 | -37.848478 | -37.856744 |
| **N⁻** | -54.547386 | -54.513941 | -54.548334 |
| **O⁻** | -75.078617 | -75.066755 | -75.079299 |
| **F⁻** | -99.808251 | -99.795227 | -99.808662 |
| **Ne⁻** | -128.715650 | -128.706785 | -128.716994 |
| **Na⁻** | -162.185998 | -162.182532 | -162.187635 |
| **Mg⁻** | -199.937942 | -199.933725 | -199.939794 |
| **Al⁻** | -242.257631 | -242.248783 | -242.257735 |
| **Si⁻** | -289.286880 | -289.278679 | -289.287300 |
| **P⁻** | -341.146492 | -341.136762 | -341.147075 |
| **S⁻** | -398.029485 | -398.019323 | -398.030102 |
| **Cl⁻** | -460.105380 | -460.096453 | -460.105992 |
| **Ar⁻** | -527.271836 | -527.264585 | -527.273199 |
| **K⁻** | -599.718893 | -599.715351 | -599.721303 |
| **Ca⁻** | -677.341371 | -677.336237 | -677.343948 |
| **Sc⁻** | -760.406151 | -760.371326 | -760.405252 |
| **Ti⁻** | -849.156566 | -849.122011 | -849.155865 |
| **V⁻** | -943.713743 | -943.679668 | -943.712950 |
| **Cr⁻** | -1044.241299 | -1044.209742 | -1044.240189 |
| **Mn⁻** | -1150.768351 | -1150.736710 | -1150.752466 |
| **Fe⁻** | -1263.469111 | -1263.428400 | -1263.467813 |
| **Co⁻** | -1382.558195 | -1382.503162 | -1382.556363 |
| **Ni⁻** | -1508.121119 | -1508.061741 | -1508.118868 |
| **Cu⁻** | -1640.339993 | -1640.289002 | -1640.337514 |
| **Zn⁻** | -1779.173810 | -1779.137583 | -1779.173005 |
| **Ga⁻** | -1924.580624 | -1924.556108 | -1924.579261 |
| **Ge⁻** | -2076.701911 | -2076.681506 | -2076.701091 |
| **As⁻** | -2235.552845 | -2235.533328 | -2235.552348 |
| **Se⁻** | -2401.254587 | -2401.236272 | -2401.254295 |
| **Br⁻** | -2573.902377 | -2573.886548 | -2573.902225 |
| **Kr⁻** | -2753.355758 | -2753.342025 | -2753.356355 |

**Table 13.** IPs computed with PBE using alternative densities, and density sensitivity of IPs in eV.

| Atom | IP PBE(SVWN) | IP PBE(HF) | IP PBE(PBE0) | $S^{IP}_{PBE(SVWN,HF)}$ | $S^{IP}_{PBE(SVWN,PBE0)}$ |
|---|---|---|---|---|---|
| H | 13.59 | 13.59 | 13.60 | 0.0035 | 0.0100 |
| He | 24.46 | 24.45 | 24.46 | 0.0159 | 0.0016 |
| Li | 5.58 | 5.57 | 5.59 | 0.0074 | 0.0032 |
| Be | 9.00 | 8.99 | 9.00 | 0.0101 | 0.0003 |
| B | 8.67 | 8.63 | 8.67 | 0.0396 | 0.0063 |
| C | 11.54 | 11.51 | 11.54 | 0.0313 | 0.0003 |
| N | 14.73 | 14.71 | 14.72 | 0.0184 | 0.0040 |
| O | 14.05 | 13.99 | 14.05 | 0.0640 | 0.0009 |
| F | 17.65 | 17.58 | 17.64 | 0.0688 | 0.0092 |
| Ne | 21.66 | 21.60 | 21.65 | 0.0579 | 0.0124 |
| Na | 5.35 | 5.33 | 5.35 | 0.0278 | 0.0005 |
| Mg | 7.61 | 7.59 | 7.61 | 0.0157 | 0.0004 |
| Al | 6.07 | 6.04 | 6.08 | 0.0376 | 0.0080 |
| Si | 8.19 | 8.19 | 8.20 | 0.0029 | 0.0034 |
| P | 10.49 | 10.52 | 10.49 | 0.0329 | 0.0005 |
| S | 10.42 | 10.37 | 10.42 | 0.0566 | 0.0024 |
| Cl | 12.97 | 12.96 | 12.97 | 0.0109 | 0.0029 |
| Ar | 15.71 | 15.74 | 15.70 | 0.0359 | 0.0044 |
| K | 4.44 | 4.42 | 4.44 | 0.0272 | 0.0011 |
| Ca | 6.07 | 6.05 | 6.07 | 0.0153 | 0.0023 |
| Sc | 6.27 | 6.38 | 6.25 | 0.1089 | 0.0144 |
| Ti | 6.55 | 6.57 | 6.54 | 0.0215 | 0.0144 |
| V | 6.28 | 5.96 | 6.24 | 0.3152 | 0.0388 |
| Cr | 7.26 | 7.08 | 7.25 | 0.1819 | 0.0128 |
| Mn | 7.16 | 7.04 | 7.15 | 0.1232 | 0.0038 |
| Fe | 8.12 | 7.61 | 7.80 | 0.5156 | 0.3198 |
| Co | 7.85 | 6.89 | 7.01 | 0.9644 | 0.8436 |
| Ni | 8.01 | 7.72 | 8.00 | 0.2944 | 0.0143 |
| Cu | 8.14 | 7.83 | 8.13 | 0.3150 | 0.0173 |
| Zn | 9.38 | 9.19 | 9.37 | 0.1828 | 0.0101 |
| Ga | 6.00 | 5.94 | 6.00 | 0.0571 | 0.0005 |
| Ge | 7.92 | 7.91 | 7.92 | 0.0108 | 0.0002 |
| As | 9.91 | 9.94 | 9.91 | 0.0342 | 0.0010 |
| Se | 9.67 | 9.61 | 9.67 | 0.0635 | 0.0005 |
| Br | 11.82 | 11.81 | 11.82 | 0.0085 | 0.0014 |
| Kr | 14.05 | 14.09 | 14.05 | 0.0403 | 0.0015 |

**Table 14.** EAs computed with PBE using alternative densities, and density sensitivity of EAs in eV.

| Atom | EA PBE(SVWN) | EA PBE(HF) | EA PBE(PBE0) | $S^{EA}_{PBE(SVWN,HF)}$ | $S^{EA}_{PBE(SVWN,PBE0)}$ |
|---|---|---|---|---|---|
| H | 0.77 | 0.73 | 0.75 | 0.0394 | 0.0147 |
| He | -1.07 | -1.08 | -1.07 | 0.0091 | 0.0060 |
| Li | 0.53 | 0.51 | 0.52 | 0.0202 | 0.0053 |
| Be | 0.17 | 0.08 | 0.18 | 0.0813 | 0.0104 |
| B | 0.69 | 0.57 | 0.67 | 0.1182 | 0.0198 |
| C | 1.62 | 1.50 | 1.60 | 0.1126 | 0.0198 |
| N | 0.38 | -0.43 | 0.37 | 0.8126 | 0.0154 |
| O | 1.81 | 1.62 | 1.78 | 0.1854 | 0.0260 |
| F | 3.67 | 3.49 | 3.64 | 0.1808 | 0.0296 |
| Ne | -4.02 | -4.07 | -4.02 | 0.0488 | 0.0018 |
| Na | 0.54 | 0.52 | 0.54 | 0.0198 | 0.0001 |
| Mg | -0.30 | -0.33 | -0.30 | 0.0282 | 0.0025 |
| Al | 0.65 | 0.62 | 0.64 | 0.0339 | 0.0098 |
| Si | 1.51 | 1.50 | 1.51 | 0.0097 | 0.0074 |
| P | 0.92 | 0.84 | 0.91 | 0.0805 | 0.0059 |
| S | 2.18 | 2.13 | 2.17 | 0.0452 | 0.0116 |
| Cl | 3.66 | 3.65 | 3.65 | 0.0093 | 0.0130 |
| Ar | -1.91 | -1.92 | -1.90 | 0.0131 | 0.0059 |
| K | 0.46 | 0.45 | 0.46 | 0.0028 | 0.0053 |
| Ca | 0.03 | 0.00 | 0.03 | 0.0312 | 0.0056 |
| Sc | 0.30 | -0.29 | 0.27 | 0.5913 | 0.0296 |
| Ti | 1.04 | 0.54 | 1.03 | 0.4925 | 0.0115 |
| V | 1.22 | 0.98 | 1.23 | 0.2413 | 0.0087 |
| Cr | 0.45 | 0.18 | 0.43 | 0.2735 | 0.0187 |
| Mn | 0.50 | 0.22 | 0.07 | 0.2897 | 0.4306 |
| Fe | 0.52 | 0.47 | 0.81 | 0.0491 | 0.2820 |
| Co | 1.08 | 1.22 | 1.88 | 0.1405 | 0.8021 |
| Ni | 1.22 | 0.60 | 1.19 | 0.6189 | 0.0295 |
| Cu | 1.29 | 0.96 | 1.26 | 0.3322 | 0.0258 |
| Zn | -0.08 | -0.08 | -0.07 | 0.0055 | 0.0076 |
| Ga | 0.56 | 0.51 | 0.54 | 0.0413 | 0.0113 |
| Ge | 1.46 | 1.44 | 1.45 | 0.0170 | 0.0077 |
| As | 0.91 | 0.82 | 0.90 | 0.0882 | 0.0086 |
| Se | 2.14 | 2.10 | 2.13 | 0.0418 | 0.0099 |
| Br | 3.50 | 3.50 | 3.49 | 0.0014 | 0.0095 |
| Kr | -1.34 | -1.34 | -1.33 | 0.0048 | 0.0084 |

Table 15. Mean signed error (MSE), mean absolute error (MAE) standard deviation (STD) of signed error (SE), maximum error (Max), minimum error (Min), slope, intercept and $R^2$ between the computed IPs of the listed methods and the experimental IPs of Table S1.

| Method | MSE | MAE | STD of SE | Max | Min | slope | intercept | $R^2$ |
|---|---|---|---|---|---|---|---|---|
| B97-1 | -0.03 | 0.12 | 0.17 | 0.24 (O) | -0.65 (Co) | 0.990 | 0.127 | 0.999 |
| B98 | 0.01 | 0.13 | 0.17 | 0.27 (B) | -0.60 (Co) | 0.987 | 0.124 | 0.999 |
| BMK | -0.05 | 0.13 | 0.17 | 0.37 (O) | -0.50 (Be) | 0.981 | 0.244 | 0.999 |
| X3LYP | 0.07 | 0.13 | 0.16 | 0.46 (O) | -0.25 (Be) | 0.991 | 0.022 | 0.999 |
| PW6B95D3 | 0.00 | 0.13 | 0.17 | 0.30 (O) | -0.46 (V) | 0.981 | 0.185 | 0.999 |
| CAM-B3LYP | 0.07 | 0.15 | 0.18 | 0.48 (O) | -0.35 (V) | 0.986 | 0.072 | 0.999 |
| MN15-L | -0.09 | 0.15 | 0.19 | 0.48 (He) | -0.57 (V) | 0.985 | 0.233 | 0.999 |
| wB97XD | -0.07 | 0.15 | 0.20 | 0.26 (B) | -0.64 (Co) | 0.980 | 0.268 | 0.998 |
| MN15 | -0.06 | 0.15 | 0.22 | 0.32 (He) | -0.89 (V) | 0.976 | 0.292 | 0.998 |
| B97-2 | -0.07 | 0.15 | 0.21 | 0.26 (Mn) | -0.81 (Co) | 0.982 | 0.251 | 0.998 |
| O3LYP | -0.06 | 0.15 | 0.22 | 0.27 (He) | -0.86 (V) | 0.979 | 0.268 | 0.998 |
| B3LYP | 0.11 | 0.16 | 0.17 | 0.52 (O) | -0.22 (V) | 0.988 | 0.009 | 0.999 |
| B1B95 | -0.12 | 0.16 | 0.19 | 0.14 (B) | -0.68 (V) | 0.982 | 0.297 | 0.999 |
| VSXC | -0.01 | 0.16 | 0.23 | 0.35 (N) | -0.66 (Ti) | 0.976 | 0.253 | 0.998 |
| HSE06 | -0.03 | 0.16 | 0.21 | 0.36 (B) | -0.62 (V) | 0.995 | 0.080 | 0.998 |
| RevPBE0 | -0.02 | 0.16 | 0.22 | 0.36 (B) | -0.63 (V) | 0.994 | 0.083 | 0.998 |
| PBE0 | -0.04 | 0.16 | 0.22 | 0.35 (B) | -0.67 (V) | 0.994 | 0.103 | 0.998 |
| APFD | -0.01 | 0.17 | 0.22 | 0.38 (B) | -0.68 (V) | 0.990 | 0.115 | 0.998 |
| TPSS | 0.00 | 0.17 | 0.23 | 0.47 (B) | -0.58 (V) | 0.988 | 0.125 | 0.998 |
| N12-SX | -0.07 | 0.17 | 0.24 | 0.32 (F) | -0.86 (Co) | 0.976 | 0.302 | 0.998 |
| BHandHLYP | -0.12 | 0.17 | 0.19 | 0.28 (B) | -0.71 (Co) | 0.992 | 0.197 | 0.998 |
| TPSSh | -0.04 | 0.17 | 0.24 | 0.46 (B) | -0.73 (V) | 0.986 | 0.176 | 0.997 |
| BLYP | 0.04 | 0.18 | 0.22 | 0.53 (O) | -0.35 (Be) | 0.992 | 0.043 | 0.998 |
| mPW1PW91 | -0.02 | 0.18 | 0.24 | 0.39 (B) | -0.71 (V) | 0.988 | 0.140 | 0.997 |
| tHCTHhyb | 0.09 | 0.18 | 0.20 | 0.52 (Mn) | -0.54 (Co) | 0.985 | 0.053 | 0.998 |
| PBE | 0.05 | 0.18 | 0.25 | 0.50 (Cr) | -0.73 (Ti) | 0.994 | 0.014 | 0.997 |
| M06 | -0.04 | 0.18 | 0.22 | 0.53 (Fe) | -0.53 (V) | 0.981 | 0.223 | 0.998 |
| RevTPSS | -0.03 | 0.18 | 0.26 | 0.46 (B) | -0.82 (Ti) | 0.988 | 0.156 | 0.997 |
| mPW3PBE | 0.03 | 0.19 | 0.23 | 0.42 (B) | -0.66 (V) | 0.987 | 0.103 | 0.998 |
| RevPBE | 0.07 | 0.19 | 0.26 | 0.56 (Cr) | -0.68 (Ti) | 0.995 | -0.015 | 0.997 |
| wPBEhPBE | 0.07 | 0.19 | 0.26 | 0.56 (Cr) | -0.68 (Ti) | 0.995 | -0.015 | 0.997 |
| B97D | 0.05 | 0.19 | 0.28 | 0.87 (Mn) | -0.54 (Co) | 0.984 | 0.111 | 0.996 |
| B3P86 | 0.13 | 0.19 | 0.19 | 0.48 (B) | -0.45 (V) | 0.989 | -0.017 | 0.998 |
| OP86 | 0.02 | 0.20 | 0.27 | 0.34 (N) | -0.87 (Ti) | 0.979 | 0.185 | 0.997 |
| OLYP | -0.14 | 0.20 | 0.28 | 0.20 (He) | -1.09 (Co) | 0.978 | 0.358 | 0.997 |
| G96PBE | -0.06 | 0.22 | 0.33 | 0.45 (Cr) | -0.96 (Ti) | 0.978 | 0.281 | 0.995 |
| M06-L | -0.14 | 0.23 | 0.27 | 0.32 (B) | -0.74 (Ti) | 0.979 | 0.350 | 0.997 |
| BP86 | 0.16 | 0.23 | 0.24 | 0.64 (Cr) | -0.55 (Ti) | 0.989 | -0.054 | 0.997 |
| OPBE | -0.13 | 0.25 | 0.35 | 0.31 (N) | -1.20 (Ti) | 0.975 | 0.377 | 0.995 |
| BHandH | -0.26 | 0.26 | 0.19 | 0.04 (B) | -0.76 (Co) | 1.007 | 0.194 | 0.998 |
| HCTH407 | 0.20 | 0.27 | 0.26 | 0.73 (Fe) | -0.65 (Co) | 0.984 | -0.036 | 0.997 |
| tHCTH | 0.14 | 0.27 | 0.35 | 1.06 (B) | -0.68 (Co) | 0.978 | 0.087 | 0.995 |
| LC-wHPBE | -0.04 | 0.27 | 0.35 | 0.51 (O) | -1.14 (V) | 0.966 | 0.377 | 0.995 |
| SVWN5 | 0.19 | 0.29 | 0.30 | 0.70 (Cr) | -0.57 (H) | 0.990 | -0.085 | 0.996 |
| M11-L | 0.05 | 0.30 | 0.37 | 0.69 (Kr) | -0.83 (Ti) | 0.998 | -0.032 | 0.993 |
| SLYP | -0.24 | 0.30 | 0.34 | 0.36 (Cu) | -1.34 (He) | 1.018 | 0.062 | 0.995 |
| B2PLYP | -0.32 | 0.34 | 0.22 | 0.14 (B) | -0.86 (Co) | 1.004 | 0.287 | 0.998 |
| M11 | -0.16 | 0.37 | 0.49 | 0.62 (O) | -1.41 (V) | 0.945 | 0.704 | 0.992 |
| SVWN | 0.65 | 0.65 | 0.32 | 1.17 (Ne) | -0.09 (H) | 0.982 | -0.457 | 0.995 |
| OVWN | 0.77 | 0.79 | 0.37 | 1.75 (He) | -0.31 (Ti) | 0.945 | -0.181 | 0.997 |

Table 16. Mean signed error (MSE), mean absolute error (MAE) standard deviation (STD) of signed error (SE), maximum error (Max), minimum error (Min), slope, intercept and $R^2$ between the computed EAs of the listed methods and the experimental EAs of Table S1.

| Method | MSE | MAE | STD of SE | Max | Min | slope | intercept | $R^2$ |
|---|---|---|---|---|---|---|---|---|
| B98 | 0.05 | 0.12 | 0.16 | 0.47 (Co) | -0.27 (Sc) | 0.975 | -0.018 | 0.975 |
| B97-1 | 0.04 | 0.13 | 0.19 | 0.53 (Ti) | -0.31 (Sc) | 0.991 | -0.029 | 0.964 |
| PW6B95D3 | 0.02 | 0.13 | 0.19 | 0.52 (Ti) | -0.36 (Sc) | 0.972 | 0.013 | 0.967 |
| APFD | 0.02 | 0.13 | 0.18 | 0.50 (V) | -0.36 (Cr) | 0.984 | -0.005 | 0.967 |
| TPSS | 0.09 | 0.13 | 0.21 | 0.88 (Ti) | -0.30 (Cr) | 1.002 | -0.088 | 0.955 |
| X3LYP | 0.08 | 0.14 | 0.19 | 0.57 (Ti) | -0.27 (Sc) | 0.979 | -0.055 | 0.967 |
| BMK | -0.09 | 0.14 | 0.21 | 0.41 (Co) | -0.87 (Sc) | 0.906 | 0.183 | 0.965 |
| RevTPSS | 0.10 | 0.15 | 0.23 | 0.87 (Ti) | -0.26 (Cr) | 1.027 | -0.132 | 0.950 |
| B97-2 | 0.03 | 0.15 | 0.21 | 0.61 (Ti) | -0.26 (Sc) | 0.997 | -0.026 | 0.956 |
| TPSSh | 0.05 | 0.15 | 0.23 | 0.76 (Ti) | -0.34 (Cr) | 1.014 | -0.068 | 0.949 |
| O3LYP | 0.04 | 0.15 | 0.24 | 0.79 (Ti) | -0.25 (Ni) | 0.991 | -0.028 | 0.943 |
| CAM-B3LYP | 0.03 | 0.15 | 0.19 | 0.43 (Ti) | -0.51 (Sc) | 0.943 | 0.031 | 0.968 |
| wB97XD | 0.06 | 0.15 | 0.24 | 0.88 (Fe) | -0.49 (Sc) | 0.963 | -0.017 | 0.943 |
| PBE0 | 0.02 | 0.16 | 0.22 | 0.62 (Ti) | -0.37 (Cr) | 1.001 | -0.025 | 0.954 |
| RevPBE0 | 0.03 | 0.16 | 0.21 | 0.63 (Ti) | -0.36 (Cr) | 1.001 | -0.033 | 0.955 |
| HSE06 | 0.03 | 0.16 | 0.22 | 0.63 (Ti) | -0.36 (Cr) | 1.003 | -0.037 | 0.954 |
| B1B95 | -0.07 | 0.16 | 0.19 | 0.47 (Ti) | -0.41 (Sc) | 0.978 | 0.091 | 0.963 |
| mPW1PW91 | 0.03 | 0.16 | 0.22 | 0.66 (Ti) | -0.40 (Cr) | 0.995 | -0.025 | 0.950 |
| M06-L | -0.02 | 0.17 | 0.22 | 0.59 (Ti) | -0.34 (F) | 1.044 | -0.025 | 0.953 |
| B3LYP | 0.13 | 0.17 | 0.20 | 0.64 (Ti) | -0.20 (Sc) | 0.976 | -0.098 | 0.963 |
| mPW3PBE | 0.10 | 0.17 | 0.23 | 0.75 (Ti) | -0.31 (Cr) | 0.986 | -0.084 | 0.949 |
| M06 | 0.00 | 0.17 | 0.25 | 0.77 (Co) | -0.45 (Fe) | 0.965 | 0.039 | 0.939 |
| VSXC | 0.11 | 0.17 | 0.24 | 0.73 (Ti) | -0.40 (Fe) | 0.960 | -0.065 | 0.947 |
| G96PBE | 0.04 | 0.17 | 0.26 | 0.76 (Ti) | -0.59 (Cr) | 0.968 | -0.002 | 0.934 |
| BLYP | 0.10 | 0.18 | 0.22 | 0.70 (Ti) | -0.26 (Ca) | 0.969 | -0.067 | 0.952 |
| MN15 | 0.13 | 0.18 | 0.24 | 0.79 (V) | -0.30 (Sc) | 1.016 | -0.150 | 0.945 |
| OLYP | 0.03 | 0.18 | 0.27 | 0.84 (Ti) | -0.23 (Cu) | 0.980 | -0.011 | 0.926 |
| OPBE | 0.04 | 0.18 | 0.29 | 0.93 (Ti) | -0.43 (Cr) | 0.987 | -0.022 | 0.919 |
| LC-wHPBE | -0.01 | 0.19 | 0.26 | 0.53 (Ti) | -0.50 (Ni) | 0.941 | 0.077 | 0.935 |
| N12-SX | -0.07 | 0.19 | 0.27 | 0.94 (Fe) | -0.44 (Sc) | 0.940 | 0.127 | 0.933 |
| SLYP | 0.00 | 0.20 | 0.25 | 0.59 (Ti) | -0.40 (Ca) | 0.928 | 0.082 | 0.941 |
| PBE | 0.17 | 0.20 | 0.24 | 0.94 (Ti) | -0.25 (Cr) | 0.978 | -0.143 | 0.942 |
| BHandHLYP | -0.19 | 0.21 | 0.18 | 0.11 (Ti) | -0.78 (Sc) | 0.975 | 0.214 | 0.968 |
| B3P86 | 0.20 | 0.21 | 0.18 | 0.68 (V) | -0.07 (Sc) | 1.001 | -0.200 | 0.967 |
| tHCTHhyb | 0.19 | 0.21 | 0.24 | 0.82 (Ti) | -0.09 (Ca) | 0.989 | -0.179 | 0.943 |
| wPBEhPBE | 0.19 | 0.22 | 0.24 | 0.94 (Ti) | -0.24 (Cr) | 0.977 | -0.157 | 0.942 |
| RevPBE | 0.19 | 0.22 | 0.24 | 0.94 (Ti) | -0.24 (Cr) | 0.977 | -0.157 | 0.942 |
| M11 | -0.09 | 0.23 | 0.37 | 0.29 (Br) | -0.98 (Cu) | 0.830 | 0.261 | 0.901 |
| OP86 | 0.20 | 0.23 | 0.25 | 1.06 (Ti) | -0.13 (Cu) | 0.996 | -0.198 | 0.937 |
| B97D | 0.21 | 0.25 | 0.29 | 0.86 (Cr) | -0.15 (Ca) | 1.006 | -0.222 | 0.915 |
| BHandH | -0.25 | 0.26 | 0.19 | 0.08 (Ti) | -0.86 (Sc) | 0.955 | 0.288 | 0.967 |
| MN15-L | 0.13 | 0.30 | 0.43 | 0.86 (V) | -1.65 (Sc) | 0.849 | 0.057 | 0.843 |
| BP86 | 0.25 | 0.30 | 0.29 | 1.04 (Ti) | -0.66 (Fe) | 0.915 | -0.132 | 0.925 |
| tHCTH | 0.30 | 0.31 | 0.32 | 1.18 (Ti) | -0.04 (Ga) | 0.996 | -0.299 | 0.899 |
| HCTH407 | 0.23 | 0.33 | 0.35 | 1.15 (Co) | -0.57 (Ti) | 0.872 | -0.063 | 0.898 |
| SVWN5 | 0.35 | 0.36 | 0.27 | 1.03 (Ti) | -0.05 (Ca) | 0.904 | -0.211 | 0.936 |
| B2PLYP | -0.37 | 0.37 | 0.21 | 0.00 (Ti) | -0.91 (Sc) | 1.012 | 0.360 | 0.957 |
| M11-L | 0.25 | 0.53 | 0.61 | 1.41 (Ti) | -1.36 (K) | 0.753 | 0.083 | 0.709 |
| SVWN | 0.76 | 0.76 | 0.30 | 1.42 (Ti) | 0.31 (Ca) | 0.870 | -0.521 | 0.933 |
| OVWN | 0.63 | 0.87 | 0.81 | 1.60 (Ti) | -3.22 (Fe) | 0.588 | 0.080 | 0.683 |

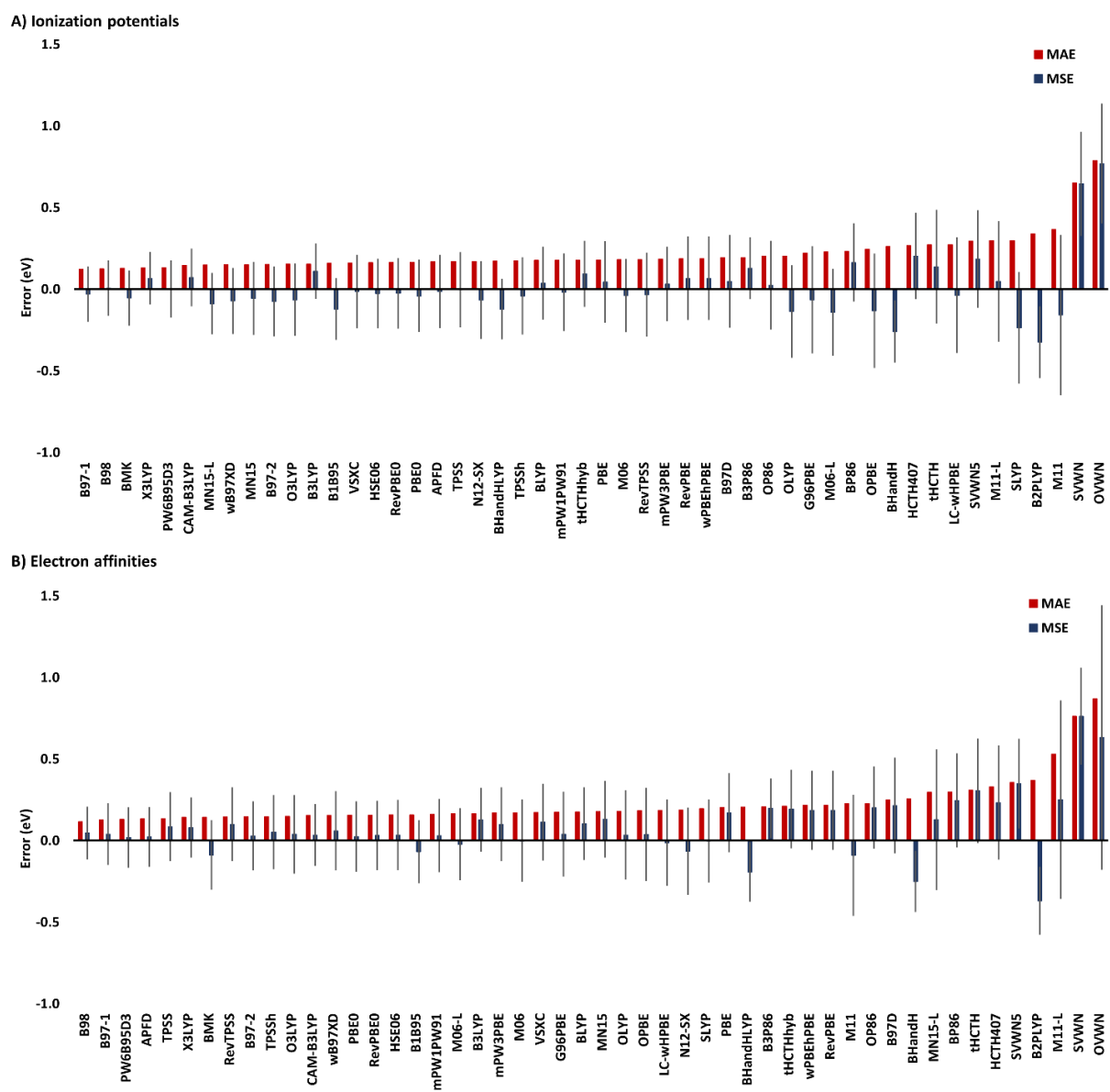

**Figure S1. Errors in computed ionization potentials (A) and electron affinities (B) for the 50 studied density functionals.** Red bars indicate MAEs, black bars indicate MSEs, and thin bars indicate the standard deviations of the errors.

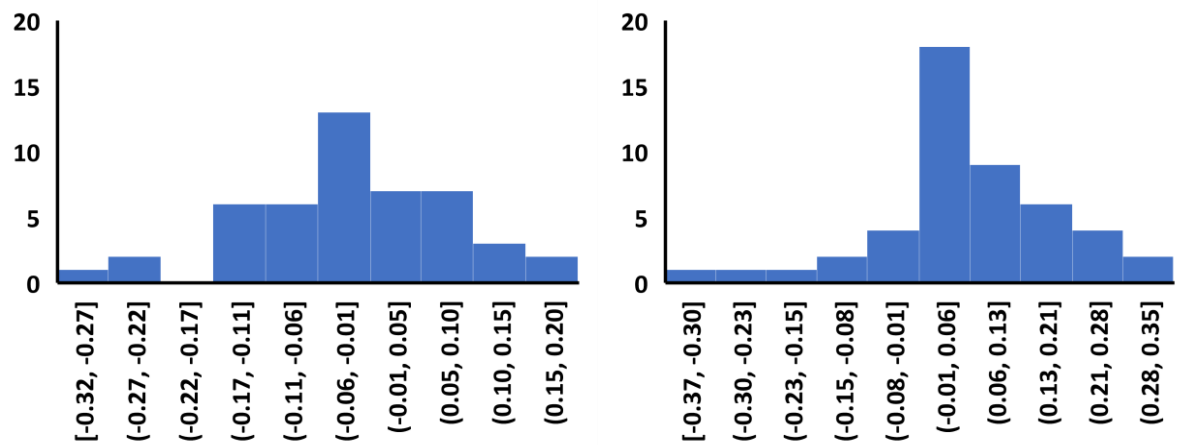

**Figure S2. Distribution of mean signed errors for all functionals except SVWN and OVWN (in eV).** Left: ionization potentials. Right: Electron affinities.

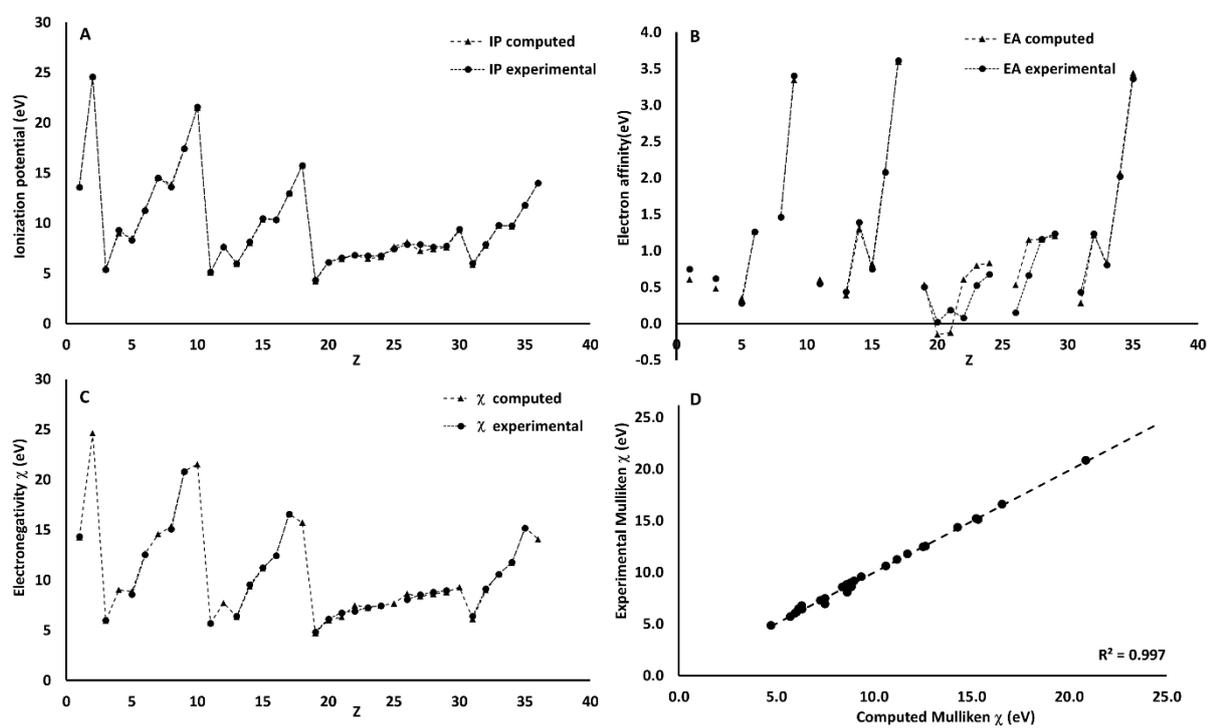

**Figure S3. Performance of B97-1 as example of one of the best functionals. (A)** Computed vs. experimental IPs. **(B)** Computed vs. experimental EAs. **(C)** Computed vs. experimental χ. **(D)** Computed vs. experimental η. All values are in eV.

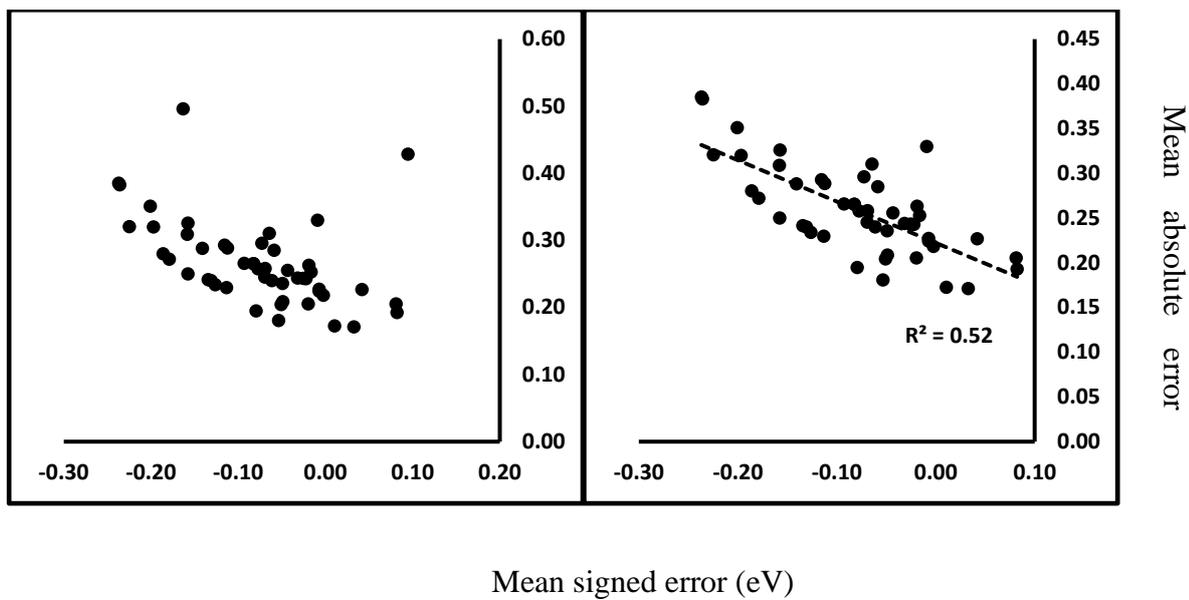

**Figure S4. MAEs plotted against MSEs for hardness.** Left: All functionals. Right: leaving out M11-L and OVWN.

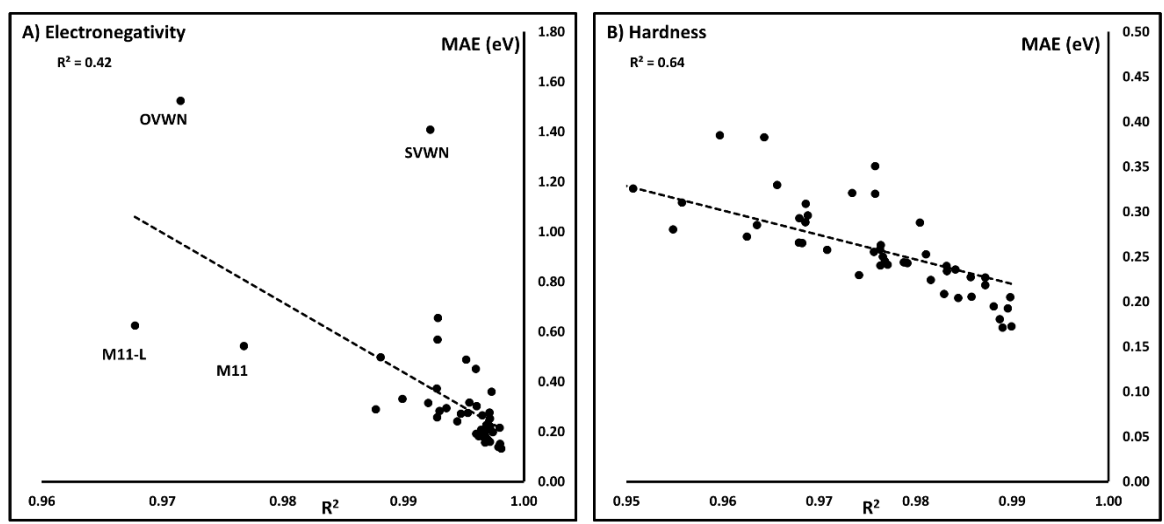

**Figure S5. Trend accuracy vs. absolute accuracy.** A) Relation between trend accuracy ($R^2$ of linear regression) and numerical accuracy (MAE, in eV) for the 50 studied functionals. A) electronegativity; B) absolute hardness.